\documentclass[prd,twocolumn,nofootinbib,superscriptaddress,amsmath,amssymb,groupedaddress]{revtex4-1}
\usepackage{mathtools}
\usepackage{graphics}
\usepackage{graphicx}
\usepackage{dcolumn}
\usepackage{bm}
\usepackage{dsfont} 
\usepackage{amsmath,amssymb}
\usepackage{hyperref}
\usepackage{tabularx}
\usepackage{epsf,epsfig}
\usepackage[normalem]{ulem}
\usepackage[usenames]{color}
\usepackage{multirow}
\usepackage{makecell}
\usepackage{diagbox}
\usepackage[abs]{overpic}
\allowdisplaybreaks
\hypersetup{
    colorlinks=true,
    linkcolor=blue,
    filecolor=magenta,      
    urlcolor=blue,
    citecolor=blue
}
\newcommand{\EdGB}{{\mbox{\tiny EdGB}}}
\newcommand{\NC}{{\mbox{\tiny NC}}}
\newcommand{\ST}{{\mbox{\tiny ST}}}
\newcommand{\MDR}{{\mbox{\tiny MDR}}}
\newcommand{\GR}{{\mbox{\tiny GR}}}
\newcommand{\dCS}{{\mbox{\tiny dCS}}}
\newcommand{\MR}{{\mbox{\tiny MR}}}
\newcommand{\II}{{\mbox{\tiny I}}}

\definecolor{red(ncs)}{rgb}{0.77, 0.01, 0.2}

\usepackage{array}
\newcolumntype{C}[1]{>{\centering\let\newline\\\arraybackslash\hspace{0pt}}m{#1}}

\newcolumntype{C}[1]{>{\centering\arraybackslash}m{#1}}

\begin{document}

\title{Parameterized and inspiral-merger-ringdown consistency tests of gravity with multi-band gravitational wave observations}

\author{Zack Carson}
\author{Kent Yagi}

\affiliation{Department of Physics, University of Virginia, Charlottesville, Virginia 22904, USA}

\date{\today}


\begin{abstract}
The gravitational wave observations of colliding black holes have opened a new window into the unexplored \emph{extreme gravity} sector of physics, where the gravitational fields are immensely strong, non-linear, and dynamical.
10 binary black hole merger events observed so far can be used to test Einstein's theory of general relativity, which has otherwise been proven to agree with observations from several sources in the weak- or static-field regimes.
One interesting future possibility is to detect gravitational waves from GW150914-like stellar-mass black hole binaries with both ground-based and space-based detectors.
We here demonstrate the power of testing extreme gravity with such \emph{multi-band} gravitational-wave observations.
In particular, we consider two theory-agnostic methods to test gravity using gravitational waves.
The first test is the parameterized test where we introduce generic non-Einsteinian corrections to the waveform, which can easily be mapped to parameters in known example theories beyond general relativity.
The second test is the inspiral-merger-ringdown consistency test where one derives the mass and spin of a merger remnant from the inspiral and merger-ringdown independently assuming general relativity is correct, and then check their consistency.
In both cases, we use Fisher analyses and compare the results with Bayesian ones wherever possible.
Regarding the first test, we find that multi-band observations can be crucial in probing certain modified theories of gravity, including those with gravitational parity violation.
Regarding the second test, we show that future single-band detections can improve upon the current tests by roughly three orders of magnitude, and further 7-10 times improvement may be realized with multi-band observations.

\end{abstract}

\maketitle


\section{Introduction}\label{sec:intro}

Einstein's theory of General Relativity (GR), which elegantly relates the geometries of spacetime to the manifestation of gravity, has remained at its post as the prevailing theory of gravity for over 100 years.
Throughout this era, GR has been subject to a plethora of tests in search for minute deviations which may point to alternative theories of gravity.
As pointed out by Popper~\cite{popper}, scientific theories such as GR can never be entirely proven, however alternative theories may be constrained.
When subject to observations on the solar-system scale where gravity is weak and approximately static, 
such as photon-deflection, Shapiro time-delay, perihelion advance of Mercury, and the Nordvedt effect~\cite{Will_SolarSystemTest}, 
GR has passed the tests with flying colors.
Observations concerning the strong-field, static systems of binary pulsar systems~\cite{Stairs_BinaryPulsarTest,Wex_BinaryPulsarTest} similarly proved GR to be entirely consistent.
On the large-scale side, cosmological observations~\cite{Ferreira_CosmologyTest,Clifton_CosmologyTest,Joyce_CosmologyTest,Koyama_CosmologyTest,Salvatelli_CosmologyTest} have also proven Einstein to be correct.
See also Ref.~\cite{Ishak:2018his} for a review on testing general relativity on cosmological scales.

Most recently, a new window into gravity was opened upon the first observation of gravitational waves (GWs) from the coalescence of two black holes (BHs), dubbed GW150914~\cite{GW150914} by the LIGO and Virgo Collaboration (LVC).
Events such as GW150914 allows for the unique opportunity to probe gravity in the strong-field, non-linear, dynamical region (\emph{extreme-gravity}) of the phase-space in question, and indeed when put to the test of GR, this event and the following 10~\cite{GW_Catalogue} have similarly identified no significant deviations from Einstein's theory~\cite{Abbott_IMRcon2,Abbott_IMRcon}.

With such a convincing case for the correctness of GR, why should we continue to search for divergences?
While GR has been successful in its predictions of many observations, there are still multiple unsolved problems in physics which could potentially be solved with alternative theories of gravity.
The unification of GR with quantum mechanics, dark matter and the rotation curves of galaxies, dark energy and the expansion of the universe, and the inflation of the early universe are all prevalent examples of unexplained phenomena which could be attributed to an alternative theory of gravity.
Such a theory could then reduce to GR in the weak-field limit, while being active in the extreme-gravity regime.
For this reason, binary BH inspirals with immense gravitational fields and velocities reaching $\sim50\%$ the speed of light could very well prove to be vital in probing such a theory's untested side.

With such a large array of proposed modified gravity alternatives, how does one go about determining which one most accurately describes nature?
As always, we must rule them out one at a time with experimental observations.
For example, weak-field observations of the solar system and binary pulsar systems have placed very stringent constraints on several scalar-tensor theories~\cite{Jordan1959,Brans1961,Damour1993_2,Damour1993,Damour1996,Damour1992}, as well as the spontaneous scalarization of neutron stars~\cite{Sampson2014,Anderson2016}.
There are some remaining theories that have not yet been strongly constrained -- for example, theories with gravitational parity violation~\cite{Jackiw:2003pm,Yunes_dcs,Alexander_cs} which may not be activated in the weak-curvature systems currently studied.
In this investigation, we study various modified theories of gravity, all of which affect the gravitational waveform with different dependence on the relative velocities of binary constituents, or gravitational wave frequencies.

We currently live in a very exciting era of gravitational wave astronomy.  
With great success in both the past and on-going observing runs, many new ground-based GW interferometers are planned: several upgrades to the current LIGO infrastructure (Advanced LIGO, LIGO A\texttt{+}, LIGO Voyager)~\cite{Ap_Voyager_CE}, as well as new third-generation detectors like Cosmic Explorer (CE)~\cite{Ap_Voyager_CE} and the Einstein Telescope (ET)~\cite{ET}, each with improved sensitivity in the $1-10^4$ Hz range.
While such detectors have been designed with incredible sensitivities that are able to observe millions of events per year, with signal-to-noise ratios (SNRs) on the order of $10^3$~\cite{Zack:URrelations}, they can not probe the sub-unity frequency bands dominated by compact binary early inspirals, supermassive BH binaries, white-dwarf binaries, etc.
Space-based detectors such as LISA~\cite{LISA}, TianQin~\cite{TianQin}, B-DECIGO~\cite{B-DECIGO}, and DECIGO~\cite{DECIGO} on the other hand, have long Mm- to Gm-scale arms which allow them to accurately probe the low-frequency $10^{-4}-1$ Hz portion of the GW spectrum.
While ground-based detectors are proficient at observing the late, high-frequency, high-velocity, merger-ringdown portion of stellar-mass binary BH gravitational waveforms, space-based detectors can probe the early, low-frequency, low velocity inspiral portion.

Shortly after the observation of GW150914, Sesana~\cite{Sesana:2016ljz} pointed out that joint \emph{multi-band} detections of GW150914-like events\footnote{Multi-band GW observations are also possible for more massive binary BHs~\cite{AmaroSeoane:2009ui,Cutler:2019krq} and binary neutron stars~\cite{Isoyama:2018rjb}.} could be made using both LISA and ground-based detectors, with multi-band detection rates on the order of $\mathcal{O}(1)$~\cite{Gerosa:2019dbe,Sesana:2016ljz}.
Such events would first be observed in their early inspiral stage by space-based telescopes, until leaving the space-band at $1$ Hz for LISA or TianQin for several months before entering the ground-band to eventually merge at $\sim300$ Hz.
The early detections by space-based interferometers could give alert to electromagnetic telescopes~\cite{Sesana:2016ljz} for follow-up observations, as well as ground-based detectors, allowing for potential sensitivity optimizations which could be used to improve upon tests of GR~\cite{Tso:2018pdv}.
Similarly, ground-based observations will allow one to revisit sub-threshold space-based data, effectively lowering the detection threshold SNR from 15 to 9~\cite{Wong:2018uwb,Moore:2019pke}, and enhance the overall number of detections~\cite{Moore:2019pke,Cutler:2019krq}.
Additionally, multi-band GW observations can improve upon the measurement accuracy of many binary parameters, specifically the masses and sky positions~\cite{Nair:2015bga,Nair:2018bxj,Vitale:2016rfr,Cutler:2019krq}. 

In this analysis, we demonstrate the improvements one can gain with multi-band observations for different tests of GR~\cite{Nair:2015bga,Barausse:2016eii,Vitale:2016rfr,Carson_multiBandPRL,Gnocchi:2019jzp}.
In particular, this paper is a follow-up of~\cite{Carson_multiBandPRL}, which studied two theory-agnostic ways of testing GR with GWs: (i) parameterized waveform tests~\cite{Yunes:2009ke,Abbott_IMRcon2,Abbott_IMRcon}, and (ii) inspiral-merger-ringdown (IMR) consistency tests~\cite{Ghosh_IMRcon,Ghosh_IMRcon2,Abbott_IMRcon,Abbott_IMRcon2}. 
We extend~\cite{Carson_multiBandPRL} by explaining our analysis in more detail and studying a variety of example theories of gravity, including
Einstein-dilaton Gauss-Bonnet (EdGB) gravity, dynamical Chern-Simons (dCS) gravity, scalar-tensor theories, noncommutative gravities, theories with time-varying $G$, time-varying BH mass or modified dispersion relations.


\subsection{Executive summary of results}\label{sec:summary}

\renewcommand{\arraystretch}{1.2}
\begin{table*}
\centering
\resizebox{\textwidth}{!}{%
\begin{tabular}{|c|c|c||c|c|c|c|c||c|}
\hline
&&&&&&&&\\[-1em]
\multirow{2}{*}{Theory} & \multirow{2}{*}{PN ($b$)} & \multirow{2}{*}{Parameter} & \multirow{2}{*}{CE~\cite{Ap_Voyager_CE}} & TianQin~\cite{TianQin} & LISA~\cite{LISA} & B-DECIGO~\cite{B-DECIGO} & DECIGO~\cite{DECIGO} & \multirow{2}{*}{Current}\\
&&&&&&&&\\[-1em]
 &  & & & \textbf{(+CE)} & \textbf{(+CE)} & \textbf{(+CE)} & \textbf{(+CE)} & \\
\hline
\hline
&&&&&&&&\\[-1em]
\multirow{2}{*}{EdGB~\cite{Maeda:2009uy,Yagi_EdGBmap}} &  \multirow{2}{*}{$-1$ ($-7$)} & \multirow{2}{*}{$\sqrt{\alpha_\EdGB}$ \lbrack km\rbrack} & \multirow{2}{*}{$4.2$} & $0.55$ & $0.72$ & $0.29$ & $0.12$ & \multirow{1}{*}{$10^7$~\cite{Amendola_EdGB}}\\
&&&&&&&& $2-6$~\cite{Kanti_EdGB,Pani_EdGB,Yagi_EdGB,Nair_dCSMap,Yamada:2019zrb,Tahura:2019dgr} \\[-1em]
 & & &  & ($\bm{0.30}$) & ($\bm{0.37}$) & ($\bm{0.25}$) & ($\bm{0.11}$) & \\
\hline
&&&&&&&&\\[-1em]
\multirow{2}{*}{dCS~\cite{Jackiw:2003pm,Alexander_cs,Yagi:2012vf,Nair_dCSMap}} &  \multirow{2}{*}{+2 ($-1$)} & \multirow{2}{*}{$\sqrt{\alpha_\dCS}$ \lbrack km\rbrack} & \multirow{2}{*}{--} & -- & -- & -- & $23$ & \multirow{2}{*}{$10^8$~\cite{AliHaimoud_dCS,Yagi_dCS}}\\
&&&&&&&&\\[-1em]
 & & & & ($\bm{31}$) & ($\bm{31}$) & ($\bm{29}$) & ($\bm{18}$) & \\
\hline
&&&&&&&&\\[-1em]
\multirow{2}{*}{Scalar-tensor~\cite{Horbatsch_STcosmo,Jacobson_STcosmo}} &  \multirow{2}{*}{$-1$ ($-7$)} & \multirow{2}{*}{$\dot \phi$ \lbrack sec$^{-1}$\rbrack} & \multirow{2}{*}{$340$} & $6.1$ & $10.8$ & $1.6$ & $0.25$ & \multirow{2}{*}{$10^{-6}$~\cite{Horbatsch_STcosmo}}\\
&&&&&&&&\\[-1em]
 & & &  & ($\bm{1.8}$) & ($\bm{2.7}$) & ($\bm{1.2}$) & ($\bm{0.21}$) & \\
\hline
&&&&&&&&\\[-1em]
\multirow{2}{*}{Noncommutative~\cite{Harikumar:2006xf,Kobakhidze:2016cqh}}  & \multirow{2}{*}{+2 ($-1$)} & \multirow{2}{*}{$\sqrt{\Lambda}$} & \multirow{2}{*}{$0.74$} & $1.5$ & $1.5$ & $0.83$ & $0.57$ & \multirow{2}{*}{3.5~\cite{Kobakhidze:2016cqh}}\\
&&&&&&&&\\[-1em]
 & & &  & ($\bm{0.66}$) & ($\bm{0.66}$) & ($\bm{0.64}$) & ($\bm{0.50}$) & \\
\hline
&&&&&&&&\\[-1em]
\multirow{2}{*}{Varying $G$~\cite{Will_SEP,Yunes_GdotMap,Tahura_GdotMap}} &  \multirow{2}{*}{$-4$ ($-13$)} & \multirow{2}{*}{$\dot G$ \lbrack yr$^{-1}$\rbrack} & \multirow{2}{*}{$990$} & $1.6\times 10^{-7}$ & $2.3\times 10^{-6}$ & $9.4\times 10^{-8}$ & $3.4\times 1.9\times10^{-9}$ & \multirow{1}{*}{$10^{-13}-10^{-12}$}\\
&&&&&&&& \cite{Bambi_Gdot,Copi_Gdot,Manchester_Gdot,Konopliv_Gdot} \\[-1em]
 & & &  & ($\bm{3.0\times 10^{-8}}$) & ($\bm{5.5\times 10^{-7}}$) & ($\bm{7.2\times 10^{-8}}$) & ($\bm{1.5\times 10^{-9}}$) & \\
\hline
&&&&&&&&\\[-1em]
\multirow{2}{*}{Varying $M$~\cite{Yagi_EDmap,Berti_ModifiedReviewSmall}} & \multirow{2}{*}{$-4$ ($-13$)} & \multirow{2}{*}{$\dot{M}$ \lbrack M$_\odot$ yr$^{-1}$\rbrack} & \multirow{2}{*}{$1.9\times 10^4$} & $5.8\times10^{-6}$ & $9.0\times10^{-5}$ & $3.8\times10^{-6}$ & $7.2\times10^{-8}$ & \multirow{2}{*}{--}\\
&&&&&&&&\\[-1em]
 & &  & & ($\bm{1.2\times 10^{-6}}$) & ($\bm{2.0\times 10^{-5}}$) & ($\bm{2.8\times 10^{-6}}$) & ($\bm{5.9\times 10^{-8}}$) & \\
\hline
&&&&&&&&\\[-1em]
\multirow{1}{*}{Massive graviton~\cite{Zhang:2017jze,Finn:2001qi}} &\multirow{2}{*}{$-3$ ($-11$)} & \multirow{2}{*}{$m_g$ \lbrack eV\rbrack} & \multirow{2}{*}{$6.7\times10^{-16}$} & $1.9\times10^{-19}$ & $3.1\times10^{-19}$ & $6.6\times10^{-20}$ & $9.5\times10^{-21}$ & \multirow{1}{*}{$10^{-14}$~\cite{Chung:2018dxe}}\\
 (dynamical) &&&&&&&& $5.2\times10^{-21}$~\cite{Miao:2019nhf} \\[-1em]
 & & & & ($\bm{9.2\times10^{-20}}$) & ($\bm{1.1\times10^{-19}}$) & ($\bm{5.6\times10^{-20}}$) & ($\bm{8.5\times10^{-21}}$) & \\
\hline \hline
&&&&&&&&\\[-1em]
\multirow{1}{*}{Massive graviton~\cite{Mirshekari_MDR}} & \multirow{2}{*}{+1 ($-3$)} & \multirow{2}{*}{$m_g$ \lbrack eV\rbrack} & \multirow{2}{*}{$1.3\times10^{-24}$} & $1.6\times10^{-22}$ & $1.6\times10^{-22}$ & $1.4\times10^{-23}$ & $3.4\times10^{-24}$ & \multirow{2}{*}{$5\times10^{-23}$~\cite{Abbott_IMRcon}}\\
 (propagation)&&&&&&&&\\[-1em]
 & & &  & ($\bm{6.1\times10^{-25}}$) & ($\bm{6.3\times10^{-25}}$) & ($\bm{2.9\times10^{-25}}$) & ($\bm{1.8\times10^{-25}}$) & \\
\hline 
\end{tabular}
}
\caption{
Tabulated list of modified theories of gravity considered in this paper.
The first column displays the modified theory of gravity in question, the second column indicates the post-Newtonian (PN) order (or ppE exponent $b$) at which the effect enters the gravitational waveform phase, and the third column identifies the appropriate parameters associated with the theory.
The fourth column tabulates the estimated constraints on the above-mentioned theoretical parameter as if a GW150914-like event were detected on ground-based detector CE, while the fifth to eighth columns likewise display the same bound as observed by space-based detectors LISA, TianQin, B-DECIGO, and DECIGO (top), and again for the multi-band GW observations in combination with CE (bottom).
Finally, the last column displays the current constraints on the theoretical parameters as found in the literature.
Entries with a horizontal dash (in the fifth to eighth columns) correspond to bounds on parameters which do not satisfy the small-coupling approximation, indicating that no meaningful bounds may be placed.
All GW bounds are derived from GW generation mechanism bounds, except for the last row, which comes from bounds on the GW propagation correction.
We note that such bounds are obtained via an initial LISA detection exactly four years prior to merger corresponding to the LISA missions lifetime.
This assumption is investigated for validity in Sec.~\ref{sec:results}.
}\label{tab:theories}
\end{table*}

Here we summarize our final results for busy readers.
We begin by finding constraints on the parameterized post-Einsteinian (ppE) magnitude parameter $\beta$~\cite{Yunes:2009ke}, which describes the strength of a generalized deviation from the GR waveform $A_\GR e^{i\Psi_\GR} \rightarrow A_\GR e^{i(\Psi_\GR+\beta u^{b})}$, where $A_\GR$ and $\Psi_\GR$ are the amplitude and phase of gravitational waveforms in the Fourier domain predicted from GR, $u$ is the effective velocity of the binary BH system, and $b$ categorizes the power of velocity at which the modified theory of gravity affects the waveform.
Using Fisher analysis techniques~\cite{Poisson:Fisher,Berti:Fisher,Yagi:2009zm}, we estimate the maximum magnitude $\beta$ can take while remaining consistent with the statistical detector noises 
for GW150914-like~\cite{GW150914} events observed on a ground-based detector (CE~\cite{Ap_Voyager_CE}), space-based detectors (LISA~\cite{LISA}~\cite{LISA}, TianQin~\cite{TianQin}, B-DECIGO~\cite{B-DECIGO}, and DECIGO~\cite{DECIGO}), and the multi-band combinations thereof.
We find that, as expected, space-based detectors which are sensitive to low-frequencies (or small binary velocities) are most proficient at constraining $\beta$ at small values of $b$, while ground-based detectors are most proficient at large values of $b$, which is consistent with e.g.~\cite{Chamberlain:2017fjl}.
When combining measurements from both types of detector, we find improvements across all values of $b$.

Following this, constraints on $\beta$ can be mapped to the associated parameters of various theories of gravity identified by their value of $b$, as summarized in Table~\ref{tab:theories}.
We see that EdGB, dCS, noncommutative gravity, and massive graviton (both dynamical and weak-field) theories can provide stronger constraints than the current best bounds found in the literature, displayed in the last column of Table~\ref{tab:theories}. For dCS in particular, multi-band GW observations are crucial in most cases to place meaningful bounds, which are more stringent than the existing bounds by $\sim 7$ orders of magnitude.

\begin{figure}
\includegraphics[width=\columnwidth]{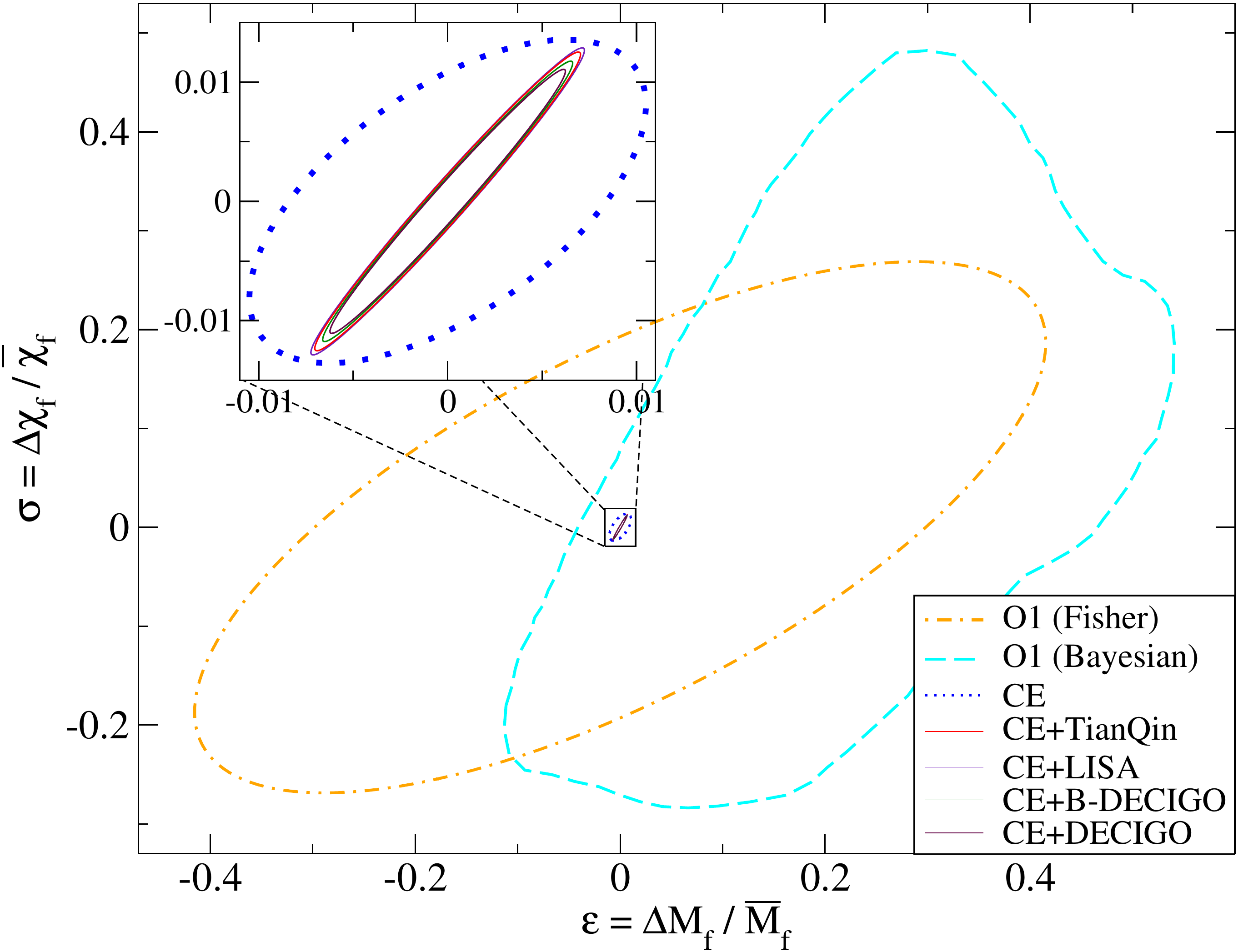}
\caption{
90\% credible region contours of the transformed probability distributions in the $\epsilon-\sigma$ plane (see Eqs.~\eqref{eq:epsilon}--\eqref{eq:transform}), describing the difference in the remnant mass and spin predictions between the inspiral and merger-ringdown estimate for GW150914-like event using the GR templates. 
Here we display the results for LIGO O1 (Fisher and Bayesian~\cite{Abbott_IMRcon} for comparison), CE, and the multi-band observations of CE and LISA, TianQin, B-DECIGO, and DECIGO.
The areas of such confidence regions show the following: (i) good agreement within 10\% between the Fisher and Bayesian analyses, (ii) three orders-of-magnitude improvement from LIGO O1 to CE, and (iii) up to an additional order-of-magnitude improvement with multi-band observations.
}\label{fig:IMRDconsistencyTransformed}
\end{figure}

Lastly, we offer a simplified (Fisher-analysis-based), predictive IMR consistency test~\cite{Ghosh_IMRcon,Ghosh_IMRcon2,Abbott_IMRcon,Abbott_IMRcon2} for future GW150914-like events.
This is done by computing the Gaussian posterior probability distributions between the remnant BH's mass $M_f$ and spin $\chi_f$ obtained independently from both the inspiral (I) and merger-ringdown (MR) signals.
The consistency between such final mass and spin parameters obtained independently from the inspiral and merger-ringdown signals could then tell one something about the underlying theory of gravity.
In particular, if such predictions disagree with each other to a statistically significant level, evidence of non-GR behavior emergent within the signal can be presented.
Such distributions are then combined into the joint-probability distribution  between non-GR parameters $\Delta M_f\equiv M_f^\II-M_f^\MR$ and $\Delta \chi_f\equiv \chi_f^\II-\chi_f^\MR$, with $(\Delta M_f,\Delta \chi_f)|_\GR=(0,0)$ being the GR value.
We estimate the effective \emph{size} of the 90\% confidence region in Fig.~\ref{fig:IMRDconsistencyTransformed}
for ground based detectors, as well as the combination with different space-based detectors.
The areas of such posterior probability distributions can be used to predict the amount of ``wiggle" room a given non-GR theory of gravity will have to become the correct theory of gravity with future detectors, as any separation of the two posteriors could indicate deviations from GR.
Following suit with the previous analysis, we find that the ground-based detector is optimal at measuring the merger-ringdown portion of the signal, the space-based detectors are efficient at measuring the inspiral portion of the signal, and the combination of the two\footnote{Note that space-based detectors can not observe the merger-ringdown phase of GW150914-like events. Thus, signals can \emph{only} be combined for the inspiral portion of the signal.} proves to reduce the posterior sizes by up to an order of magnitude.

The outline of the paper is as follows.
We begin in Sec.~\ref{sec:technical} with a discussion on 
the gravitational waveform model, GW interferometers and the Fisher analysis techniques used in our analysis.
In Sec.~\ref{sec:parameterized}, we review the formulation of parameterized tests of GR with GWs and display our results for the constraints on the parameters of example theories which impact the generation and propagation of GWs.
Sec.~\ref{sec:IMRDconsistency} provides an analysis on the IMR consistency test of GR.
Finally, we conclude and offer avenues of future direction in Sec.~\ref{sec:conclusion}.
Throughout this paper, we have adopted the geometric units of $G=1=c$, unless otherwise stated.


\section{Fisher Analysis}\label{sec:technical}
GWs fill the observable universe, embedded with information describing their various sources.
Detecting such waves however, is exceedingly difficult due to their weak interactions with matter.
With strains on the order of $10^{-21}$~\cite{GW150914}, these signals easily become lost in the detectors' noise arising from various technical details.
One strategy commonly used to reveal these signals is known as \emph{matched filtering}, where one effectively maximizes the correlations between the total signal output from the detector, and a chosen template waveform weighted by the detectors' spectral noise.
In this section, we begin by discussing the gravitational waveform template in GR, followed by a description of the various future detectors and GW sources we consider, finished off with a description of the Fisher analysis techniques used to approximate parameter uncertainties.

\subsection{Gravitational waveform models and detectors} 
\label{sec:waveform}
Accurate models of the gravitational waveform are crucial in the matched filtering process used to detect such signals, as well as in the extraction of observables from the signals.
In this document, we consider the non-precessing \emph{IMRPhenomD}~\cite{PhenomDI,PhenomDII} model in GR, which describes the IMR phases of the coalescence of BHs in a binary system.
These waveform templates were developed from numerical relativity (NR) simulations described comprehensively in~\cite{PhenomDI,PhenomDII}, and can be written in their general form as
\begin{equation}
\tilde{h}_\GR(f)=A_\GR(f)e^{i\Psi_\GR(f)}.
\end{equation}
Here $\tilde h$ is the Fourier-transformed waveform, $f$ is the GW frequency, and $A_\GR$ and $\Psi_\GR$ are the amplitude and the phase respectively.

The parameters $\theta^a$ of the ppE waveform template (we consider sky-averaged waveforms in order to neglect the polarization and sky-location angles in our analysis) when considering phase modifications are as follows:
\begin{equation}
\theta^a_\GR=(\ln A, \phi_c, t_c, \ln \mathcal{M},\ln \eta, \chi_s, \chi_a),
\end{equation} 
Here $A\equiv \mathcal{M}^{5/6}/(\sqrt{30}\pi^{2/3}D_L)$ is a normalized amplitude factor with $D_L$ being the luminosity distance to the event, and $\mathcal{M}\equiv M \eta^{3/5}$ is the chirp mass with $M\equiv m_1+m_2$ and $\eta\equiv m_1m_2/M$ being the total mass and symmetric mass ratio, $\phi_c$, $t_c$ are the phase and time at coalescence, and $\chi_{s,a}=\frac{1}{2}(\chi_1 \pm \chi_2)$ are the symmetric and anti-symmetric dimensionless spin parameters. 

\begin{table*}
\centering
\begin{tabular}{c|c|c|c|c|c|c}
Detector & Location & GW150914 $f_{\text{low}}$ (Hz) & GW150914 $f_{\text{high}}$ (Hz) & GW150914 SNR & Arm length & interferometers\\
\hline
CE~\cite{Ap_Voyager_CE} & Ground & {1} & {400} & $3.36 \times 10^3$ & $40$ km & 1\\
LISA~\cite{LISA} & Space & {0.02} & {1} & $9.30$ & 2.5 Gm & 2\\
TianQin~\cite{TianQin} & Space & {0.02} & {1} & 10.7 & 0.173 Gm & 2\\
B-DECIGO~\cite{B-DECIGO} & Space & {0.02} & {100} & $6.07 \times 10^2$ & 100 km & 2\\
DECIGO~\cite{DECIGO} & Space & {0.02} & {100} & $1.53 \times 10^4$ & 1,000 km & 8\\
\end{tabular}
\caption{
Tabulated information for the ground-based detector CE and all 4 space-based detectors LISA, TianQin, B-DECIGO, and DECIGO considered in our analysis.
The frequency integration ranges $f_{\text{low}}$-$f_{\text{high}}$ are computed using Eqs.~(\ref{eq:groundFreq}) and~(\ref{eq:spaceFreq}) for GW150914.
The lower ground-based and upper space-based frequency limits correspond to the detector limits $f_{\text{low-cut}}$ and $f_{\text{high-cut}}$, while the upper ground-based and lower space-based limits correspond to an arbitrary value such that the gravitational wave spectrum is sufficiently small compared to the detector sensitivity, and the GW frequency four years prior to merger.
The GW150914 SNR is computed via Eq.~(\ref{eq:SNR}).
}\label{tab:detectors}
\end{table*}

In the following analysis, we consider the observation of GW150914-like events on the third generation ground-based detector CE~\cite{Ap_Voyager_CE}  in conjunction with the various proposed space-based design sensitivities of LISA~\cite{LISA}\footnote{We note here that currently, the lifetime of the LISA mission will be four years. In the following analysis, we assume the best-case scenario in which all four years of the mission will be able to observe the same event.}, TianQin~\cite{TianQin}, B-DECIGO~\cite{B-DECIGO}, and DECIGO~\cite{DECIGO}. 
The noise spectral density $S_n(f)$ of these interferometers can be found in e.g. Fig.~1 of~\cite{Carson_multiBandPRL}.
Table~\ref{tab:detectors} compares the various properties of each detector (low and high cut-off frequencies $f_{\text{low-cut}}$ and $f_{\text{high-cut}}$ and SNR for GW150914-like events, arm length, and number of independent interferometers\footnote{The number of independent interferometers (e.g. 2 for LISA) has been accounted for by directly modifying the spectral noise density $S_n(f)$ by $N_{\text{detectors}}^{-1}$. Similarly, any detector geometry that is not $90^{\circ}$ obtains an additional factor of $1/\sin{\theta_{\text{arm}}}$ applied to the spectral noise density.}). 

\subsection{Parameter estimation}
The most reliable, comprehensive method used to extract parameters from a given signal $s=h_t + n$ (the sum of the true gravitational waveform $h_t$ with true parameters $\theta^a_t$ and noise $n$), with known GW template $h$, is through a full \emph{Bayesian analysis}.
In such analyses, one reconstructs the full posterior probability distributions for parameters $\theta^a$, given a signal $s$.
With such a large parameter space, this form of analysis proves to be quite computationally expensive, and infeasible when many samples are required. 
However, for large enough SNRs~\cite{Vallisneri:FisherSNR,Vallisneri:FisherSNR2}, a \emph{Fisher analysis}~\cite{Cutler:Fisher,Poisson:Fisher,Berti:Fisher,Yagi:2009zm} may be used as a reliable approximation to the Bayesian analysis. 
Assuming that we have a perfect template ($h=h_t$), the SNR is given by the inner product of the waveform with itself, weighted by the spectral noise density $S_n(f)$ of the detector:
\begin{equation}\label{eq:SNR}
\rho\equiv\sqrt{(h|h)},
\end{equation}
where the inner product is defined as
\begin{equation}
(a|b) \equiv 2 \int^{f_{\text{high}}}_{f_{\text{low}}}\frac{\tilde{a}^*\tilde{b}+\tilde{b}^*\tilde{a}}{S_n(f)}df.
\label{eq:overlap}
\end{equation}

We choose the limiting frequencies $f_{\text{low}}$ and $f_{\text{high}}$ from Eq.~\eqref{eq:overlap} as follows.
For ground-based detectors, we choose the upper and lower integration frequencies as
\begin{equation}\label{eq:groundFreq}
f_{\text{low}}^{\text{ground}}=f_{\text{low-cut}},\quad
f_{\text{high}}^{\text{ground}}=400 \text{ Hz},
\end{equation}
where $f_{\text{low-cut}}$ is the detector lower cut-off frequency while the upper limit of $400$ Hz was chosen such that the gravitational wave spectrum $2\sqrt{f}|\tilde{h}|$ is sufficiently small compared to the detector sensitivity $S_n$ for GW150914-like events.
Similarly, for space-based detectors, we choose
\begin{equation}\label{eq:spaceFreq}
f_{\text{low}}^{\text{space}}=f_{\text{4yrs}}, \quad
f_{\text{high}}^{\text{space}}=f_{\text{high-cut}},
\end{equation}
where $f_{\text{high-cut}}$ is the detector higher cut-off frequency while
\begin{equation}\label{eq:f4yr}
f_{T_\mathrm{obs}} = 1.84\times10^{-2} \left( \frac{\mathcal{M}}{28\text{ M}_\odot} \right)^{-5/8} \left( \frac{T_{\text{obs}}}{1\text{ yr}} \right)^{-3/8}\,,
\end{equation}
is the frequency $T_\mathrm{obs}$ prior to merger. $f_{\text{high}}$ and $f_{\text{low}}$ are tabulated in Table~\ref{tab:detectors}.

In a Fisher analysis, we make the assumptions that the detector noise is stationary and Gaussian.
Following Refs.~\cite{Cutler:Fisher,Yagi:2009zm}, the noise follows a probability distribution of the form
\begin{equation}
p(n) \propto \exp \left\lbrack  -\frac{1}{2}(n|n) \right\rbrack,
\end{equation}
which, given a detected signal $s=h_t+n$, is transformed to
\begin{equation}
p(\theta^a_t|s) \propto p^{(0)} \exp \left\lbrack (h_t|s) - \frac{1}{2} (h_t|h_t) \right\rbrack,
\end{equation}
where $p^{(0)}$ are the prior distributions on parameters $\theta^a$.
The maximum-likelihood parameters $\hat{\theta}^a$ can be determined by maximizing the above distribution, resulting in
\begin{equation}\label{eq:fisherPDF}
p(\theta^a|s) \propto p^{(0)}_{\theta^a} \exp \left\lbrack -\frac{1}{2} \Gamma_{ij} \Delta \theta^i \Delta \theta^j \right\rbrack,
\end{equation}
where $\Delta \theta^i \equiv \theta^i-\hat{\theta}^i$, and the \emph{Fisher matrix} $\Gamma_{ij}$ is determined to be
\begin{equation}
\Gamma_{ij}\equiv(\partial_i h | \partial_j h),
\end{equation}
with $\partial _i \equiv \partial/\partial \theta^i$.
When the prior distributions are absent, Eq.~\eqref{eq:fisherPDF} is reminiscent of the multivariate Gaussian distribution about parameters $\hat{\theta}^i$ with variance-covariance matrix given by $\Gamma_{ij}^{-1}$.

In this paper, we follow~\cite{Cutler:Fisher,Poisson:Fisher,Berti:Fisher} and assume that the priors are Gaussian. 
Then, we define the effective Fisher matrix as
\begin{equation}
\tilde \Gamma_{ij} \equiv \Gamma_{ij} +\frac{1}{\left(\sigma^{(0)}_{\theta^a}\right)^2}\delta_{ij},
\end{equation}
where $\sigma^{(0)}_{\theta^a}$ are the prior root-mean-square estimates of parameters $\theta^a$. The corresponding $1\sigma$ root-mean-square errors on parameters $\hat{\theta}^i$ can be written as 
\begin{equation}
\Delta\theta_i=\sqrt{\tilde \Gamma^{-1}_{ii}}.
\end{equation}
Correspondingly, if one utilizes information from multiple detectors A and B, the resulting Fisher matrix becomes
\begin{equation}
\tilde \Gamma_{ij}^{\text{total}}=\Gamma_{ij}^{\text{A}}+ \Gamma_{ij}^{\text{B}}+\frac{1}{\left(\sigma^{(0)}_{\theta^a}\right)^2}\delta_{ij}.
\end{equation}

In the following investigation, we consider only GW150914-like~\cite{GW150914} events, with masses $(m_1,m_2)=(35.8\text{ M}_\odot,29.1\text{ M}_\odot)$ and spins $(\chi_1,\chi_2)=(0.15,0)$.
Such spins are taken to be non-vanishing so that the spin-dependent BH scalar charges are non-zero in dCS gravity, yet still small enough to be consistent with the LVC's measurement of the effective spin~\cite{GW150914}.
The luminosity distance is scaled such that an SNR of $\rho_{\text{O2}}=25.1$ would be achieved on the sensitivity for the LIGO/Virgo's 2nd Observing Run (O2)~\cite{aLIGO}. 
We also note that we assume the initial LISA detection of GW150914-like events to take place exactly four years prior to their merger, corresponding to the expected lifetime of the LISA mission.
Such an assumption is considered for its validity in further detail in upcoming Sec.~\ref{sec:results}.
Fiducial values used for the other parameters are $\phi_c=t_c=0$.
Finally, priors on the BH spins of $|\chi_{s,a}|<1$ $(\sigma^{(0)}_{\chi_{s,a}} = 1)$ are imposed.

\subsection{Detectability of GW150914-like events}\label{sec:SNRs}
In this section, we discuss the feasibility of detecting GW150914-like events using the space-based GW interferometer LISA\footnote{We found that space-based detector TianQin observes very similar, yet slightly louder ($\rho=10.7$ for GW150914-like events) results to that of LISA. 
Additionally, DECIGO and B-DECIGO can detect strong GW150914-like signals with SNRs of $10^2-10^4$.}.
As described in Refs.~\cite{Wong:2018uwb}, the standard threshold SNR of $\rho_\text{th} \sim 8$ can be reduced to $\rho_\text{th} \sim 4-5$ for LISA by revisiting sub-threshold events in prior LISA data with information from high-SNR events in the ground-based bands of e.g.~CE. 
Moore \textit{et al.}~\cite{Moore:2019pke} later pointed out that a template-based search for LISA requires a much larger SNR threshold of $\rho_\text{th} \sim 15$, which can be further reduced to $\rho_\text{th} \sim 9$ in combination with ground-based detectors.
However, such an estimation may be pessimistic, as non-template-based approaches may bring this threshold down further.

\begin{figure}
\includegraphics[width=\columnwidth]{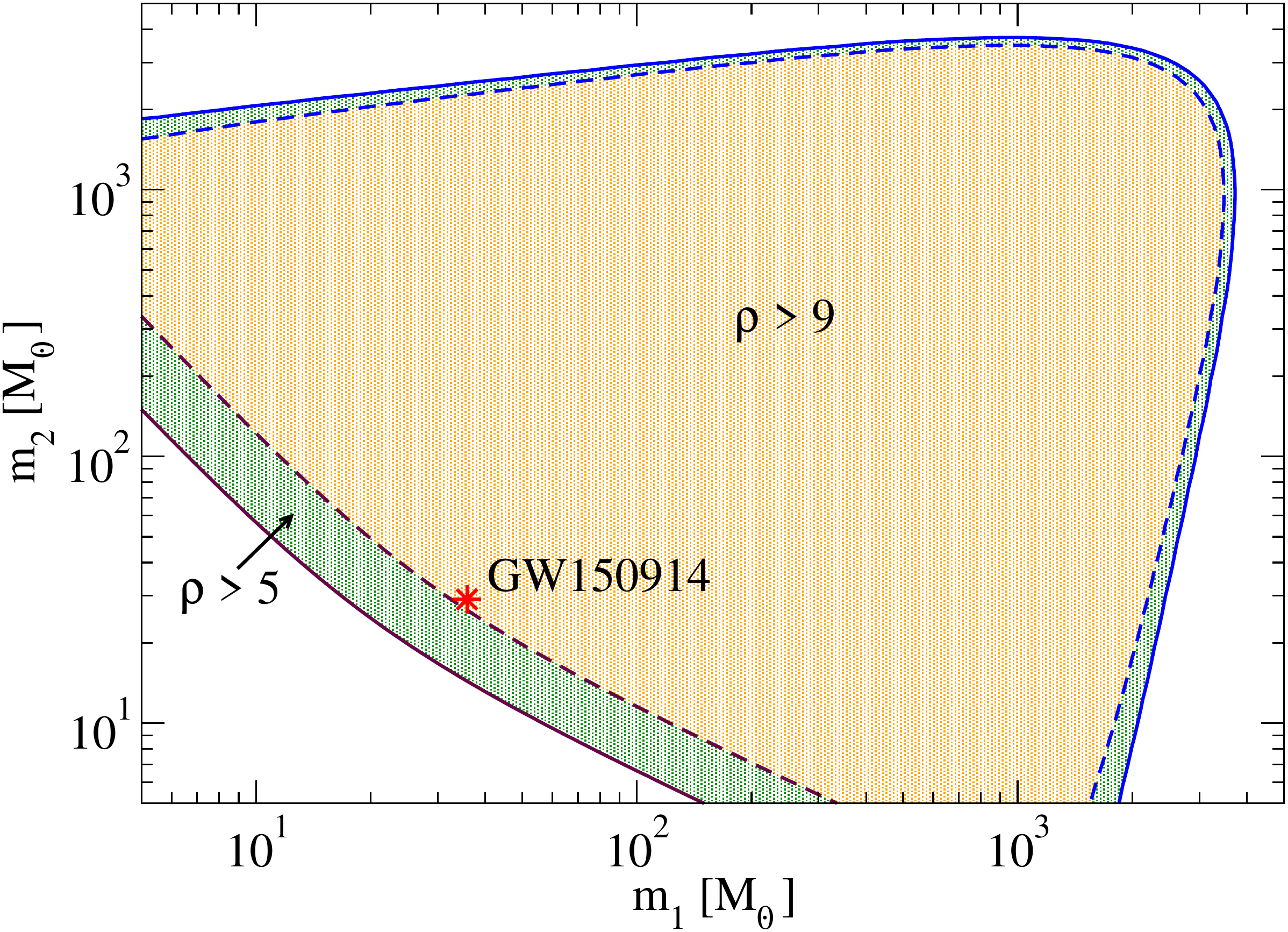}
\caption{Multi-band detectability region as a function of the constituent BH masses $m_1$ and $m_2$.
This region is formed by SNRs in agreement with the condition $\rho>\rho_\text{th}$ for $\rho_\text{th}=5$ (optimistic) and $\rho_\text{th}>9$~\cite{Wong:2018uwb,Moore:2019pke} (pessimistic) for events detected by both the ground-based detector CE, and space-based detector LISA. 
Such SNRs have been computed with the assumption of non-spinning BHs at luminosity distances of $410$ Mpc.
The upper-right edge (blue) of the region corresponds to CE's $\rho_\text{th}$ contour, while the lower-left edge (maroon) is formed by LISA's contour. 
The SNRs are computed following Eq.~\eqref{eq:SNR}.
Additionally shown as a red star is the event GW150914 with $(m_1,m_2)=(35.8\text{ M}_\odot,29.1\text{ M}_\odot)$.
Observe how GW150914-like events marginally fall within LISAs larger observational SNR threshold of $\rho_{\text{th}}=9$.
Alternatively, such events exceed both ground-based SNR thresholds by more than 2 orders-of-magnitude.
}\label{fig:SNRs}
\end{figure}

To demonstrate how well such events can be detected in either case, Fig.~\ref{fig:SNRs} displays the region in the $(m_1,m_2)$ parameter space where SNRs exceeds the threshold value of $\rho_\text{th} = 5$ or $9$ for both CE and LISA.
Observe how in both cases, GW150914-like events with $(m_1,m_2)=(35.8\text{ M}_\odot,29.1\text{ M}_\odot)$ fall within the multiband detectability region defined by both $\rho>5$ or $\rho>9$.
For LISA observations of GW150914-like events, the SNR of $\rho=9.3$ only marginally falls within the larger threshold of $9$, while the CE observation well exceeds both thresholds by $\sim2$ orders-of-magnitude.
We note that for the following analysis, we consider nearly non-rotating GW150914-like events that satisfy such detectability criteria.
We refer to the discussion by Jani \emph{et al.}~\cite{Jani:2019ffg} for a more in-depth analysis into the multi-band detection between third-generation detectors and LISA.

\section{Parameterized Tests of GR}\label{sec:parameterized}

In this section, we study the first theory-agnostic test of GR, namely the parameterized tests. Several methods have been proposed to construct parameterized waveforms to capture non-GR effects. Here, we follow the \emph{parameterized post-Einsteinian} (ppE) formalism~\cite{Yunes:2009ke}.

\subsection{Formalism}

While the waveform template explained in Sec.~\ref{sec:waveform} does not allow for deviations from GR, we modify it to allow for general phase and amplitude modifications entering at different \emph{post-Newtonian} (PN) orders\footnote{An $n$-PN order term is proportional to $(u/c)^{2n}$ relative to the leading-order term in the GR waveform.}.
The ppE waveform template in Fourier space can therefore be written as
\begin{equation}
\tilde{h}_{\text{ppE}}(f)=A_\GR(f)(1+\alpha u^a)e^{i[\Psi_\GR(f)+\beta u^b]},
\end{equation}
where $u=(\pi \mathcal{M} f)^{1/3}$ is the effective relative velocity of the gravitating bodies in a binary, $(a,b)$ characterize the velocity dependence at which non-GR modifications of magnitude $(\alpha,\beta)$ enter the waveform in the amplitude and phase, respectively.
For modifications to generation mechanisms, the ppE correction is only included in the inspiral portion of the waveform, while for those to propagation mechanisms, the correction is included in the IMR waveform throughout.
The exponents $a$ and $b$ can be mapped to the familiar PN order $n$ as
\begin{equation}
b=2n-5, \hspace{5mm} a=2n.
\end{equation}

\begin{figure*}
\includegraphics[width=.8\textwidth]{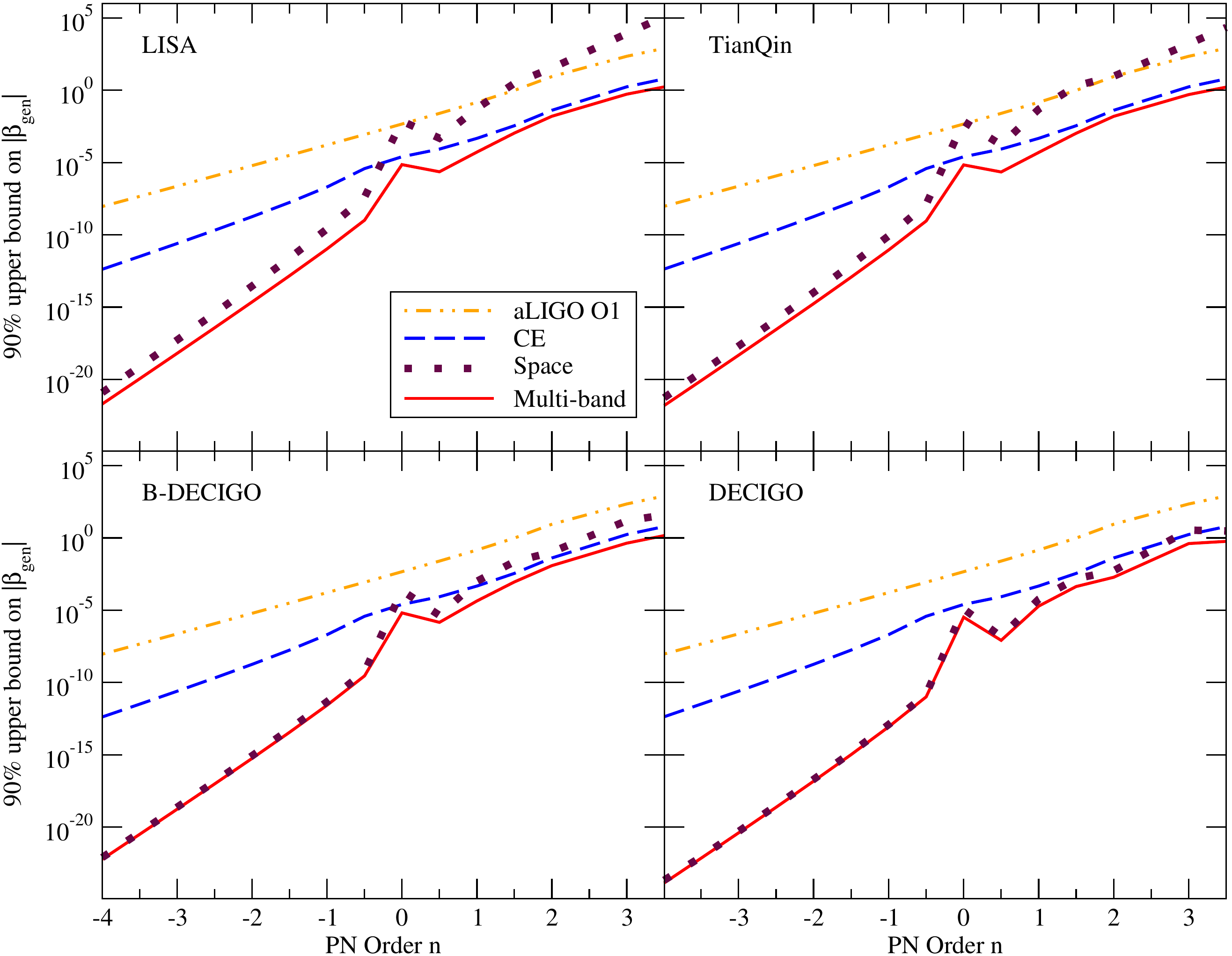}
\caption{
Constraints on the generalized non-GR phase parameter for generation effects $|\beta_{\text{gen}}|$ as a function of PN order for GW150914-like events observed on various space- and ground-based detectors individually.
Observe how space-based detectors are most proficient at probing effects that enter at negative-PN orders, with ground-based detectors more suited to probing positive-PN effects.
The combination of the two (multi-band) results in improved bounds across all PN orders.
}\label{fig:betaBoundsGen}
\end{figure*}

The ppE formalism is highly advantageous, as it allows one to constrain the effects of any generic modification to GR into one parameter (\emph{i.e.} $\beta$), controlled by the chosen power of velocity ($b$).
By choosing the power $b$ or $a$ corresponding to the modified theory of gravity one wishes to study, the size of the effect ($\beta$ or $\alpha$) can be mapped to the corresponding theoretical constants. 
Below we primarily consider corrections in the phase because amplitude corrections are usually subdominant~\cite{Tahura:2019dgr}.
In App.~\ref{sec:theory}, we list all of the example theories considered in this paper in detail,  together with the mapping between $\beta$ and the theoretical constants.

Below, we carry out Fisher analyses as explained in Sec.~\ref{sec:technical} assuming that GR is the true theory of gravity in nature (i.e. choosing the fiducial value of the ppE parameter as $\beta =0$) to estimate statistical errors on $\beta$ for observing GW150914-like events with different detectors.
We note here that for the following analysis, we consider constraints on $\beta$ from a non-rotating BH and a slowly-rotating BH with non-vanishing scalar charges in dCS gravity, that are consistent with the LVC's effective spin measurement.
Reference~\cite{Yunes_ModifiedPhysics} showed that such analyses give very similar results to those from Bayesian analyses.
We consider bounds on corrections to GW generation and propagation mechanisms in turn. 
We also map such bounds on generic ppE parameters to those on parameters in example modified theories of gravity

\subsection{Results}\label{sec:results}

We next discuss our findings. 
We first show results for bounds on GW generation mechanisms, followed by those on GW propagation mechanisms.

\subsubsection{GW generation mechanisms}

Figure~\ref{fig:betaBoundsGen} presents 90\% upper credible level bounds on the generalized non-GR phase parameter $|\beta_\mathrm{gen}|$ for generation mechanisms\footnote{While certain theories of gravity correspond to different signs of the ppE parameter $\beta$, in this analysis we only constrain the modulus of the parameter to remain as generic and theory-agnostic as possible. In future, more directed analyses with a Bayesian approach, priors of $\beta<0$ or $\beta>0$ can be imposed to improve constraints. 
}, and we show bounds on the amplitude parameter $\alpha_\mathrm{gen}$ in App.~\ref{app:amplitude}.
We consider CE, LISA, TianQin, B-DECIGO, DECIGO, and the combinations of each space-based detector with CE.
Bounds are obtained for each half-integer PN order between $-4$PN and +3.5PN, with the exception of +2.5PN which observes complete degeneracies with the coalescence phase $\phi_c$.
Additionally, the bound from O1 (LVC's 1st observing run) is taken from Ref.~\cite{Yunes_ModifiedPhysics} for comparison.

\begin{figure*}
\includegraphics[width=.9\columnwidth]{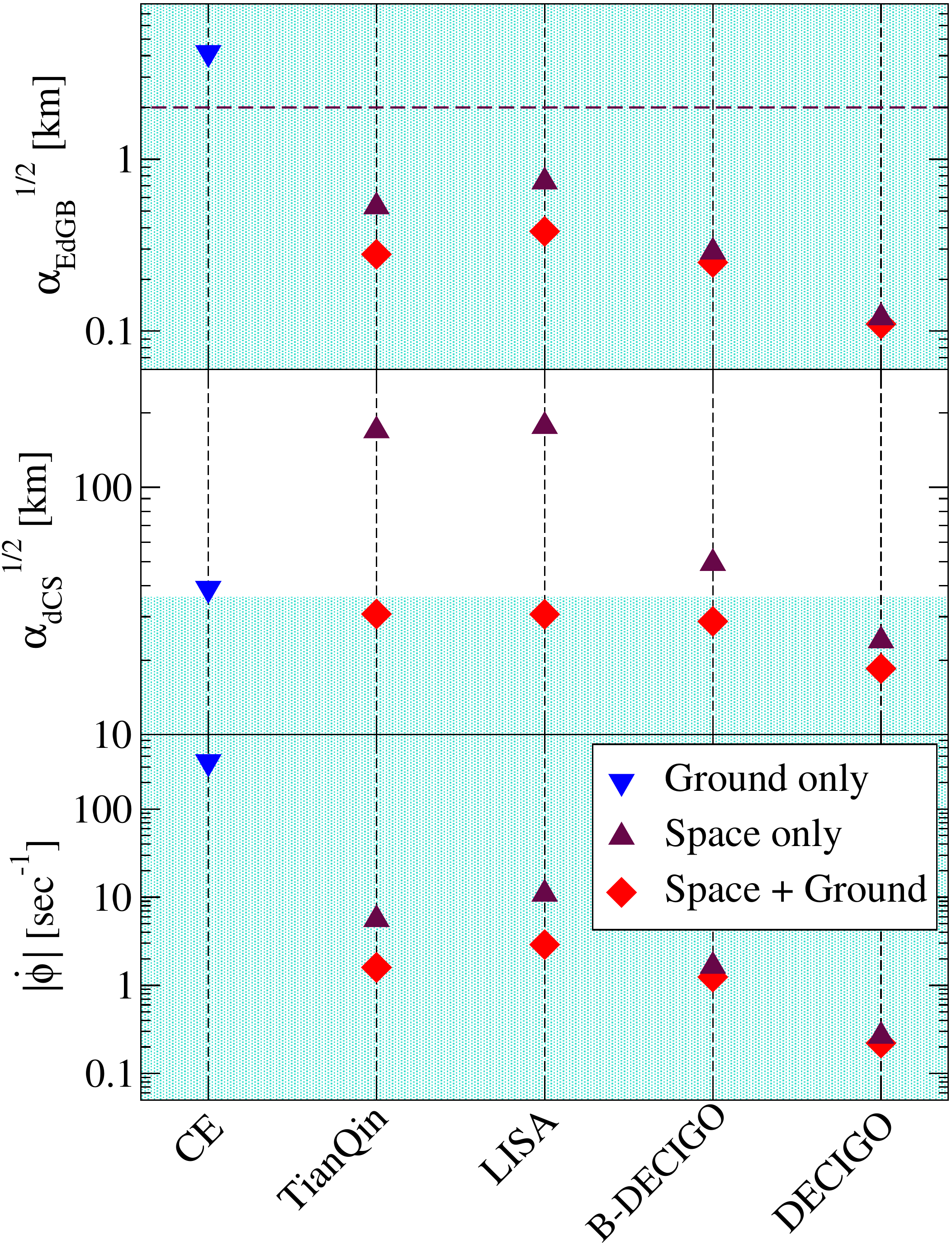}
\includegraphics[width=.9\columnwidth]{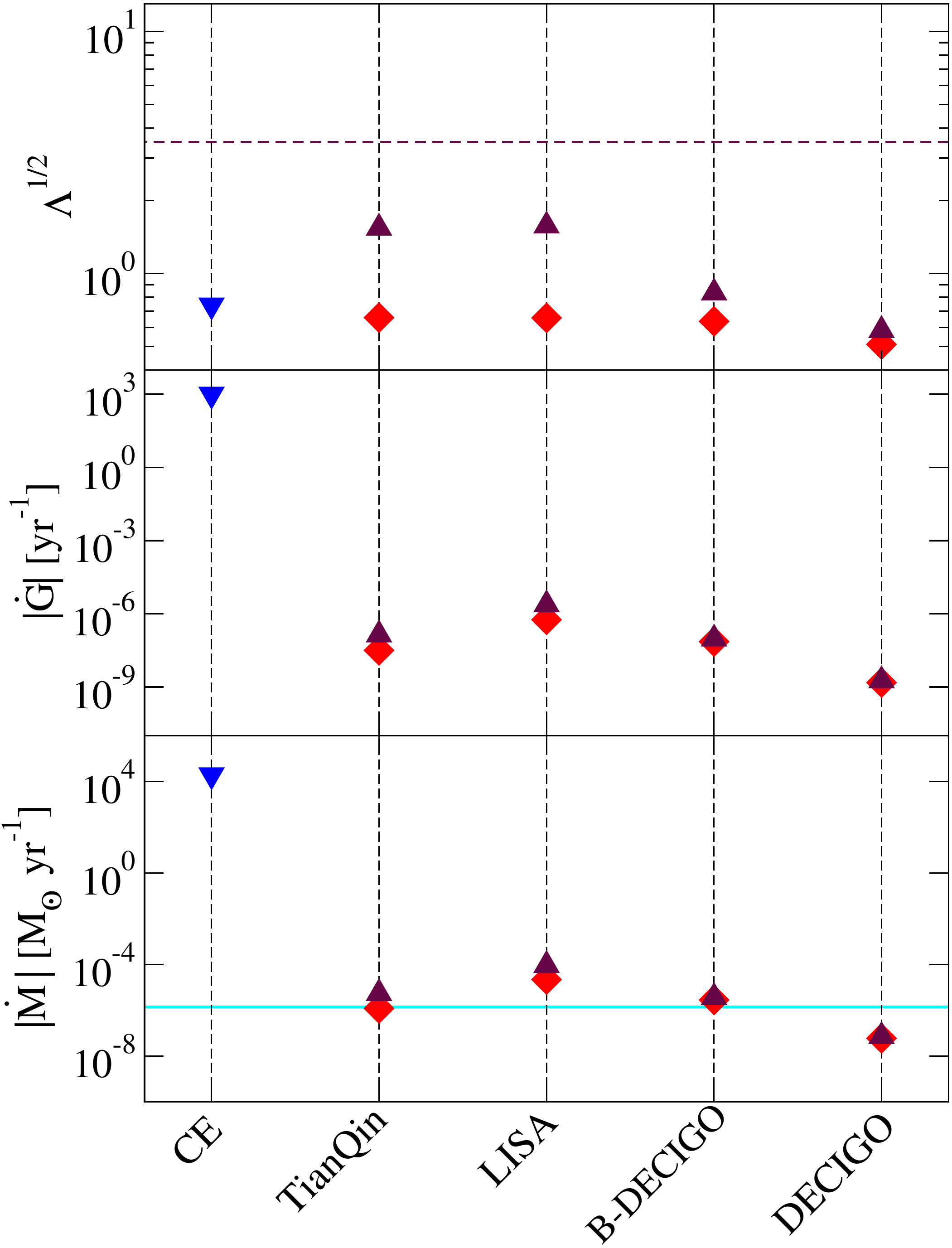}
\caption{
90\% upper-bound credible level constraints on the theoretical parameters representative of 6 of the modified theories of gravity considered in this analysis for GW150914-like events.
Bounds are presented for: EdGB gravity ($\sqrt{\alpha_\EdGB}$, $-1$PN order), dCS gravity ($\sqrt{\alpha_\dCS}$, +2PN order), scalar tensor theories ($\dot{\phi}$, $-1$PN order), noncommutative gravity ($\sqrt{\Lambda}$, +2PN order), varying-G theories ($\dot{G}$, $-4$PN order), and BH mass-varying theories ($\dot{M}$, $-4$PN order), and are additionally tabulated in Table~\ref{tab:theories}.
The blue shaded regions correspond to the region such that the small coupling approximations $\zeta_\EdGB \ll 1$, $\zeta_\dCS \ll 1$, and $m\dot{\phi} \ll 1$ are valid (the definition of the dimensionless coupling constants $\zeta$ can be found in App.~\ref{sec:theory}), and the dashed maroon lines correspond to the current bounds in the literature, also tabulated in Table~\ref{tab:theories}.
The cyan line in the bottom right panel corresponds to the Eddington accretion rate for GW150914-like events.
}\label{fig:modifiedBounds}
\end{figure*}

We can make the following observations from the figure. 
First, notice that non-GR effects entering the gravitational waveform at negative-PN orders can be constrained most stringently by space-based detectors, while positive-PN effects are most proficiently constrained by ground-based detectors.
The obvious exception being DECIGO, which bridges the gap between the two frequency bands and provides the strongest bounds at both positive and negative PN orders.
Second, when one considers multi-band observations by combining both space- and ground-based detectors, we see large improvements of up to a factor of 40~\cite{Carson_multiBandPRL} across all PN orders.

Here we briefly discuss the effect of LISA's mission lifetime on observations of the theory-agnostic parameter $\beta$.
In particular, for the above calculations we assumed the best-case scenario in which all four years of LISA's mission can be used to observe the same GW signal from a GW150914-like event.
For comparison, we instead consider an estimate of the same effect (at $-4$PN order for the largest effect possible) given that only the last three, two, or one years of LISA's lifetime will be able to observe the GW150914-like signal.
We find that the resulting constraints on $\beta$ are weakened by ratios of $1.8$, $4.3$, and $16$ respectively, compared to the best-case four-year scenario.
Thus we conclude that such effects change our results on the order of an order of magnitude for the worst-case scenario of only one-year observation by LISA. 
Additionally, we find such weakened observations to have SNRs of $9.2$, $8.2$, $6.7$, and $5$ for four-, three-, two-, and one-year observations respectively.
All such SNRs still remain within the multi-band detectibility region found in Fig.~\ref{fig:SNRs} for the best-case threshold SNR.

Now we consider the cases in which LISA can observe the GW signal more than four years prior to the coalescence.
In particular, we consider the two new scenarios in which (i) LISA observes the early inspiral signal from six years to two years prior to merger before going offline, and then CE observes the merger two years later, and (ii) LISA observes the early inspiral signal from ten years to six years prior to merger before going offline, and then CE observes the merger 6 years later.
The above to cases have LISA SNRs of 8.8 and 7.7 respectively, each above the best-case SNR threshold.
Relative to the scenario considered in this paper in which LISA begins observing four years prior to the merger, we find such constraints to be \emph{strengthened} by factors of 1.7, and 3.1 respectively at $-4$PN order.
Such constraints have been improved because the $-4$PN order correction is the best-case scenario for observing the earlier inspiral.
In the worst-case scenario of $3.5$PN order corrections, we find constraints on $\beta$ to \emph{weaken} by factors of 1.04, and 1.12 respectively.
Such changes are insignificant to our analysis, and for almost all PN orders give way to improved ppE constraints, making our estimates conservative.

Now that constraints on the agnostic non-GR parameter $\beta$ have been obtained, we apply them to the specific theories of gravity reviewed in App.~\ref{sec:theory} to constrain their theoretical parameters using various single-band and multi-band observations.
Such bounds are obtained by simply selecting the constraints on $|\beta_{\text{gen}}|$ corresponding to the PN order associated with the desired modified theory of gravity, and finally mapping them to the theoretical parameters with the ppE expressions found in App.~\ref{sec:theory}.
When obtaining constraints on theory-specific parameters from the theory-agnostic ppE parameter $\beta$, we assume a fiducial value of $\beta=0$ corresponding to GR. 
The resulting root-mean-square variance on $\beta$ describes the statistical variance $\beta$ is allowed to observe within the detector noise.
For this reason, under consideration of propagation of uncertainties when transforming $\sigma_\beta$ to $\sigma_\epsilon$ for some theory-specific parameter $\epsilon$, all terms containing measurement errors on intrinsic template parameter vanish due to their proportionalitiy with $\beta\rightarrow0$.
See Ref.~\cite{Perkins:2018tir} for a more in-depth discussion on this topic.

\begin{figure}
\includegraphics[width=\columnwidth]{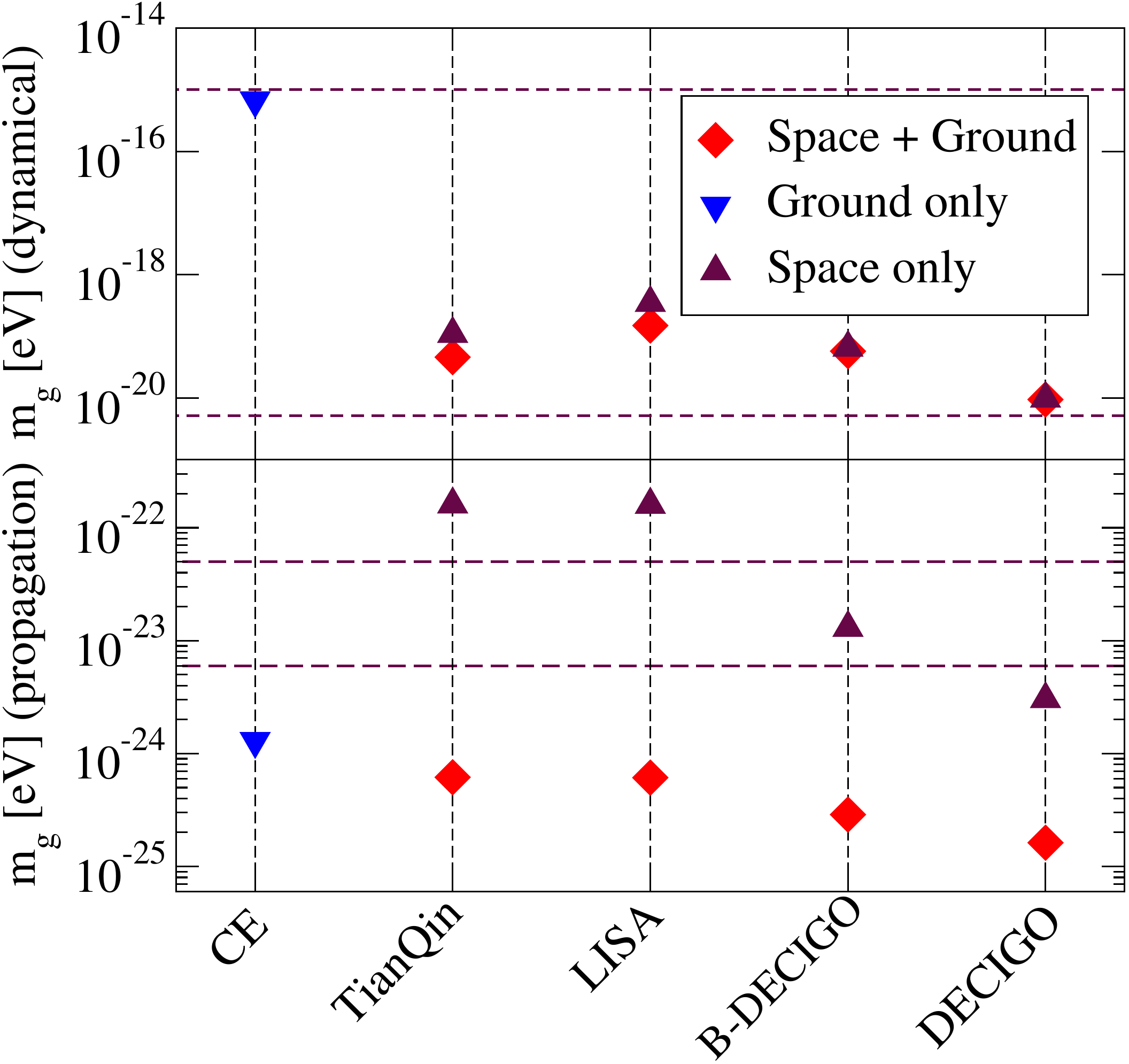}
\caption{Similar to Fig.~\ref{fig:modifiedBounds}, but for constraints on the graviton mass in the dynamical, generation mechanism regime (top), and in the weak-field, propagation mechanism regime (bottom).
}\label{fig:gravitonMassBounds}
\end{figure}

Figure~\ref{fig:modifiedBounds} and the top panel of Fig.~\ref{fig:gravitonMassBounds} display the 90\% upper credible level limit on the associated parameters for theories that modify the generation of GWs: EdGB, dCS, scalar-tensor, noncommutative, varying-$G$, varying-$M$, and massive graviton theories with dynamical effects for both single- and multi-band detections of GW150914-like events.
Additionally, the constraints are tabulated in Table~\ref{tab:theories} for each detector analyzed, along with the current strongest bounds from the literature. We summarize below our findings for each theory.

\begin{itemize}

\item \emph{EdGB, scalar-tensor:}  
In both theories, corrections to the waveform were derived within the small-coupling approximations, in which non-GR corrections are assumed to be always smaller than the GR contribution. Bounds on the theoretical constants (square-root of the coupling constant $\sqrt{\alpha_\EdGB}$ for EdGB gravity and the time-variation of the scalar field $\dot \phi$ for scalar-tensor theories) both satisfy the small-coupling approximations for every detection scenario, however only $\sqrt{\alpha_\EdGB}$ can improve upon the current strongest bound of 2--6 km~\cite{Kanti_EdGB,Pani_EdGB,Yagi_EdGB,Nair_dCSMap,Yamada:2019zrb,Tahura:2019dgr}, for both space-based and multi-band detections.

\item \emph{dCS:} Similar to EdGB and scalar-tensor theories, corrections to the waveform have been derived within the small-coupling approximation. 
Constraints placed on the parity-violation constant $\sqrt{\alpha_\dCS}$ with CE, LISA, TianQin, and B-DECIGO fall short of the small-coupling approximation, and thus are not reliable for GW150914-like events. One can place valid constraints only when multi-band detections are made, improving upon the current constraint of $10^8$ km~\cite{AliHaimoud_dCS,Yagi_dCS} by $\sim$7 orders-of-magnitude.

\item \emph{noncommutative:} Bounds on the noncommutative parameter $\sqrt{\Lambda}$ can slightly be improved upon the current most stringent bound of 3.5~\cite{Kobakhidze:2016cqh} with multi-band observations.

\item \emph{varying-$M$:} Constraints on the time-variation of the total black hole mass $\dot M$ (motivated not only by astrophysical gas accretion, but also by a classical evaporation in a braneworld scenario~\cite{Emparan:2002px,Tanaka:2002rb} or dark energy accretion~\cite{Babichev:2014lda,Babichev:2005py,Babichev:2004yx}) are below that of the Eddington accretion rate for BHs in GW150914-like events for B-DECIGO(+CE), DECIGO(+CE), and LISA+CE multi-band detections.
See also a recent analysis~\cite{Caputo:2020irr} in which the impact of gas accretion on the orbital evolution of BH binaries, and thus the GW emission.

\item \emph{varying-$G$:} Space and multi-band observations can improve significantly over ground-based ones, though the former bounds on the time-variation in $G$ are still much weaker than other existing bounds~\cite{Bambi_Gdot,Copi_Gdot,Manchester_Gdot,Konopliv_Gdot}.\footnote{Space-based bounds can be comparable with current strongest bound for GW observations of supermassive BH binaries~\cite{Yunes_GdotMap}.}

\item \emph{massive graviton (dynamical):} CE bounds (from modifications in the inspiral) on the mass of the graviton are comparable to GW150914-bounds from ringdown~\cite{Chung:2018dxe} while bounds from space-based detectors can be comparable to those from binary pulsars~\cite{Miao:2019nhf}.

\end{itemize}

Finally, we comment that, similar to the constraints on $\beta$, theories that modify GR at negative-PN orders (EdGB, scalar-tensor, varying-$G$, varying-$M$, and dynamical massive graviton) are more strongly constrained by space-based detectors, while positive-PN theories (dCS and noncommutative) observe stronger bounds with ground-based detectors.

\subsubsection{GW propagation mechanisms}

\begin{figure*}
\includegraphics[width=.8\textwidth]{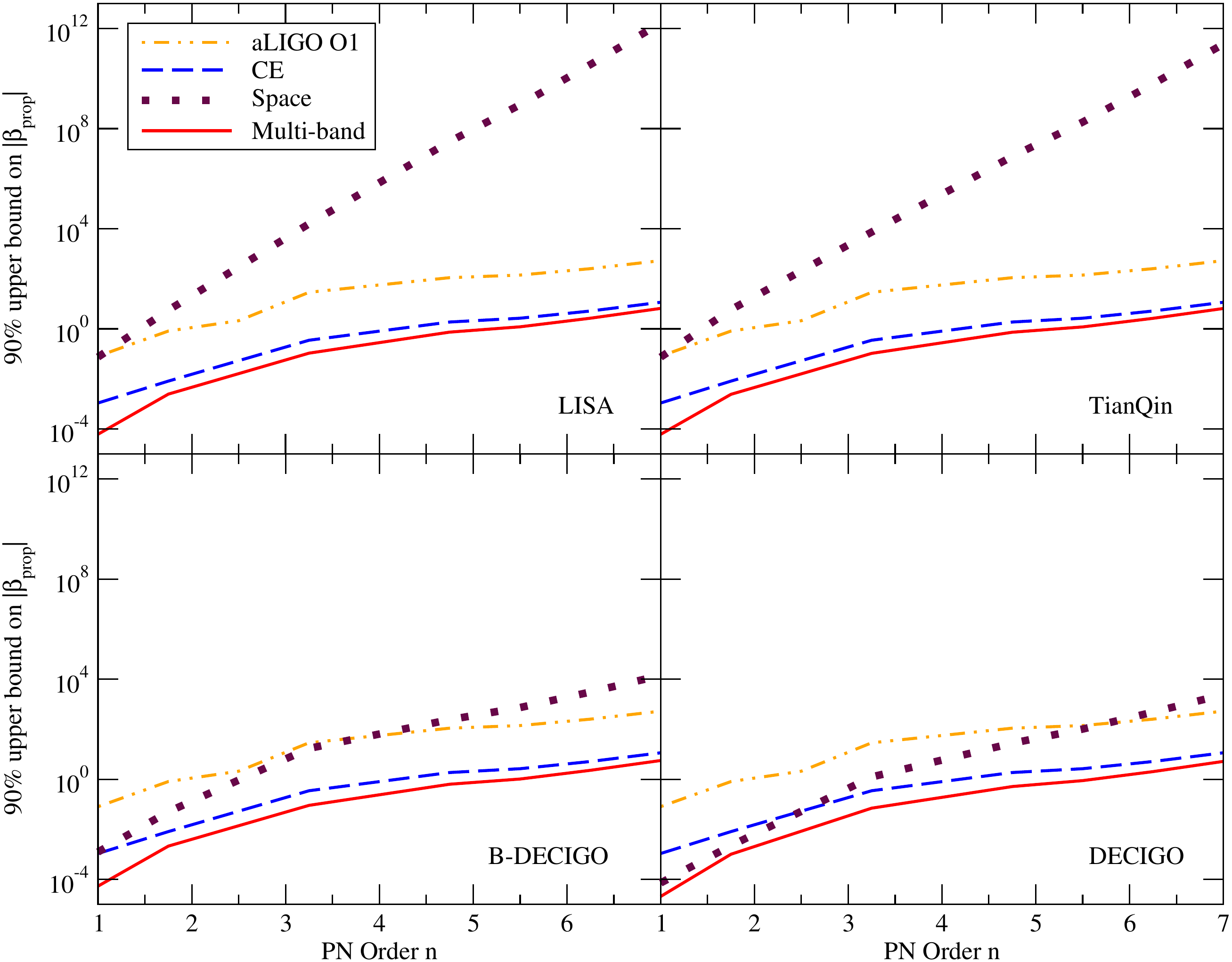}
\caption{
Similar to Fig.~\ref{fig:betaBoundsGen}, but for propagation effects $\beta_{\text{prop}}$ entering the GW phase.
Observe how (with the exception of DECIGO), CE gives stronger constraints on $\beta_{\text{prop}}$ than space-based detectors, while the multi-band case can be even stronger by up to an order-of-magnitude.
}\label{fig:betaBoundsProp}
\end{figure*}

We next move on to studying bounds on the ppE phase parameter $\beta_{\text{prop}}$ from the GW propagation mechanisms (bounds on the ppE amplitude parameter $\alpha_{\text{prop}}$ are shown in App.~\ref{sec:amp}). 
Figure~\ref{fig:betaBoundsProp} presents bounds on $\beta_{\text{prop}}$ against each PN order at which the correction enters. 
We only show bounds on positive PN orders because all of the example theories discussed in App.~\ref{sec:theory_prop} show such a feature. 
Observe that bounds placed with CE dominate those by space-based detectors, with little improvement via multi-band observations. 
These bounds on $\beta_{\text{prop}}$ can easily be mapped to those on the magnitude $\mathbb{A}$ of the correction to the graviton dispersion relation using Eq.~\eqref{eq:MDR}, as shown in Fig.~\ref{fig:Abounds} in App.~\ref{app:amplitude}.

While these constraints may be used to compute bounds on a variety of propagation mechanism non-GR effects, we here focus on the case of massive gravitons with $\mathbb{A}=m_g^2$ and $a_\MDR=0$, now in the weak-field regime.
See App.~\ref{sec:theory} for a more thorough description of modifications to the propagation of gravitational waves.
The bottom panel of Fig.~\ref{fig:gravitonMassBounds} displays such bounds for each detector considered.
We observe that CE places the strongest constraints out of all single-band observations\footnote{The bound becomes much stronger for observing supermassive BH binaries with space-based detectors~\cite{Will_mg,Berti:Fisher,Yagi:2009zm,Berti:2011jz,Cornish:2011ys,Keppel:2010qu,Chamberlain:2017fjl}.}.
When combined to make multi-band detections, we see an improvement on the graviton mass bound, with more than an order-of-magnitude reduction from the current solar system bound of $6\times10^{-24}$ eV~\cite{Will:2018gku}.

We finally consider a comparison of different space-based detectors' ability to test GR.
In particular, we compare constraints on coupling parameters found in a selected few modified theories of gravity investigated with each space-based GW detector: LISA, TianQin, B-DECIGO, and DECIGO.
For EdGB gravity, we find constraints on $\sqrt{\alpha_\EdGB}$ to be $0.7$ km, $0.6$ km, $0.3$ km, and $0.1$ km respectively.
In dCS gravity, we find constraints on $\sqrt{\alpha_\dCS}$ to be respectively $169$ km, $176$ km, $49$ km, and $24$ km.
Finally, for the propagation effect of massive gravitons, we find constraints of $1.6\times10^{-22}$ eV, $1.6\times10^{-22}$ eV, $1.3\times10^{-23}$ eV, and $1.6\times10^{-24}$ eV respectively.
We see that in general, space-based detector DECIGO forms the strongest constraints on all theories of gravity by nearly an order of magnitude in some cases, while B-DECIGO places similar, let slightly weaker bounds.
When comparing similar space-based detectors LISA and TianQin, we see comparable constraints that differ by less than $\sim10\%$.
In particular, we see that TianQin can place stronger constraints for theories that enter at higher PN orders, and vice-versa for LISA.




\section{Inspiral-merger-ringdown consistency tests}\label{sec:IMRDconsistency}

Let us now move onto the second theory-agnostic test of GR with GWs, namely the IMR consistency tests of the remnant BH.

\subsection{Formalism}
\label{sec:theory_IMRD}

While two GW150914-like stellar-mass BHs in a binary system inspiral together via GW radiation, space-based GW interferometers can effectively probe the inspiral portion of the waveform, occurring at low frequencies.
Once the separation distance between the bodies become close enough, they fall into a plunging orbit until finally they merge, forming a common horizon which will settle down via radiation of quasi-normal modes~\cite{Chandrasekhar_QNM,Vishveshwara_QNM} - a high-frequency merger-ringdown signal which is best observed by ground-based GW detectors.
The remnant BH with mass $M_f=M_f(m_1,m_2,\chi_1,\chi_2)$ and spin $\chi_f=\chi_f(m_1,m_2,\chi_1,\chi_2)$ (provided by NR fits in Refs.~\cite{PhenomDII}) can then be entirely described by the same two parameters, in accordance with the BH no-hair theorems.

Using only the inspiral portion of the signal, the final mass and spin of the remnant BH can be uniquely estimated, thanks to numerical relativity simulations, using measurements of the initial mass and spin parameters $m_1$, $m_2$, $\chi_1$, and $\chi_2$, while having no information about the merger-ringdown portion.
The opposite is also true: the final mass and spin may be predicted from the merger-ringdown portion of the signal with no accompanying information about the inspiral\footnote{We follow the same calculation as in the inspiral portion. Namely, we utilize the Fisher analysis method to predict a four-dimensional probability distribution between $m_1$, $m_2$, $\chi_1$, and $\chi_2$, which can be transformed into the two-dimensional probability distribution between $M_f$ and $\chi_f$ using fits from numerical-relativity simulations.}.
Assuming the SNR is sufficiently large\footnote{GW150914 was observed with total SNR of 25.1, which is assumed throughout the analysis.} for both the inspiral and merger-ringdown waveforms, the estimates of $(M_f^\II,\chi_f^\II)$ should agree with those of $(M_f^\MR,\chi_f^\MR)$ within the statistical errors, assuming that GR is the correct theory of gravity.
This test, known as the \emph{IMR consistency test}~\cite{Ghosh_IMRcon,Ghosh_IMRcon2,Abbott_IMRcon,Abbott_IMRcon2} enables one to detect emergent modified theories of gravity, manifesting themselves as a difference between the remnant BH parameters $(M_f,\chi_f)$, as computed from the inspiral, and merger-ringdown portions of the waveform individually.
Such a test can be performed by computing the two-dimensional posterior probability distributions $P_\II(M_f,\chi_f)$ and $P_\MR(M_f,\chi_f)$ from each section of the waveform.
The overlap of such distributions can determine how well GR describes the observed signal.

The agreement between the two above distributions is typically measured by transforming to the new parameters $\epsilon$ and $\sigma$, describing the departures $\Delta M_f$ and $\Delta \chi_f$ from the GR predictions of final mass and spin from the inspiral and merger-ringdown, normalized by the averages between the two $\bar{M}_f$ and $\bar{\chi}_f$~\cite{Ghosh_2017}
\begin{align}\label{eq:epsilon}
\epsilon &\equiv \frac{\Delta M_f}{\bar{M}_f} \equiv 2 \frac{M_f^\II-M_f^\MR}{M_f^\II+M_f^\MR},\\
\label{eq:sigma} \sigma &\equiv \frac{\Delta \chi_f}{\bar{\chi}_f} \equiv 2 \frac{\chi_f^\II-\chi_f^\MR}{\chi_f^\II+\chi_f^\MR}.
\end{align}
The probability distributions $P_\II(M_f,\chi_f)$ and $P_\MR(M_f,\chi_f)$ can be transformed to $P(\epsilon,\sigma)$ by following the Appendix of Ref.~\cite{Ghosh_2017}, resulting in the final expression given by:
\begin{align}\label{eq:transform}
P(\epsilon,\sigma)&=\int\limits^1_0 \int\limits^{\infty}_0 P_\II \left( \left\lbrack 1+\frac{\epsilon}{2} \right\rbrack \bar{M}_f , \left\lbrack 1+\frac{\sigma}{2} \right\rbrack \bar{\chi}_f \right)\\
\nonumber &\times P_\MR \left( \left\lbrack 1-\frac{\epsilon}{2} \right\rbrack \bar{M}_f , \left\lbrack 1-\frac{\sigma}{2} \right\rbrack \bar{\chi}_f \right) \bar{M}_f \bar{\chi}_f d\bar{M}_f d\bar{\chi}_f.
\end{align}
Finally, the consistency of the posterior probability distribution with the GR value of $(\epsilon,\sigma)|_\GR\equiv(0,0)$ will determine the agreement of the signal with GR.
Any statistically significant deviations from the GR value may uncover evidence  of modified theories of gravity present in any portion of the GW signal.
To date, all observed GW signals have been found to be consistent with GR~\cite{Abbott_IMRcon,Abbott_IMRcon2,Ghosh_IMRcon,Ghosh_IMRcon2,Ghosh_2017}.

While most similar tests are performed through a Bayesian statistical analysis~\cite{Abbott_IMRcon,Abbott_IMRcon2,Ghosh_IMRcon,Ghosh_IMRcon2,Ghosh_2017}, here we offer a new method using the Fisher analysis techniques described in Sec.~\ref{sec:technical} that is computationally less expensive. 
Namely, for each of the inspiral and merger-ringdown portions of the waveform, we first derive posterior distributions of parameters 
\begin{equation}
\theta^a_\GR=(\ln A, \phi_c, t_c, m_1,m_2, \chi_1, \chi_2),
\end{equation} 
using the Fisher analysis method. 
Next, we marginalize over the first three parameters to find the posterior distributions for $(m_1,m_2,\chi_1,\chi_2)$.
Marginalization over a given parameter is typically accomplished by integration over the full range of values, or in the case of multi-variate Gaussian distributions by simply removing the corresponding row and column from the covariance matrix $\Sigma_{ij}\equiv\Gamma_{ij}^{-1}$.
Finally, using the Jacobian transformation matrix and the NR fits provided in Ref.~\cite{PhenomDII}, the two-dimensional Gaussian probability distributions $P_\II(M_f,\chi_f)$ and $P_\MR(M_f,\chi_f)$ are constructed.

What are the limitations of the Fisher analysis? 
Below, we will only use the GR gravitational waveform, which corresponds to injecting the GR waveform and also recovering it with the GR waveform. 
Such a method does not allow us to estimate the systematic errors, and thus the final distribution is always centered around the true GR value, which is not the case in a real analysis~\cite{Abbott_IMRcon2,Abbott_IMRcon}. 
Moreover, the posterior distribution from the Fisher method is always Gaussian, and thus a 90\% credible contour in a two-dimensional parameter space is always given by an ellipse, which is also not true in reality.

However, what a Fisher analysis can accurately describe is the \emph{size} and  direction of correlation of the posterior distributions for $(\epsilon,\sigma)$, which are of high value when predicting the future resolving power from the GR value of $(0,0)$.
Throughout this analysis, we consider the \emph{area} of the 90\% confidence region as a metric of the discriminatory power one can gain upon use of this test with ground-based, space-based, and multi-band detections.
Such information may be used to gain valuable insight on how well future observations can discern GR effects from non-GR effects.
In this investigation, we choose $132$ Hz to be the separating frequency between the inspiral and merger-ringdown portions of the signal, as was chosen in Ref.~\cite{Abbott_IMRcon2}.

\subsection{Results}

\begin{figure}
\includegraphics[width=\columnwidth]{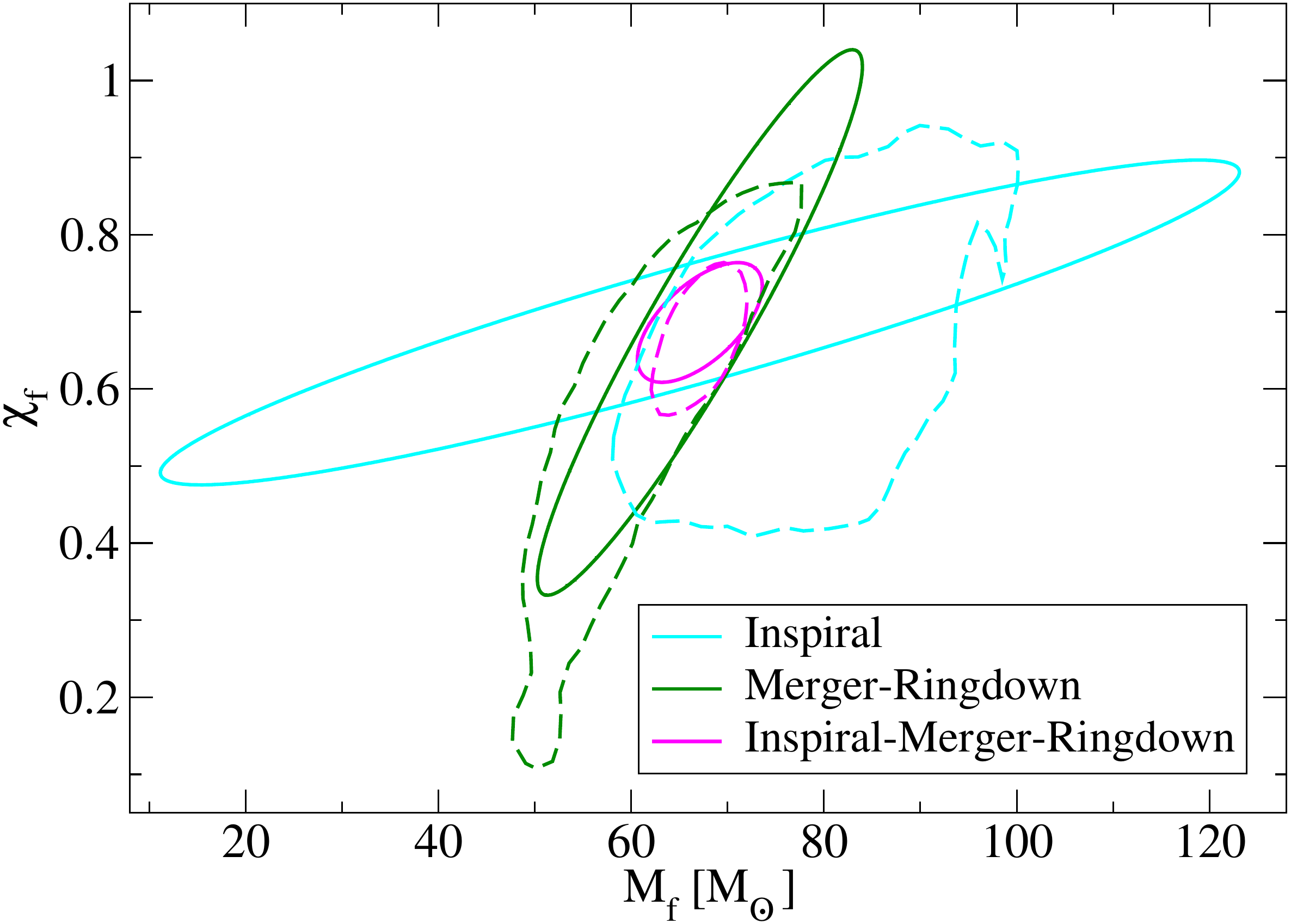}
\caption{90\% credible region contours of the inspiral, merger-ringdown, and complete waveform posterior distributions in the $M_f-\chi_f$ plane, for GW150914-like events observed on the LIGO O1 detector.
We present both the Fisher analysis results (solid) discussed here and the Bayesian results of Ref.~\cite{Abbott_IMRcon} (dashed) for comparison.
We observe good agreement between the two analyses in both the direction of correlation, and in the overall areas, which agree to within 10\% for all three distributions.
}\label{fig:IMRDconsistency}
\end{figure}

Figure~\ref{fig:IMRDconsistency} displays the 90\% confidence regions of the remnant mass and spin predictions from the inspiral, merger-ringdown, and full waveforms as detected on LIGO O1, in comparison with the Bayesian results of Ref.~\cite{Abbott_IMRcon}.
Here we see good agreement between the probability distributions, in both the direction of correlation, and the area of the 90\% confidence regions -- the latter agreeing to within 10\% for all contours considered.
We note that the agreement between the inspiral and merger-ringdown probability distributions indicates the degree of consistency with GR.

Next, we follow Ref.~\cite{Ghosh_2017} to transform the individual inspiral and merger-ringdown probability distributions into the joint probability distribution between new parameters $(\epsilon,\sigma)$ via Eq.~\eqref{eq:transform}.
These quantities determine the remnant mass and spin (predictions assuming GR) discrepancies $\Delta M_f$ and $\Delta \chi_f$ between the inspiral and merger-ringdown waveforms, normalized by the averages between the two $\bar{M}_f$ and $\bar{\chi}_f$.
Figure~\ref{fig:IMRDconsistencyTransformed} displays the estimated 90\% credible regions in the $\epsilon-\sigma$ plane for GW150914-like events observed on the following detectors: LIGO O1 (Fisher and Bayesian\footnote{Such Bayesian results are extracted from the \texttt{IMRPhenomPv2} results of Ref.~\cite{Abbott_IMRcon}. Similar results were found with the non-precessing \texttt{SEOBNRv4} model presented there.}~\cite{Abbott_IMRcon}), CE, and the multi-band observations between CE and TianQin, LISA, B-DECIGO, and DECIGO\footnote{As the merger-ringdown portion of the signal begins beyond the observing capacity of all space-based detectors for GW150914-like events, the IMR consistency test may not be performed solely by space-based detectors for such events. However, Ref.~\cite{Hughes:2004vw} showed that supermassive BH binaries are compatible with these observations.}.
The consistency of such distributions with the GR value of $(\sigma,\epsilon)|_\GR=(0,0)$ gives insight into how well the entire waveform agrees with GR, while any statistically significant deviations may indicate non-GR effects.

\begin{figure}
\includegraphics[width=\columnwidth]{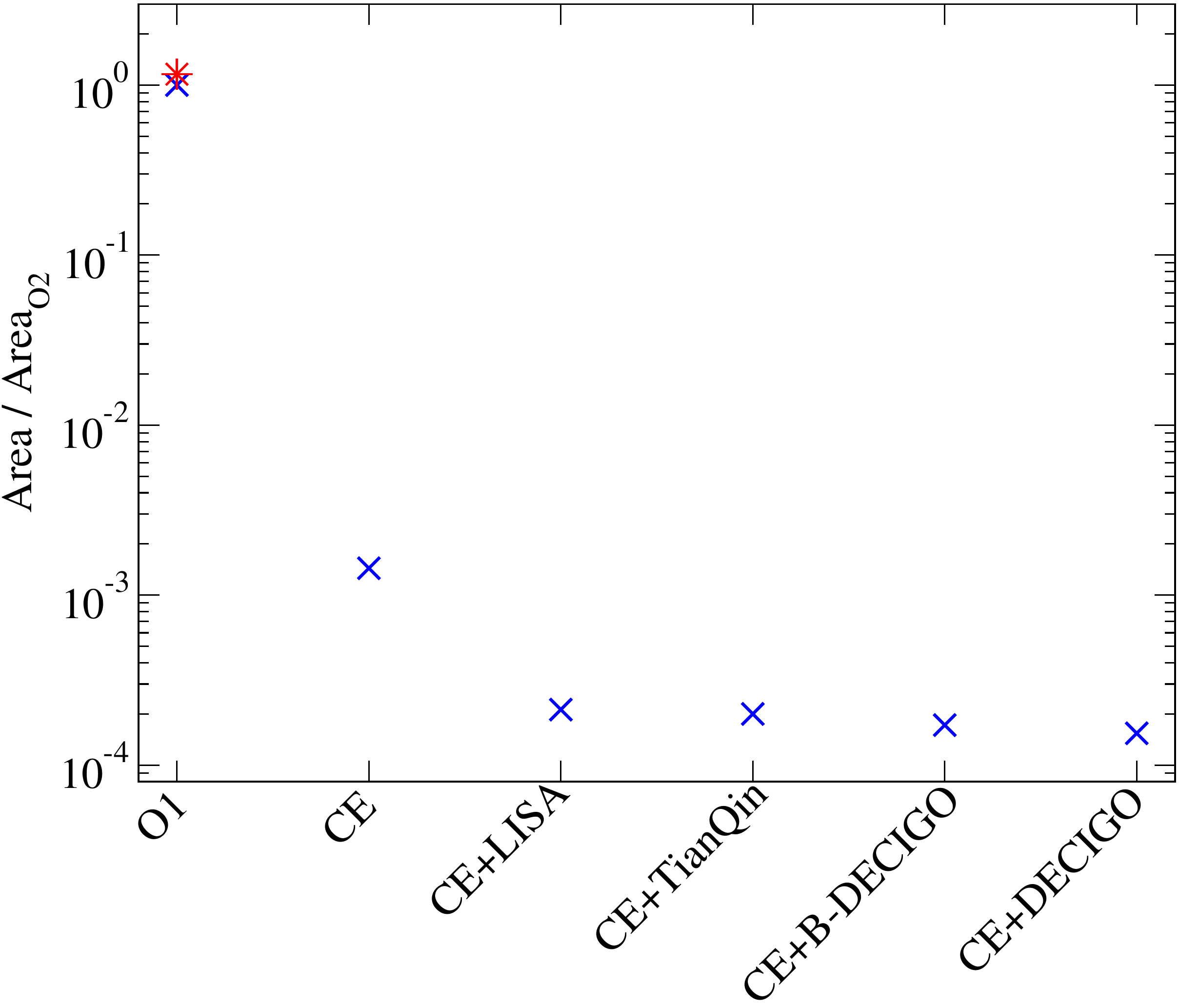}
\caption{Ratios of the areas of the 90\% credible regions relative to that found with the LIGO O1 detector (Fisher result) shown in Fig.~\ref{fig:IMRDconsistency} for GW150914-like events, obtained from a Fisher analysis (blue cross).
We report good agreement within 10\% between the LIGO O1 Fisher and Bayesian~\cite{Abbott_IMRcon} (red star) results.
We also observe up to three orders-of-magnitude of improvement from the results of LIGO O1 to CE, and a further improvement of 7-10 times upon the use of multi-band observations.
}\label{fig:IMRDareas}
\end{figure}

Now we quantify the resolving power gained for each single-band and multi-band observation, describing how effectively one can discriminate between GR and non-GR effects.
To do this, we compute and compare the areas of the 90\% confidence regions as a metric towards this resolution.
Figure~\ref{fig:IMRDareas} presents the ratio of such areas for the LIGO O1 (Fisher and Bayesian~\cite{Abbott_IMRcon}) detector relative to CE, and to the multi-band observations with CE and TianQin, LISA, B-DECIGO, and DECIGO.
Here we observe three important features.
First, the results obtained here for LIGO O1 agree very well (within 10\%) with the Bayesian analysis of Ref.~\cite{Abbott_IMRcon}, showing good validity of our Fisher-estimated analysis.
Second, we observe almost a three-order-of-magnitude improvement upon the use of CE from the results of LIGO O1.
Third, we see additional gains in resolving power by a factor of 7-10 upon the use of multi-band observations.

\section{Conclusion and Discussion}\label{sec:conclusion}
In this paper, we have highlighted the power in utilizing multi-band observations of GWs to test GR.
We began by performing parameterized tests of GR by considering generalized modifications to the GW phase, finding that multi-band observations can provide constraints reaching up to 40 times stronger than their single-band counterparts.
Such constraints were applied to the specific cases of EdGB, dCS, scalar-tensor, noncommutative, varying-$G$, varying-$M$, and massive graviton theories of gravity, resulting in constraints on the theories' associated parameters.
In particular, we find that constraints placed on the EdGB, dCS, noncommutative, and massive graviton non-GR effects show improvement upon the current constraints (by up to seven orders-of-magnitude for dCS gravity) found in the literature when making use of multi-band detections.

We next investigated the consistency between the inspiral and merger-ringdown portions of the gravitational waveform, in the so called ``IMR consistency test".
We demonstrated the resolving power gained upon the use of multi-band observations, finding up to an order-of-magnitude improvement relative to the single-band detections made by ground-based detectors alone.
Such an improvement gives way to the enhanced opportunity to shed light on even the most minuscule deviations from GR in the extreme gravity regime.

While this work demonstrated the large gains one can make on testing GR upon the use of multi-band observations, the analysis can be improved in numerous ways.
One example of such improvements would be to consider a full Bayesian analyses rather than the Fisher analysis used here -- although it was found that our results agree well with their Bayesian counterparts.
Additionally, one can simulate the multi-band event rates discussed in Refs.~\cite{Sesana:2016ljz,Cutler:2019krq,Gerosa:2019dbe} to combine the signals and further reduce the systematic errors residing in our bounds on non-GR theoretical parameters.
Finally, we can use the results in App.~\ref{sec:amp} to consider alternative theories of gravity which modify the GW amplitude rather than phase, such as the GW amplitude birefringence in parity-violating gravity~\cite{Alexander:2007kv,Yunes:2008bu,Yunes:2010yf,Yagi:2017zhb}.


\section*{Acknowledgments}\label{acknowledgments}
We thank Katerina Chatziioannou and Carl-Johan Haster for providing valuable comments on the comparisons between Fisher and Bayesian analyses. 
We also thank Enrico Barausse and Jian-dong Zhang for pointing us to the correct noise curve for TianQin.
Z.C. and K.Y. acknowledge support from NSF Award PHY-1806776 and the Ed Owens Fund. K.Y. would like to also acknowledge support by a Sloan Foundation Research Fellowship, the COST Action GWverse CA16104 and JSPS KAKENHI Grants No. JP17H06358.

\appendix

\section{Modified theories of gravity}\label{sec:theory}

In this appendix we describe the several modified theories of gravity considered in this paper, which can be thought of as breaking or deforming the fundamental pillars of GR.
First, the \emph{strong equivalence principle} (SEP) pillar~\cite{Will_SEP}, which states that the trajectories of free-falling and self-gravitating bodies are independent of their internal structure.
Second, the \emph{Lorentz invariance} (LI) pillar, which tells us that there is no preferred direction in our universe.
And last, the \emph{four dimensional spacetime} (4D) pillar, which conveys that the universal spacetime is composed of only four dimensions: 3 spatial and 1 temporal.
Finally, we consider the principle of \emph{massless gravitons} ($m_g$) as a result of GR, which describes gravity as being mediated by massless bosons traveling at the speed of light.
Theories which violate these fundamental pillars of GR can be broadly cataloged into two groups:
\begin{itemize}
\item Modifications to GW generation mechanisms: These modifications to GR alter how GWs are formed, and are active only during the coalescence event, with non-zero time derivatives of the source multipole moments.
Because of this, these theories depend only on the local properties of the source, such as the masses and spins.
\item Modifications to GW propagation mechanisms: These modifications alter the speed or dispersion relations of GWs themselves, and are only active  during their travel between their source and Earth.
Because of this, such theories depend on global properties such as the luminosity distance $D_L$ to the event.
\end{itemize} 
In the following subsections, we provide a brief description of the modified theories of gravity considered in this investigation, together with the mapping between the ppE parameter $\beta$ and the theoretical parameters.
We point the reader towards the more comprehensive analyses of Refs.~\cite{Berti_ModifiedReviewLarge,Berti_ModifiedReviewSmall,Yunes_ModifiedPhysics} for more complete descriptions of each theory.

\subsection{Generation mechanism modifications}\label{sec:theory_gen}

\subsubsection{EdGB gravity}
EdGB gravity is a string theory inspired gravity in which an additional scalar field $\phi$ non-minimally couples to a quadratic curvature term in the action~\cite{Maeda:2009uy}.
In this SEP-violating theory, scalar monopole charge is then accumulated on BHs -- inducing  scalar dipole radiation and ultimately accelerating the rate of inspiral between the gravitating bodies.
The primary coupling factor in this theory is $\alpha_\EdGB$. 
The mapping between $\beta_\EdGB$ and the coupling parameter $\alpha_\EdGB$ is given by~\cite{Yagi_EdGBmap}
\begin{equation}
\beta_\EdGB=-\frac{5}{7168}\frac{16\pi\alpha_\EdGB^2}{M^4}\frac{(m_1^2s_2^\EdGB-m_2^2s_1^\EdGB)^2}{M^4\eta^{18/5}}.
\end{equation}
In the above expression, $s_i^\EdGB\equiv2(\sqrt{1-\chi_i^2}-1+\chi_i^2)/\chi_i^2$ are the dimensionless EdGB  BH scalar charges normalized by the masses~\cite{Berti_ModifiedReviewSmall,Prabhu:2018aun}, where $\chi_i \equiv |\vec{S_i}|/m_i^2$ are the dimensionless spins of BHs with spin angular momentum $\vec{S}_i$.
The ppE exponent is $b=-7$ in this theory, which means that the leading correction enters at $-1$PN order.
We also note that here, it is assumed that the small coupling approximation $\zeta_\EdGB \equiv 16\pi\alpha_\EdGB^2/M^4 \ll 1$ is satisfied, else meaningful constraints on $\alpha_\EdGB$ may not be placed.
In particular, modifications to the gravitational waveform in EdGB gravity were derived under the assumption of the small-coupling approximation, thus the ppE framework is only valid within it.
In addition,  beyond the small-coupling approximation corresponds to large couplings between the scalar field and the curvature, which have largely been ruled out with observations.
Current constraints on $\sqrt{\alpha_\EdGB}$ are $10^7$ km~\cite{Amendola_EdGB} (solar system) and $2$ km~\cite{Kanti_EdGB,Pani_EdGB,Yagi_EdGB,Nair_dCSMap,Yamada:2019zrb,Tahura:2019dgr} (theoretical; low-mass X-ray binaries; GWs).

\subsubsection{dCS gravity}
Similar to EdGB gravity, dCS gravity is a SEP-violating effective field theory which modifies the Einstein-Hilbert action with a quadratic curvature term called the Pontryagin density, which violates parity, and is non-minimally coupled to a scalar field~\cite{Jackiw:2003pm,Alexander_cs}.
Scalar dipole charge is accumulated on the BHs, inducing scalar quadrupole radiation which in turn accelerates the inspiral.
The magnitude of the correction is proportional to the dCS coupling parameter $\alpha_\dCS$. 
The mapping between $\beta_\dCS$ and $\alpha_\dCS$ can be written as~\cite{Nair_dCSMap}
\begin{align}
\nonumber \beta_\dCS=&-\frac{5}{8192}\frac{16\pi\alpha_\dCS^2}{M^4\eta^{14/5}}\frac{(m_1 s_2^\dCS-m_2 s_1^\dCS)^2}{M^2}\\
\nonumber &+\frac{15075}{114688}\frac{16\pi\alpha_\dCS^2}{M^4\eta^{14/5}} \\ 
&\times \left( \frac{m_1^2 \chi_1^2 +m_2^2 \chi_2^2}{M^2}-\frac{305}{201}\eta \chi_1 \chi_2  \right),
\end{align}
where the dimensionless BH scalar charge can be written as $s_i^\dCS=(2+2\chi_i^4-2\sqrt{1-\chi_i^2}-\chi_i^2 \lbrack 3-2\sqrt{1-\chi_i^2} \rbrack)/2\chi_i^3$~\cite{Yagi:2012vf}. The ppE exponent is $b = -1$, which corresponds to a $+2$PN correction.
We also note that once again the small coupling approximation $\zeta_\dCS \equiv 16\pi\alpha_\dCS^2/M^4 \ll 1$ must be satisfied in order to place meaningful constraints on $\sqrt{\alpha_\dCS}$.
Current constraints on $\sqrt{\alpha_\dCS}$ are obtained from solar system and table-top experiments as $10^8$ km~\cite{AliHaimoud_dCS,Yagi_dCS}.

\subsubsection{Scalar-Tensor theories}
Scalar-tensor theories which violate the SEP include a coupling into the Einstein Hilbert action, where the Ricci scalar $R$ is multiplied by some function of the scalar field $\phi$.
If such a scalar field is time-dependent with a growth rate of $\dot\phi$ (for example, from a cosmological background~\cite{Jacobson_STcosmo,Horbatsch_STcosmo,Berti_STcosmo}), BHs will accumulate scalar charges which accelerate the inspiral.
The mapping between $\beta_\ST$ and $\dot\phi$ is given by~\cite{Horbatsch_STcosmo,Jacobson_STcosmo}
\begin{equation}
\beta_\ST=-\frac{5}{1792}\dot{\phi}^2\eta^{2/5}(m_1 s_1^\ST-m_2s_2^\ST)^2,
\end{equation}
where the dimensionless BH scalar charges $s_i^\ST$ are given by $s_i^\ST\equiv(1+\sqrt{1-\chi_i^2})/2$~\cite{Horbatsch_STcosmo}.
The ppE exponent $b = -7$ is the same as the EdGB case and the correction enters in the waveform at $-1$PN order.
We also note here that once again the small coupling approximation $\dot \phi \, m_i \ll 1$ must be upheld for meaningful constraints to be extracted.
The current most stringent constraint on $\dot\phi$ is $10^{-6}$ s$^{-1}$~\cite{Horbatsch_STcosmo} obtained from the orbital decay of a supermassive BH binary OJ287.

\subsubsection{Noncommutative gravity (NC)}
Noncommutative theories of gravity~\cite{Harikumar:2006xf} have been proposed to quantize the spacetime coordinates, which have been promoted to operators~\cite{Snyder:QuantizedST} $\hat{x}^\mu$, in order to eliminate the quantum field theory ultraviolet divergences.
Such theories have the ultimate goal of unifying the theories of GR and quantum mechanics.
The spacetime operators as such, satisfy the familiar canonical commutation relations
\begin{equation}
\lbrack \hat{x}^\mu , \hat{x}^\nu \rbrack = i \theta^{\mu\nu},
\end{equation}
where $\theta^{\mu\nu}$ quantifies the ``fuzziness" of spacetime coordinates, similar to the reduced Planck's constant $\hbar$ in quantum mechanics.
Within this non-commutating formalism, we strive to constrain the scale of quantum spacetime.
A useful parameter to do so normalizes the magnitude of $\theta^{\mu\nu}$ to the Planck length and time scales $l_p$ and $t_p$: $\Lambda^2 \equiv \theta^{0i}\theta_{0i}/l_p^2t_p^2$.
The Lorentz-violating effects from noncommutative gravity enters the gravitational waveform at $+2$PN order ($b=-1$), and has the ppE phase correction given by~\cite{Tahura_GdotMap}
\begin{equation}
\beta_\NC=-\frac{75}{256}\eta^{-4/5}(2\eta-1)\Lambda^2.
\end{equation}
The current constraints on the scale of quantum spacetime $\sqrt{\Lambda}$ come from the GW observation of GW150914, found to be $\sqrt{\Lambda}<3.5$~\cite{Kobakhidze:2016cqh}, which is on the order of the Planck scale.

\subsubsection{Time-varying $G$ theories}
The gravitational constant $G$ may vary with time at a rate of $\dot G$, producing an anomalous acceleration of the binary system. 
In this SEP-violating theory, the mapping between $\beta_{\dot G}$ and $\dot G$ is given by~\cite{Yunes_GdotMap,Tahura_GdotMap}
\begin{align}
\nonumber \beta_{\dot G}=&-\frac{25}{851968}\eta^{3/5}\dot{G}_C\lbrack 11M\\
&+3(s_1+s_2-\delta_{\dot G})M-41(m_1s_1+m_2s_2) \rbrack,
\end{align}
with $b= - 13$ ($-4$PN order).
Here, the sensitivities are given by $s_i\equiv -\frac{G_C}{\delta G_C}\frac{\delta m_i}{m_i}$, and $\delta_{\dot G}\equiv \dot G_D - \dot G_C$ is the difference between the variation rate in the dissipative gravitational constant $G_D$ (entering in GW luminosity), and the conservative one $G_C$ (entering in Kepler's law)\footnote{See a recent paper~\cite{Wolf:2019hun} that also introduces the two different gravitational constants.}.
For simplicity, we assume $\delta_{\dot G}=0$, so that the rate of change for both dissipative and conservative gravitational constants is equivalent: $\dot G_C = \dot G_D = \dot G$.
The current strongest constraint on $|\dot G|$ is $(0.1-1)\times 10^{-12}$ yr$^{-1}$~\cite{Bambi_Gdot,Copi_Gdot,Manchester_Gdot,Konopliv_Gdot}.

\subsubsection{Time-varying BH mass theories}
Some (4D-pillar-violating) modified theories of gravity as well as astrophysical processes predict time-variation in the BH mass, $\dot{m}_A$.
Many of the string-inspired models suggest that our four-dimensional brane spacetime is embedded in larger dimensional bulks~\cite{ArkaniHamed:1998rs,ArkaniHamed:1998nn,Randall:1999ee,Randall:Braneworld,Berti_ModifiedReviewSmall}.
One example is the RS-II~\cite{Randall:Braneworld} ``braneworld" model by Randall and Sundrum, in which BHs may evaporate \emph{classically}~\cite{Emparan:2002px,Tanaka:2002rb}\footnote{This scenario is now in question given the construction of brane-localized \emph{static} BH solutions~\cite{Figueras:2011gd,Abdolrahimi:2012qi}.}. 
The evaporation rate is proportional to the size $l$ of the extra dimension, which has previously been constrained to $(10-10^3)$ $\mu$m~\cite{Adelberger:2006dh,Johannsen_ED,Johannsen_ED2,Psaltis_ED,Gnedin_ED}.
Such a modification to the BH mass enters the waveform at $-4$PN order ($b=-13$), and $\beta_{\dot{M}}$ can be mapped to $\dot{M}$ via~\cite{Yagi_EDmap}
\begin{equation}
\beta_{\dot M}=\frac{25}{851968} \dot{M}\frac{3-26\eta+34\eta^2}{\eta^{2/5}(1-2\eta)}.
\end{equation}
The evaporation rate of the binary system $\dot M$ can be written as a function of $l$~\cite{Emparan_Mdot,Berti_ModifiedReviewSmall} in the RS-II model, or mapped to any other desired model.
Alternatively, BH mass losses can be explained by cosmological effects such as the accretion of dark (or ``phantom") energy~\cite{Babichev:2014lda,Babichev:2005py,Babichev:2004yx}.
For comparison purposes, we compute the astrophysical Eddington mass accretion rate $\dot{M}_{\text{Edd}}$ at which the BH radiates the Eddington luminosity $L_{\text{Edd}}$.
For a GW150914-like binary BH, it is found to be $\dot{M}_{\text{Edd}} = 1.4\times10^{-6}\text{ M}_\odot/\text{yr}$.

\subsubsection{Dynamical graviton mass}
The ``dynamical massive graviton" theory~\cite{Zhang:2017jze} models the graviton's mass to be smaller than all current constraints in weak gravity regions (see Sec.~\ref{sec:MDR} below), while becoming much larger in dynamical, strong-field spacetimes such as in the presence of binary BH mergers.
As such, this theory enters the gravitational waveform as a generation modification, rather than the usual propagation mechanism.
Here, we offer a new ppE correction to the gravitational waveform via the fractional discrepancy between the observed and predicted decay rates of the binary system's period $\dot{P}$ found in Ref.~\cite{Finn:2001qi}. 
In particular, we focus on a class of massive gravity theories that correctly reduces to GR in the limit $m_g \to 0$ (via the cancelation of the Boulware-Deser ghost and the longitudinal mode~\cite{deRham:2016nuf}) by abandoning Lorentz invariance~\cite{Finn:2001qi}.
We found that such an effect enters the waveform at $-3$PN order ($b=-11$), with the correction given by
\begin{equation}
\beta_{m_g}=\frac{25}{19712}\frac{\mathcal{M}^2}{\hbar^2F(e)}m_g^2,
\end{equation}
where $F(e)$ is a function of the eccentricity (Eq. (4.11) of Ref.~\cite{Finn:2001qi}), taken to be $1$ for our analysis (corresponding to quasi-circular binaries).
Current constraints on the dynamical graviton mass have been found to be $5.2\times10^{-21}$ eV from binary pulsar observations~\cite{Miao:2019nhf}, and $\sim 10^{-14}$ eV from GW measurements~\cite{Chung:2018dxe}.

As we discuss in the next section, the mass of the graviton also changes the propagation of GWs. However, the amount of the graviton mass can be different between (i) in the vicinity of a BH and (ii) in the region where GWs propagate from a source to us. Thus, we treat these effects separately in this paper. We consider these effects one at a time, though one could introduce two different graviton masses, like the dynamical and propagating graviton mass, and measure these two graviton masses. However, the dynamical graviton mass that gives rise to non-GR corrections at the level of the GW generation introduces modifications to the waveform phase at $-3PN$, while those from modifications to the GW propagation enters at 1PN order. Since these two PN orders are well-separated, the amount of correlation is small, and thus we expect the bound presented here gives us a good estimate on each effect.

\subsection{Propagation mechanism modifications}\label{sec:theory_prop}
Modifications to the propagation of GWs activate during their transport between the binary coalescence source and Earth.
As such, these modifications typically violate the LI pillar of gravity as well as the massless graviton, and describe corrections to the frequency dispersion of GWs, which in turn modifies the propagation speed of GWs.
These modifications depend primarily on the distance between the binary and Earth.

\subsubsection{Modified dispersion relations (MDR)}
\label{sec:MDR}

In general, the dispersion relation for GWs with modified theories of gravity takes the following form~\cite{Mirshekari_MDR}:
\begin{equation}
E^2=p^2+\mathbb{A} \, p^{a_\MDR},
\end{equation}\label{eq:MDR}
where $E$ and $p$ are the graviton's energy and momentum, $a_\MDR$ is related to the PN order via $n=1+\frac{3}{2}a_\MDR$, and $\mathbb{A}$ corresponds to the strength of the dispersion. 
The mapping between $(\beta_\MDR,b)$ and $(\mathbb{A},a_\MDR)$ is given by~\cite{Mirshekari_MDR}
\begin{eqnarray}
\beta_\MDR&=&\frac{\pi^{2-a_\MDR}}{1-a_\MDR}\frac{D_{a}}{\lambda_{\mathbb{A}}^{2-a_\MDR}}\frac{\mathcal{M}^{1-a_\MDR}}{(1+z)^{1-a_\MDR}}, \\
b&=& 3 (a_\MDR -1).
\label{eq:MDR}
\end{eqnarray}
Here, $z$ is the redshift, $\lambda_{\mathbb{A}}\equiv h \mathbb{A}^{1/(a_\MDR-2)}$ is similar to the Compton wavelength with Plank's constant $h$, and the effective distance $D_a$ is given by~\cite{Mirshekari_MDR,Yunes_ModifiedPhysics}
\begin{eqnarray}
D_{a}&=&\frac{z}{H_0\sqrt{\Omega_M+\Omega_{\Lambda}}} \left \lbrack 1-\frac{z}{4} \left(  \frac{3 \Omega_M}{\Omega_M+\Omega_{\Lambda}}+2a_\MDR \right)  \right\rbrack \nonumber \\
&&+ \mathcal{O}(z^3),
\end{eqnarray}
where $H_0=67.9$ km sec$^{-1}$ Mpc$^{-1}$ is the local Hubble constant, and $\Omega_M=0.303$, and $\Omega_{\Lambda}=0.697$ are the energy densities of matter and dark energy~\cite{Aghanim:2018eyx}.

In this paper, we mainly investigate bounds on the specific case of the \emph{massive graviton}~\cite{deRham_mg,Hinterbichler_mg,Rubakov_mg,Will_mg} (propagation), where $\mathbb{A}=m_g^2$ and $a_\MDR=0$.
The current constraints on the graviton mass have been found to be $6\times10^{-24}$ eV~\cite{Will:2018gku,Abbott_IMRcon} from solar-system constraints (Yukawa-like corrections to the binding energy and Kepler's law), $5\times10^{-23}$ eV~\cite{Abbott_IMRcon} from the combination of GW signals from the LVC catalog (GW propagation modifications), and $\sim 10^{-14}$ eV~\cite{Chung:2018dxe} or $5 \times 10^{-21}$ eV~\cite{Miao:2019nhf}  from GW150914 and binary pulsars respectively (GW generation modifications). 
Stronger bounds have been obtained from cosmological observations (see e.g.~\cite{deRham:2016nuf,Desai:2017dwg,Gupta:2018hgm}).

Additionally, we offer general constraints on $\mathbb{A}$ in App.~\ref{sec:amp}, applicable to many alternative theories of gravity with modified dispersion relations.
Some examples of these include~\cite{Mirshekari_MDR,Yunes_ModifiedPhysics} 
\begin{itemize}
\item \emph{Double special relativity}~\cite{Magueijo_dsr,AmelinoCamelia_dsr,AmelinoCamelia_dsr2,AmelinoCamelia_dsr3} with $\mathbb{A}=\eta_{\text{dsrt}}$ and $a_\MDR=3$; 
\item \emph{Extra-dimensional theories}~\cite{Sefiedgar_edt} with $\mathbb{A}=-\alpha_{\text{edt}}$ and $a_\MDR=4$;
\item \emph{Ho\v rava-Lifshitz Gravity}~\cite{Vacaru_horava,blas_horava,Horava,Horava_2} with $\mathbb{A}=\kappa_{\text{hl}}^4\mu_{\text{hl}}^2/16$ and $a_\MDR=4$;
\item \emph{Multifractional Spacetime Theory}~\cite{Calcagni_mf,Calcagni_mf2,Calcagni_mf3,Calcagni_mf4} with $\mathbb{A}=2E_*^{2-a_\MDR}/(3-a_\MDR)$ and $a_\MDR=2-3$.
\end{itemize}

\section{Bounds on the GW amplitude and dispersion relation corrections}\label{app:amplitude}
\label{sec:amp}

\begin{figure}
\includegraphics[width=.47\textwidth]{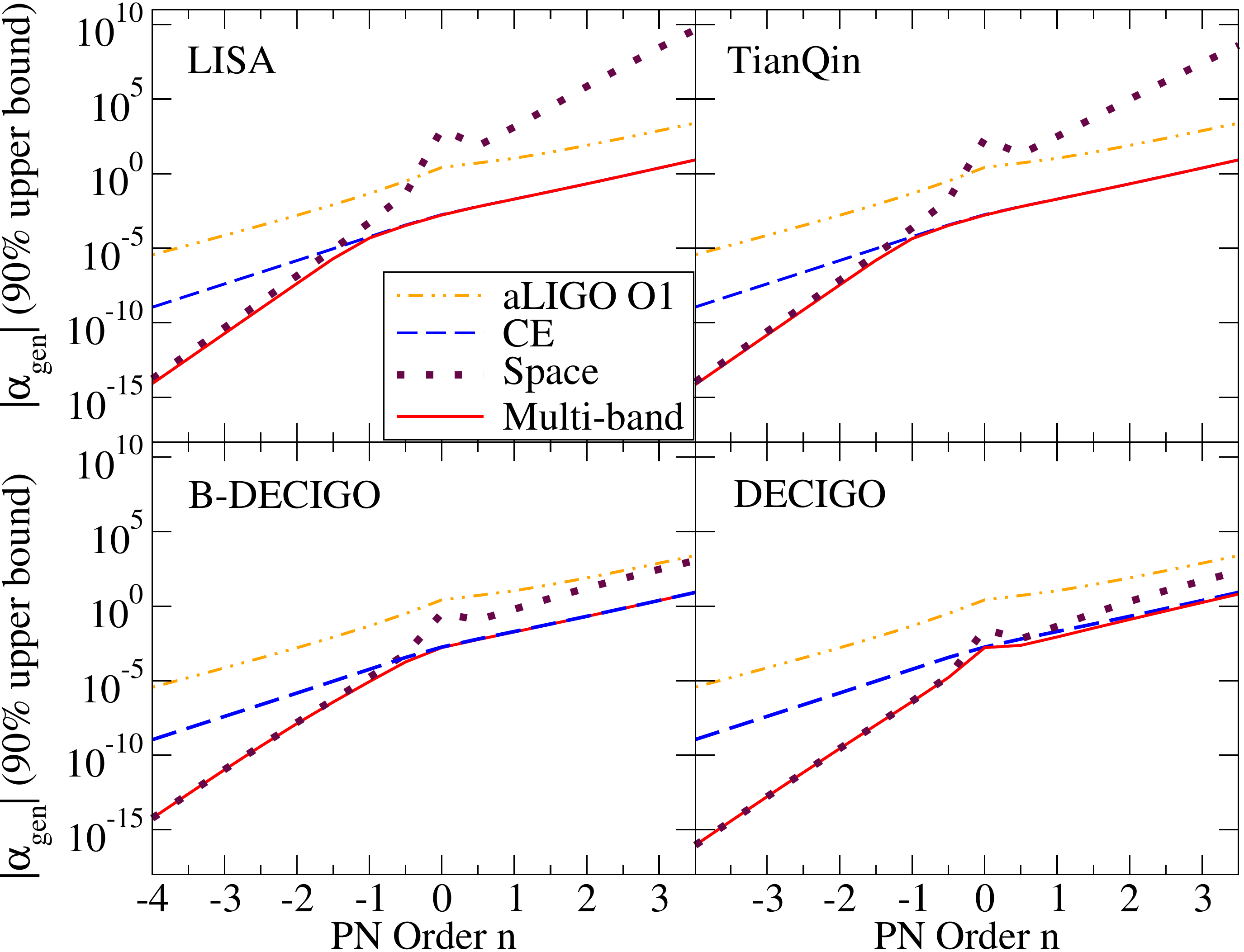}
\caption{
Similar to Fig.~\ref{fig:betaBoundsGen}, but for generation effects $\alpha_{\text{gen}}$ entering the GW amplitude.
Observe how for the case of amplitude corrections, multi-band observations do not provide for much improvement over space-based detectors for negative-PN orders, and ground-based detectors for positive-PN orders.
}\label{fig:alphaBoundsGen}
\end{figure}

\begin{figure}
\includegraphics[width=.47\textwidth]{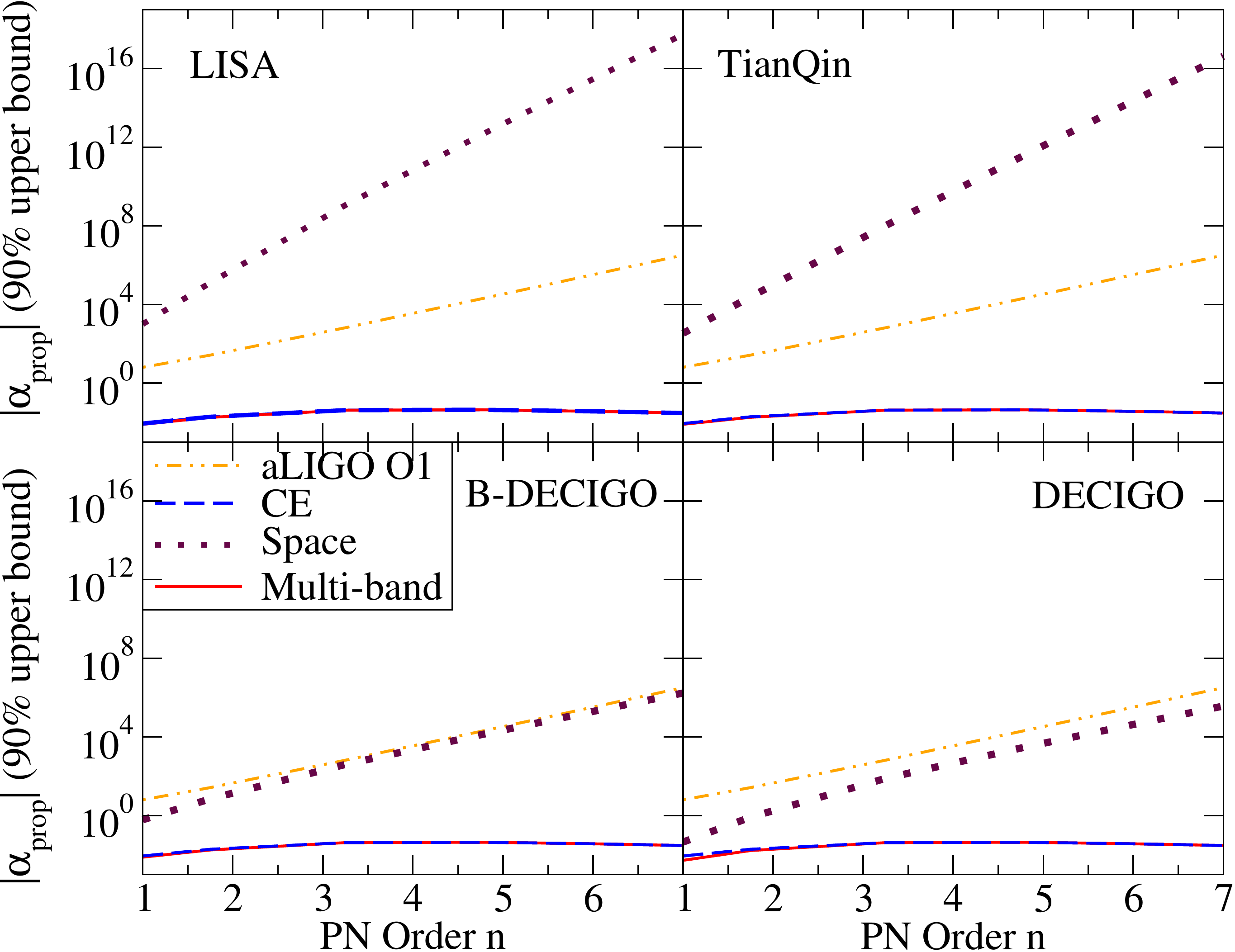}
\caption{
Similar to Fig.~\ref{fig:betaBoundsGen}, but for propagation effects $\alpha_{\text{prop}}$ entering the GW amplitude.
Observe how for the case of amplitude corrections, multi-band observations do not provide for much improvement over the constraints provided by CE.
}\label{fig:alphaBoundsProp}
\end{figure}

In this appendix, we present constraints on the ppE amplitude parameter $\alpha$, as well as on corrections to the graviton dispersion relation.
Figures~\ref{fig:alphaBoundsGen} and~\ref{fig:alphaBoundsProp} display the 90\% credible level upper limits on $|\alpha_{\text{gen}}|$ and $|\alpha_{\text{prop}}|$ for modified theories which affect GW generation, and propagation effects respectively.
Observe that the multi-band results simply follow bounds from space- (ground-)based detectors for corrections entering at negative (positive) PN orders, and do not have much improvement from single-band results.

\begin{figure}
\includegraphics[width=.49\textwidth]{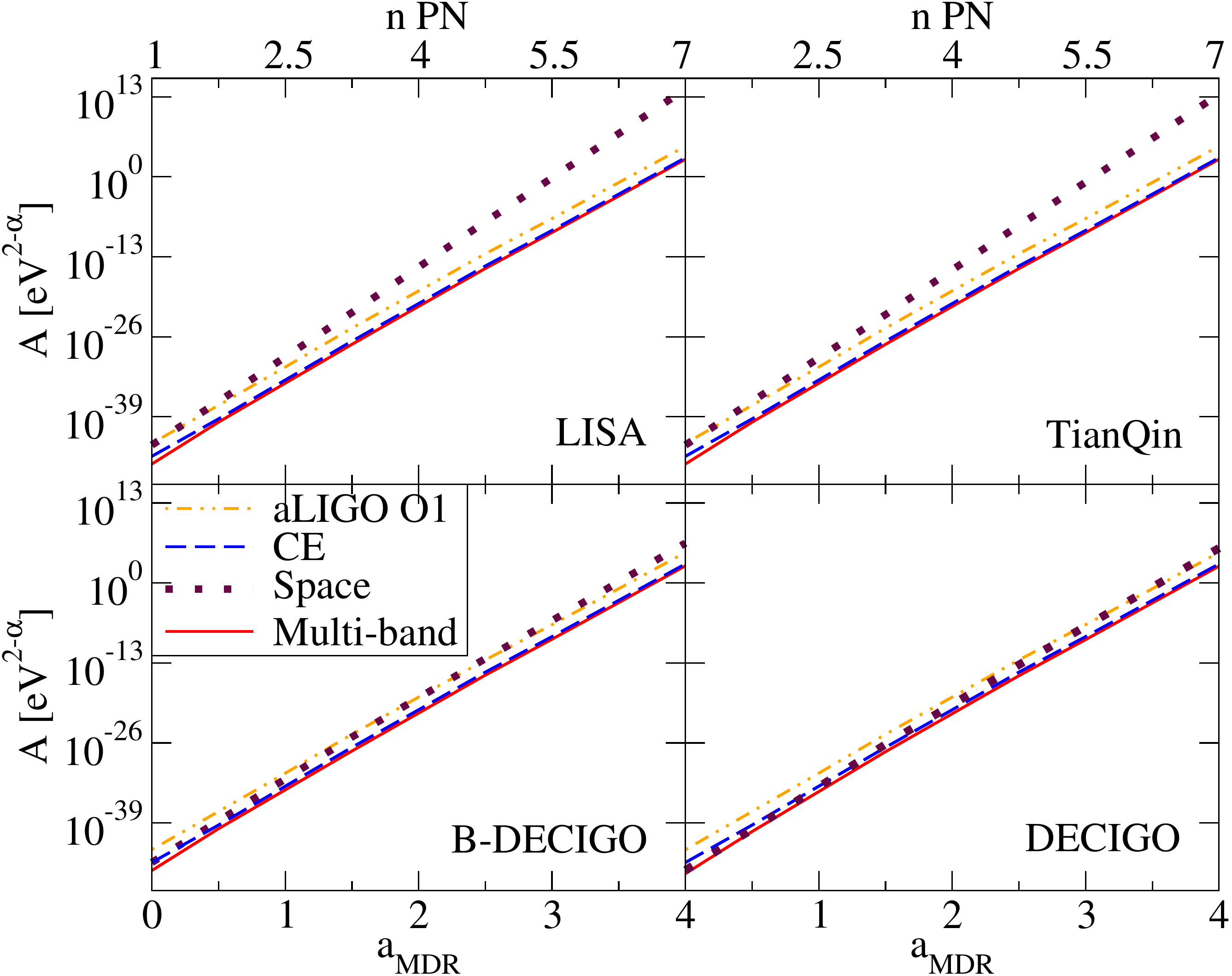}
\caption{
90\% upper-bound constraints on the magnitude of the correction to the graviton dispersion relation $\mathbb{A}$ as a function of $a_\MDR$, found in Eq.~\eqref{eq:MDR}.
Such bounds are computed from the ppE parameter for propagation mechanisms $\beta_{\text{prop}}$ for GW150914-like events on various space- and ground-based detectors, and the combination of the two.
Constraints on modifications to GR can be found by selecting a value of $\alpha$ and mapping $\mathbb{A}$ to its corresponding form, e.g. for massive gravitons with $a_\MDR=0$ and $\mathbb{A}=m_g^2$, with bounds displayed in Table~\ref{tab:theories} and Fig.~\ref{fig:gravitonMassBounds}.
}\label{fig:Abounds}
\end{figure}

Such constraints can be mapped to the desired coupling parameters of many modified theories of gravity~\cite{Tahura_GdotMap} similar to was done in Sec.~\ref{sec:parameterized}.
Figure~\ref{fig:Abounds} presents constraints on the generalized dispersion relation correction, $\mathbb{A}$.
Such bounds can be further applied to modified theories of gravity which predict modified dispersion relations, as discussed in the previous section.
Observe that the multiband bounds are very similar to those from CE, consistent with Fig.~\ref{fig:alphaBoundsProp}.

\clearpage
\bibliography{Zack}

 \newcommand{\noop}[1]{}
\begin{thebibliography}{148}%
\makeatletter
\providecommand \@ifxundefined [1]{%
 \@ifx{#1\undefined}
}%
\providecommand \@ifnum [1]{%
 \ifnum #1\expandafter \@firstoftwo
 \else \expandafter \@secondoftwo
 \fi
}%
\providecommand \@ifx [1]{%
 \ifx #1\expandafter \@firstoftwo
 \else \expandafter \@secondoftwo
 \fi
}%
\providecommand \natexlab [1]{#1}%
\providecommand \enquote  [1]{``#1''}%
\providecommand \bibnamefont  [1]{#1}%
\providecommand \bibfnamefont [1]{#1}%
\providecommand \citenamefont [1]{#1}%
\providecommand \href@noop [0]{\@secondoftwo}%
\providecommand \href [0]{\begingroup \@sanitize@url \@href}%
\providecommand \@href[1]{\@@startlink{#1}\@@href}%
\providecommand \@@href[1]{\endgroup#1\@@endlink}%
\providecommand \@sanitize@url [0]{\catcode `\\12\catcode `\$12\catcode
  `\&12\catcode `\#12\catcode `\^12\catcode `\_12\catcode `\%12\relax}%
\providecommand \@@startlink[1]{}%
\providecommand \@@endlink[0]{}%
\providecommand \url  [0]{\begingroup\@sanitize@url \@url }%
\providecommand \@url [1]{\endgroup\@href {#1}{\urlprefix }}%
\providecommand \urlprefix  [0]{URL }%
\providecommand \Eprint [0]{\href }%
\providecommand \doibase [0]{http://dx.doi.org/}%
\providecommand \selectlanguage [0]{\@gobble}%
\providecommand \bibinfo  [0]{\@secondoftwo}%
\providecommand \bibfield  [0]{\@secondoftwo}%
\providecommand \translation [1]{[#1]}%
\providecommand \BibitemOpen [0]{}%
\providecommand \bibitemStop [0]{}%
\providecommand \bibitemNoStop [0]{.\EOS\space}%
\providecommand \EOS [0]{\spacefactor3000\relax}%
\providecommand \BibitemShut  [1]{\csname bibitem#1\endcsname}%
\let\auto@bib@innerbib\@empty
\bibitem [{\citenamefont {{Popper}}(1934)}]{popper}%
  \BibitemOpen
  \bibfield  {author} {\bibinfo {author} {\bibfnamefont {K.}~\bibnamefont
  {{Popper}}},\ }\href@noop {} {\emph {\bibinfo {title} {{The Logic of
  Scientific Discovery}}}}\ (\bibinfo  {publisher} {Mohr Siebeck},\ \bibinfo
  {address} {Germany},\ \bibinfo {year} {1934})\BibitemShut {NoStop}%
\bibitem [{\citenamefont {Will}(2014{\natexlab{a}})}]{Will_SolarSystemTest}%
  \BibitemOpen
  \bibfield  {author} {\bibinfo {author} {\bibfnamefont {C.~M.}\ \bibnamefont
  {Will}},\ }\href {\doibase 10.12942/lrr-2014-4} {\bibfield  {journal}
  {\bibinfo  {journal} {Living Reviews in Relativity}\ }\textbf {\bibinfo
  {volume} {17}},\ \bibinfo {pages} {4} (\bibinfo {year}
  {2014}{\natexlab{a}})}\BibitemShut {NoStop}%
\bibitem [{\citenamefont {Stairs}(2003)}]{Stairs_BinaryPulsarTest}%
  \BibitemOpen
  \bibfield  {author} {\bibinfo {author} {\bibfnamefont {I.~H.}\ \bibnamefont
  {Stairs}},\ }\href {\doibase 10.12942/lrr-2003-5} {\bibfield  {journal}
  {\bibinfo  {journal} {Living Rev. Rel.}\ }\textbf {\bibinfo {volume} {6}},\
  \bibinfo {pages} {5} (\bibinfo {year} {2003})},\ \Eprint
  {http://arxiv.org/abs/astro-ph/0307536} {arXiv:astro-ph/0307536 [astro-ph]}
  \BibitemShut {NoStop}%
\bibitem [{\citenamefont {Wex}(2014)}]{Wex_BinaryPulsarTest}%
  \BibitemOpen
  \bibfield  {author} {\bibinfo {author} {\bibfnamefont {N.}~\bibnamefont
  {Wex}},\ }\href@noop {} {\  (\bibinfo {year} {2014})},\ \Eprint
  {http://arxiv.org/abs/1402.5594} {arXiv:1402.5594 [gr-qc]} \BibitemShut
  {NoStop}%
\bibitem [{\citenamefont {Ferreira}(2019)}]{Ferreira_CosmologyTest}%
  \BibitemOpen
  \bibfield  {author} {\bibinfo {author} {\bibfnamefont {P.~G.}\ \bibnamefont
  {Ferreira}},\ }\href@noop {} {\  (\bibinfo {year} {2019})},\ \Eprint
  {http://arxiv.org/abs/1902.10503} {arXiv:1902.10503 [astro-ph.CO]}
  \BibitemShut {NoStop}%
\bibitem [{\citenamefont {Clifton}\ \emph {et~al.}(2012)\citenamefont
  {Clifton}, \citenamefont {Ferreira}, \citenamefont {Padilla},\ and\
  \citenamefont {Skordis}}]{Clifton_CosmologyTest}%
  \BibitemOpen
  \bibfield  {author} {\bibinfo {author} {\bibfnamefont {T.}~\bibnamefont
  {Clifton}}, \bibinfo {author} {\bibfnamefont {P.~G.}\ \bibnamefont
  {Ferreira}}, \bibinfo {author} {\bibfnamefont {A.}~\bibnamefont {Padilla}}, \
  and\ \bibinfo {author} {\bibfnamefont {C.}~\bibnamefont {Skordis}},\ }\href
  {\doibase 10.1016/j.physrep.2012.01.001} {\bibfield  {journal} {\bibinfo
  {journal} {Phys. Rept.}\ }\textbf {\bibinfo {volume} {513}},\ \bibinfo
  {pages} {1} (\bibinfo {year} {2012})},\ \Eprint
  {http://arxiv.org/abs/1106.2476} {arXiv:1106.2476 [astro-ph.CO]} \BibitemShut
  {NoStop}%
\bibitem [{\citenamefont {Joyce}\ \emph {et~al.}(2015)\citenamefont {Joyce},
  \citenamefont {Jain}, \citenamefont {Khoury},\ and\ \citenamefont
  {Trodden}}]{Joyce_CosmologyTest}%
  \BibitemOpen
  \bibfield  {author} {\bibinfo {author} {\bibfnamefont {A.}~\bibnamefont
  {Joyce}}, \bibinfo {author} {\bibfnamefont {B.}~\bibnamefont {Jain}},
  \bibinfo {author} {\bibfnamefont {J.}~\bibnamefont {Khoury}}, \ and\ \bibinfo
  {author} {\bibfnamefont {M.}~\bibnamefont {Trodden}},\ }\href {\doibase
  10.1016/j.physrep.2014.12.002} {\bibfield  {journal} {\bibinfo  {journal}
  {Phys. Rept.}\ }\textbf {\bibinfo {volume} {568}},\ \bibinfo {pages} {1}
  (\bibinfo {year} {2015})},\ \Eprint {http://arxiv.org/abs/1407.0059}
  {arXiv:1407.0059 [astro-ph.CO]} \BibitemShut {NoStop}%
\bibitem [{\citenamefont {Koyama}(2016)}]{Koyama_CosmologyTest}%
  \BibitemOpen
  \bibfield  {author} {\bibinfo {author} {\bibfnamefont {K.}~\bibnamefont
  {Koyama}},\ }\href {\doibase 10.1088/0034-4885/79/4/046902} {\bibfield
  {journal} {\bibinfo  {journal} {Rept. Prog. Phys.}\ }\textbf {\bibinfo
  {volume} {79}},\ \bibinfo {pages} {046902} (\bibinfo {year} {2016})},\
  \Eprint {http://arxiv.org/abs/1504.04623} {arXiv:1504.04623 [astro-ph.CO]}
  \BibitemShut {NoStop}%
\bibitem [{\citenamefont {Salvatelli}\ \emph {et~al.}(2016)\citenamefont
  {Salvatelli}, \citenamefont {Piazza},\ and\ \citenamefont
  {Marinoni}}]{Salvatelli_CosmologyTest}%
  \BibitemOpen
  \bibfield  {author} {\bibinfo {author} {\bibfnamefont {V.}~\bibnamefont
  {Salvatelli}}, \bibinfo {author} {\bibfnamefont {F.}~\bibnamefont {Piazza}},
  \ and\ \bibinfo {author} {\bibfnamefont {C.}~\bibnamefont {Marinoni}},\
  }\href {\doibase 10.1088/1475-7516/2016/09/027} {\bibfield  {journal}
  {\bibinfo  {journal} {JCAP}\ }\textbf {\bibinfo {volume} {1609}},\ \bibinfo
  {pages} {027} (\bibinfo {year} {2016})},\ \Eprint
  {http://arxiv.org/abs/1602.08283} {arXiv:1602.08283 [astro-ph.CO]}
  \BibitemShut {NoStop}%
\bibitem [{\citenamefont {Ishak}(2019)}]{Ishak:2018his}%
  \BibitemOpen
  \bibfield  {author} {\bibinfo {author} {\bibfnamefont {M.}~\bibnamefont
  {Ishak}},\ }\href {\doibase 10.1007/s41114-018-0017-4} {\bibfield  {journal}
  {\bibinfo  {journal} {Living Rev. Rel.}\ }\textbf {\bibinfo {volume} {22}},\
  \bibinfo {pages} {1} (\bibinfo {year} {2019})},\ \Eprint
  {http://arxiv.org/abs/1806.10122} {arXiv:1806.10122 [astro-ph.CO]}
  \BibitemShut {NoStop}%
\bibitem [{\citenamefont {Abbott}\ \emph
  {et~al.}(2016{\natexlab{a}})\citenamefont {Abbott} \emph
  {et~al.}}]{GW150914}%
  \BibitemOpen
  \bibfield  {author} {\bibinfo {author} {\bibfnamefont {B.~P.}\ \bibnamefont
  {Abbott}} \emph {et~al.} (\bibinfo {collaboration} {LIGO Scientific,
  Virgo}),\ }\href {\doibase 10.1103/PhysRevLett.116.241102} {\bibfield
  {journal} {\bibinfo  {journal} {Phys. Rev. Lett.}\ }\textbf {\bibinfo
  {volume} {116}},\ \bibinfo {pages} {241102} (\bibinfo {year}
  {2016}{\natexlab{a}})},\ \Eprint {http://arxiv.org/abs/1602.03840}
  {arXiv:1602.03840 [gr-qc]} \BibitemShut {NoStop}%
\bibitem [{\citenamefont {Abbott}\ \emph {et~al.}(2018)\citenamefont {Abbott}
  \emph {et~al.}}]{GW_Catalogue}%
  \BibitemOpen
  \bibfield  {author} {\bibinfo {author} {\bibfnamefont {B.~P.}\ \bibnamefont
  {Abbott}} \emph {et~al.} (\bibinfo {collaboration} {LIGO Scientific,
  Virgo}),\ }\href@noop {} {\  (\bibinfo {year} {2018})},\ \Eprint
  {http://arxiv.org/abs/1811.12907} {arXiv:1811.12907 [astro-ph.HE]}
  \BibitemShut {NoStop}%
\bibitem [{\citenamefont {Abbott}\ \emph
  {et~al.}(2016{\natexlab{b}})\citenamefont {Abbott} \emph
  {et~al.}}]{Abbott_IMRcon2}%
  \BibitemOpen
  \bibfield  {author} {\bibinfo {author} {\bibfnamefont {B.~P.}\ \bibnamefont
  {Abbott}} \emph {et~al.} (\bibinfo {collaboration} {LIGO Scientific,
  Virgo}),\ }\href {\doibase 10.1103/PhysRevLett.116.221101,
  10.1103/PhysRevLett.121.129902} {\bibfield  {journal} {\bibinfo  {journal}
  {Phys. Rev. Lett.}\ }\textbf {\bibinfo {volume} {116}},\ \bibinfo {pages}
  {221101} (\bibinfo {year} {2016}{\natexlab{b}})},\ \bibinfo {note} {[Erratum:
  Phys. Rev. Lett.121,no.12,129902(2018)]},\ \Eprint
  {http://arxiv.org/abs/1602.03841} {arXiv:1602.03841 [gr-qc]} \BibitemShut
  {NoStop}%
\bibitem [{\citenamefont {Abbott}\ \emph {et~al.}(2019)\citenamefont {Abbott}
  \emph {et~al.}}]{Abbott_IMRcon}%
  \BibitemOpen
  \bibfield  {author} {\bibinfo {author} {\bibfnamefont {B.~P.}\ \bibnamefont
  {Abbott}} \emph {et~al.} (\bibinfo {collaboration} {LIGO Scientific,
  Virgo}),\ }\href@noop {} {\  (\bibinfo {year} {2019})},\ \Eprint
  {http://arxiv.org/abs/1903.04467} {arXiv:1903.04467 [gr-qc]} \BibitemShut
  {NoStop}%
\bibitem [{\citenamefont {Jordan}(1959)}]{Jordan1959}%
  \BibitemOpen
  \bibfield  {author} {\bibinfo {author} {\bibfnamefont {P.}~\bibnamefont
  {Jordan}},\ }\href {\doibase 10.1007/BF01375155} {\bibfield  {journal}
  {\bibinfo  {journal} {Zeitschrift f{\"u}r Physik}\ }\textbf {\bibinfo
  {volume} {157}},\ \bibinfo {pages} {112} (\bibinfo {year}
  {1959})}\BibitemShut {NoStop}%
\bibitem [{\citenamefont {Brans}\ and\ \citenamefont
  {Dicke}(1961)}]{Brans1961}%
  \BibitemOpen
  \bibfield  {author} {\bibinfo {author} {\bibfnamefont {C.}~\bibnamefont
  {Brans}}\ and\ \bibinfo {author} {\bibfnamefont {R.~H.}\ \bibnamefont
  {Dicke}},\ }\href {\doibase 10.1103/PhysRev.124.925} {\bibfield  {journal}
  {\bibinfo  {journal} {Phys. Rev.}\ }\textbf {\bibinfo {volume} {124}},\
  \bibinfo {pages} {925} (\bibinfo {year} {1961})}\BibitemShut {NoStop}%
\bibitem [{\citenamefont {Damour}\ and\ \citenamefont
  {Nordtvedt}(1993)}]{Damour1993_2}%
  \BibitemOpen
  \bibfield  {author} {\bibinfo {author} {\bibfnamefont {T.}~\bibnamefont
  {Damour}}\ and\ \bibinfo {author} {\bibfnamefont {K.}~\bibnamefont
  {Nordtvedt}},\ }\href {\doibase 10.1103/PhysRevD.48.3436} {\bibfield
  {journal} {\bibinfo  {journal} {Phys. Rev. D}\ }\textbf {\bibinfo {volume}
  {48}},\ \bibinfo {pages} {3436} (\bibinfo {year} {1993})}\BibitemShut
  {NoStop}%
\bibitem [{\citenamefont {Damour}\ and\ \citenamefont
  {Esposito-Far\`ese}(1993)}]{Damour1993}%
  \BibitemOpen
  \bibfield  {author} {\bibinfo {author} {\bibfnamefont {T.}~\bibnamefont
  {Damour}}\ and\ \bibinfo {author} {\bibfnamefont {G.}~\bibnamefont
  {Esposito-Far\`ese}},\ }\href {\doibase 10.1103/PhysRevLett.70.2220}
  {\bibfield  {journal} {\bibinfo  {journal} {Phys. Rev. Lett.}\ }\textbf
  {\bibinfo {volume} {70}},\ \bibinfo {pages} {2220} (\bibinfo {year}
  {1993})}\BibitemShut {NoStop}%
\bibitem [{\citenamefont {Damour}\ and\ \citenamefont
  {Esposito-Far\`ese}(1996)}]{Damour1996}%
  \BibitemOpen
  \bibfield  {author} {\bibinfo {author} {\bibfnamefont {T.}~\bibnamefont
  {Damour}}\ and\ \bibinfo {author} {\bibfnamefont {G.}~\bibnamefont
  {Esposito-Far\`ese}},\ }\href {\doibase 10.1103/PhysRevD.54.1474} {\bibfield
  {journal} {\bibinfo  {journal} {Phys. Rev. D}\ }\textbf {\bibinfo {volume}
  {54}},\ \bibinfo {pages} {1474} (\bibinfo {year} {1996})}\BibitemShut
  {NoStop}%
\bibitem [{\citenamefont {Damour}\ and\ \citenamefont
  {Esposito-Farese}(1992)}]{Damour1992}%
  \BibitemOpen
  \bibfield  {author} {\bibinfo {author} {\bibfnamefont {T.}~\bibnamefont
  {Damour}}\ and\ \bibinfo {author} {\bibfnamefont {G.}~\bibnamefont
  {Esposito-Farese}},\ }\href {\doibase 10.1088/0264-9381/9/9/015} {\bibfield
  {journal} {\bibinfo  {journal} {Classical and Quantum Gravity}\ }\textbf
  {\bibinfo {volume} {9}},\ \bibinfo {pages} {2093} (\bibinfo {year}
  {1992})}\BibitemShut {NoStop}%
\bibitem [{\citenamefont {Sampson}\ \emph {et~al.}(2014)\citenamefont
  {Sampson}, \citenamefont {Yunes}, \citenamefont {Cornish}, \citenamefont
  {Ponce}, \citenamefont {Barausse}, \citenamefont {Klein}, \citenamefont
  {Palenzuela},\ and\ \citenamefont {Lehner}}]{Sampson2014}%
  \BibitemOpen
  \bibfield  {author} {\bibinfo {author} {\bibfnamefont {L.}~\bibnamefont
  {Sampson}}, \bibinfo {author} {\bibfnamefont {N.}~\bibnamefont {Yunes}},
  \bibinfo {author} {\bibfnamefont {N.}~\bibnamefont {Cornish}}, \bibinfo
  {author} {\bibfnamefont {M.}~\bibnamefont {Ponce}}, \bibinfo {author}
  {\bibfnamefont {E.}~\bibnamefont {Barausse}}, \bibinfo {author}
  {\bibfnamefont {A.}~\bibnamefont {Klein}}, \bibinfo {author} {\bibfnamefont
  {C.}~\bibnamefont {Palenzuela}}, \ and\ \bibinfo {author} {\bibfnamefont
  {L.}~\bibnamefont {Lehner}},\ }\href {\doibase 10.1103/PhysRevD.90.124091}
  {\bibfield  {journal} {\bibinfo  {journal} {Phys. Rev. D}\ }\textbf {\bibinfo
  {volume} {90}},\ \bibinfo {pages} {124091} (\bibinfo {year}
  {2014})}\BibitemShut {NoStop}%
\bibitem [{\citenamefont {Anderson}\ \emph {et~al.}(2016)\citenamefont
  {Anderson}, \citenamefont {Yunes},\ and\ \citenamefont
  {Barausse}}]{Anderson2016}%
  \BibitemOpen
  \bibfield  {author} {\bibinfo {author} {\bibfnamefont {D.}~\bibnamefont
  {Anderson}}, \bibinfo {author} {\bibfnamefont {N.}~\bibnamefont {Yunes}}, \
  and\ \bibinfo {author} {\bibfnamefont {E.}~\bibnamefont {Barausse}},\ }\href
  {\doibase 10.1103/PhysRevD.94.104064} {\bibfield  {journal} {\bibinfo
  {journal} {Phys. Rev.}\ }\textbf {\bibinfo {volume} {D94}},\ \bibinfo {pages}
  {104064} (\bibinfo {year} {2016})},\ \Eprint
  {http://arxiv.org/abs/1607.08888} {arXiv:1607.08888 [gr-qc]} \BibitemShut
  {NoStop}%
\bibitem [{\citenamefont {Jackiw}\ and\ \citenamefont
  {Pi}(2003)}]{Jackiw:2003pm}%
  \BibitemOpen
  \bibfield  {author} {\bibinfo {author} {\bibfnamefont {R.}~\bibnamefont
  {Jackiw}}\ and\ \bibinfo {author} {\bibfnamefont {S.~Y.}\ \bibnamefont
  {Pi}},\ }\href {\doibase 10.1103/PhysRevD.68.104012} {\bibfield  {journal}
  {\bibinfo  {journal} {Phys. Rev.}\ }\textbf {\bibinfo {volume} {D68}},\
  \bibinfo {pages} {104012} (\bibinfo {year} {2003})},\ \Eprint
  {http://arxiv.org/abs/gr-qc/0308071} {arXiv:gr-qc/0308071 [gr-qc]}
  \BibitemShut {NoStop}%
\bibitem [{\citenamefont {Yunes}\ and\ \citenamefont
  {Pretorius}(2009{\natexlab{a}})}]{Yunes_dcs}%
  \BibitemOpen
  \bibfield  {author} {\bibinfo {author} {\bibfnamefont {N.}~\bibnamefont
  {Yunes}}\ and\ \bibinfo {author} {\bibfnamefont {F.}~\bibnamefont
  {Pretorius}},\ }\href {\doibase 10.1103/PhysRevD.79.084043} {\bibfield
  {journal} {\bibinfo  {journal} {Phys. Rev.}\ }\textbf {\bibinfo {volume}
  {D79}},\ \bibinfo {pages} {084043} (\bibinfo {year} {2009}{\natexlab{a}})},\
  \Eprint {http://arxiv.org/abs/0902.4669} {arXiv:0902.4669 [gr-qc]}
  \BibitemShut {NoStop}%
\bibitem [{\citenamefont {Alexander}\ and\ \citenamefont
  {Yunes}(2009)}]{Alexander_cs}%
  \BibitemOpen
  \bibfield  {author} {\bibinfo {author} {\bibfnamefont {S.}~\bibnamefont
  {Alexander}}\ and\ \bibinfo {author} {\bibfnamefont {N.}~\bibnamefont
  {Yunes}},\ }\href {\doibase 10.1016/j.physrep.2009.07.002} {\bibfield
  {journal} {\bibinfo  {journal} {Phys. Rept.}\ }\textbf {\bibinfo {volume}
  {480}},\ \bibinfo {pages} {1} (\bibinfo {year} {2009})},\ \Eprint
  {http://arxiv.org/abs/0907.2562} {arXiv:0907.2562 [hep-th]} \BibitemShut
  {NoStop}%
\bibitem [{Ap_()}]{Ap_Voyager_CE}%
  \BibitemOpen
  \href {https://dcc.ligo.org/ligo-T1400316/public} {\enquote {\bibinfo {title}
  {Ligo-t1400316-v4: Instrument science white paper},}\ }\bibinfo
  {howpublished} {\url{https://dcc.ligo.org/ligo-T1400316/public}}\BibitemShut
  {NoStop}%
\bibitem [{ET()}]{ET}%
  \BibitemOpen
  \href {http://www.et-gw.eu/} {\enquote {\bibinfo {title} {The {ET} project
  website},}\ }\bibinfo {howpublished} {\url{http://www.et-gw.eu/}}\BibitemShut
  {NoStop}%
\bibitem [{\citenamefont {Carson}\ \emph {et~al.}(2019)\citenamefont {Carson},
  \citenamefont {Chatziioannou}, \citenamefont {Haster}, \citenamefont {Yagi},\
  and\ \citenamefont {Yunes}}]{Zack:URrelations}%
  \BibitemOpen
  \bibfield  {author} {\bibinfo {author} {\bibfnamefont {Z.}~\bibnamefont
  {Carson}}, \bibinfo {author} {\bibfnamefont {K.}~\bibnamefont
  {Chatziioannou}}, \bibinfo {author} {\bibfnamefont {C.-J.}\ \bibnamefont
  {Haster}}, \bibinfo {author} {\bibfnamefont {K.}~\bibnamefont {Yagi}}, \ and\
  \bibinfo {author} {\bibfnamefont {N.}~\bibnamefont {Yunes}},\ }\href
  {\doibase 10.1103/PhysRevD.99.083016} {\bibfield  {journal} {\bibinfo
  {journal} {Phys. Rev.}\ }\textbf {\bibinfo {volume} {D99}},\ \bibinfo {pages}
  {083016} (\bibinfo {year} {2019})},\ \Eprint
  {http://arxiv.org/abs/1903.03909} {arXiv:1903.03909 [gr-qc]} \BibitemShut
  {NoStop}%
\bibitem [{\citenamefont {Robson}\ \emph {et~al.}(2019)\citenamefont {Robson},
  \citenamefont {Cornish},\ and\ \citenamefont {Liu}}]{LISA}%
  \BibitemOpen
  \bibfield  {author} {\bibinfo {author} {\bibfnamefont {T.}~\bibnamefont
  {Robson}}, \bibinfo {author} {\bibfnamefont {N.}~\bibnamefont {Cornish}}, \
  and\ \bibinfo {author} {\bibfnamefont {C.}~\bibnamefont {Liu}},\ }\href
  {\doibase 10.1088/1361-6382/ab1101} {\bibfield  {journal} {\bibinfo
  {journal} {Class. Quant. Grav.}\ }\textbf {\bibinfo {volume} {36}},\ \bibinfo
  {pages} {105011} (\bibinfo {year} {2019})},\ \Eprint
  {http://arxiv.org/abs/1803.01944} {arXiv:1803.01944 [astro-ph.HE]}
  \BibitemShut {NoStop}%
\bibitem [{\citenamefont {Shi}\ \emph {et~al.}(2019)\citenamefont {Shi},
  \citenamefont {Bao}, \citenamefont {Wang}, \citenamefont {Zhang},
  \citenamefont {Hu}, \citenamefont {Sesana}, \citenamefont {Barausse},
  \citenamefont {Mei},\ and\ \citenamefont {Luo}}]{TianQin}%
  \BibitemOpen
  \bibfield  {author} {\bibinfo {author} {\bibfnamefont {C.}~\bibnamefont
  {Shi}}, \bibinfo {author} {\bibfnamefont {J.}~\bibnamefont {Bao}}, \bibinfo
  {author} {\bibfnamefont {H.}~\bibnamefont {Wang}}, \bibinfo {author}
  {\bibfnamefont {J.-d.}\ \bibnamefont {Zhang}}, \bibinfo {author}
  {\bibfnamefont {Y.}~\bibnamefont {Hu}}, \bibinfo {author} {\bibfnamefont
  {A.}~\bibnamefont {Sesana}}, \bibinfo {author} {\bibfnamefont
  {E.}~\bibnamefont {Barausse}}, \bibinfo {author} {\bibfnamefont
  {J.}~\bibnamefont {Mei}}, \ and\ \bibinfo {author} {\bibfnamefont
  {J.}~\bibnamefont {Luo}},\ }\href@noop {} {\  (\bibinfo {year} {2019})},\
  \Eprint {http://arxiv.org/abs/1902.08922} {arXiv:1902.08922 [gr-qc]}
  \BibitemShut {NoStop}%
\bibitem [{\citenamefont {Isoyama}\ \emph
  {et~al.}(2018{\natexlab{a}})\citenamefont {Isoyama}, \citenamefont {Nakano},\
  and\ \citenamefont {Nakamura}}]{B-DECIGO}%
  \BibitemOpen
  \bibfield  {author} {\bibinfo {author} {\bibfnamefont {S.}~\bibnamefont
  {Isoyama}}, \bibinfo {author} {\bibfnamefont {H.}~\bibnamefont {Nakano}}, \
  and\ \bibinfo {author} {\bibfnamefont {T.}~\bibnamefont {Nakamura}},\ }\href
  {\doibase 10.1093/ptep/pty078} {\bibfield  {journal} {\bibinfo  {journal}
  {PTEP}\ }\textbf {\bibinfo {volume} {2018}},\ \bibinfo {pages} {073E01}
  (\bibinfo {year} {2018}{\natexlab{a}})},\ \Eprint
  {http://arxiv.org/abs/1802.06977} {arXiv:1802.06977 [gr-qc]} \BibitemShut
  {NoStop}%
\bibitem [{\citenamefont {Yagi}\ and\ \citenamefont {Seto}(2011)}]{DECIGO}%
  \BibitemOpen
  \bibfield  {author} {\bibinfo {author} {\bibfnamefont {K.}~\bibnamefont
  {Yagi}}\ and\ \bibinfo {author} {\bibfnamefont {N.}~\bibnamefont {Seto}},\
  }\href {\doibase 10.1103/PhysRevD.95.109901, 10.1103/PhysRevD.83.044011}
  {\bibfield  {journal} {\bibinfo  {journal} {Phys. Rev.}\ }\textbf {\bibinfo
  {volume} {D83}},\ \bibinfo {pages} {044011} (\bibinfo {year} {2011})},\
  \bibinfo {note} {[Erratum: Phys. Rev.D95,no.10,109901(2017)]},\ \Eprint
  {http://arxiv.org/abs/1101.3940} {arXiv:1101.3940 [astro-ph.CO]} \BibitemShut
  {NoStop}%
\bibitem [{\citenamefont {Sesana}(2016)}]{Sesana:2016ljz}%
  \BibitemOpen
  \bibfield  {author} {\bibinfo {author} {\bibfnamefont {A.}~\bibnamefont
  {Sesana}},\ }\href {\doibase 10.1103/PhysRevLett.116.231102} {\bibfield
  {journal} {\bibinfo  {journal} {Phys. Rev. Lett.}\ }\textbf {\bibinfo
  {volume} {116}},\ \bibinfo {pages} {231102} (\bibinfo {year} {2016})},\
  \Eprint {http://arxiv.org/abs/1602.06951} {arXiv:1602.06951 [gr-qc]}
  \BibitemShut {NoStop}%
\bibitem [{\citenamefont {Amaro-Seoane}\ and\ \citenamefont
  {Santamaria}(2010)}]{AmaroSeoane:2009ui}%
  \BibitemOpen
  \bibfield  {author} {\bibinfo {author} {\bibfnamefont {P.}~\bibnamefont
  {Amaro-Seoane}}\ and\ \bibinfo {author} {\bibfnamefont {L.}~\bibnamefont
  {Santamaria}},\ }\href {\doibase 10.1088/0004-637X/722/2/1197} {\bibfield
  {journal} {\bibinfo  {journal} {Astrophys. J.}\ }\textbf {\bibinfo {volume}
  {722}},\ \bibinfo {pages} {1197} (\bibinfo {year} {2010})},\ \Eprint
  {http://arxiv.org/abs/0910.0254} {arXiv:0910.0254 [astro-ph.CO]} \BibitemShut
  {NoStop}%
\bibitem [{\citenamefont {Cutler}\ \emph {et~al.}(2019)\citenamefont {Cutler}
  \emph {et~al.}}]{Cutler:2019krq}%
  \BibitemOpen
  \bibfield  {author} {\bibinfo {author} {\bibfnamefont {C.}~\bibnamefont
  {Cutler}} \emph {et~al.},\ }\href@noop {} {\  (\bibinfo {year} {2019})},\
  \Eprint {http://arxiv.org/abs/1903.04069} {arXiv:1903.04069 [astro-ph.HE]}
  \BibitemShut {NoStop}%
\bibitem [{\citenamefont {Isoyama}\ \emph
  {et~al.}(2018{\natexlab{b}})\citenamefont {Isoyama}, \citenamefont {Nakano},\
  and\ \citenamefont {Nakamura}}]{Isoyama:2018rjb}%
  \BibitemOpen
  \bibfield  {author} {\bibinfo {author} {\bibfnamefont {S.}~\bibnamefont
  {Isoyama}}, \bibinfo {author} {\bibfnamefont {H.}~\bibnamefont {Nakano}}, \
  and\ \bibinfo {author} {\bibfnamefont {T.}~\bibnamefont {Nakamura}},\ }\href
  {\doibase 10.1093/ptep/pty078} {\bibfield  {journal} {\bibinfo  {journal}
  {PTEP}\ }\textbf {\bibinfo {volume} {2018}},\ \bibinfo {pages} {073E01}
  (\bibinfo {year} {2018}{\natexlab{b}})},\ \Eprint
  {http://arxiv.org/abs/1802.06977} {arXiv:1802.06977 [gr-qc]} \BibitemShut
  {NoStop}%
\bibitem [{\citenamefont {Gerosa}\ \emph {et~al.}(2019)\citenamefont {Gerosa},
  \citenamefont {Ma}, \citenamefont {Wong}, \citenamefont {Berti},
  \citenamefont {O'Shaughnessy}, \citenamefont {Chen},\ and\ \citenamefont
  {Belczynski}}]{Gerosa:2019dbe}%
  \BibitemOpen
  \bibfield  {author} {\bibinfo {author} {\bibfnamefont {D.}~\bibnamefont
  {Gerosa}}, \bibinfo {author} {\bibfnamefont {S.}~\bibnamefont {Ma}}, \bibinfo
  {author} {\bibfnamefont {K.~W.~K.}\ \bibnamefont {Wong}}, \bibinfo {author}
  {\bibfnamefont {E.}~\bibnamefont {Berti}}, \bibinfo {author} {\bibfnamefont
  {R.}~\bibnamefont {O'Shaughnessy}}, \bibinfo {author} {\bibfnamefont
  {Y.}~\bibnamefont {Chen}}, \ and\ \bibinfo {author} {\bibfnamefont
  {K.}~\bibnamefont {Belczynski}},\ }\href {\doibase
  10.1103/PhysRevD.99.103004} {\bibfield  {journal} {\bibinfo  {journal} {Phys.
  Rev.}\ }\textbf {\bibinfo {volume} {D99}},\ \bibinfo {pages} {103004}
  (\bibinfo {year} {2019})},\ \Eprint {http://arxiv.org/abs/1902.00021}
  {arXiv:1902.00021 [astro-ph.HE]} \BibitemShut {NoStop}%
\bibitem [{\citenamefont {Tso}\ \emph {et~al.}(2018)\citenamefont {Tso},
  \citenamefont {Gerosa},\ and\ \citenamefont {Chen}}]{Tso:2018pdv}%
  \BibitemOpen
  \bibfield  {author} {\bibinfo {author} {\bibfnamefont {R.}~\bibnamefont
  {Tso}}, \bibinfo {author} {\bibfnamefont {D.}~\bibnamefont {Gerosa}}, \ and\
  \bibinfo {author} {\bibfnamefont {Y.}~\bibnamefont {Chen}},\ }\href@noop {}
  {\  (\bibinfo {year} {2018})},\ \Eprint {http://arxiv.org/abs/1807.00075}
  {arXiv:1807.00075 [gr-qc]} \BibitemShut {NoStop}%
\bibitem [{\citenamefont {Wong}\ \emph {et~al.}(2018)\citenamefont {Wong},
  \citenamefont {Kovetz}, \citenamefont {Cutler},\ and\ \citenamefont
  {Berti}}]{Wong:2018uwb}%
  \BibitemOpen
  \bibfield  {author} {\bibinfo {author} {\bibfnamefont {K.~W.~K.}\
  \bibnamefont {Wong}}, \bibinfo {author} {\bibfnamefont {E.~D.}\ \bibnamefont
  {Kovetz}}, \bibinfo {author} {\bibfnamefont {C.}~\bibnamefont {Cutler}}, \
  and\ \bibinfo {author} {\bibfnamefont {E.}~\bibnamefont {Berti}},\ }\href
  {\doibase 10.1103/PhysRevLett.121.251102} {\bibfield  {journal} {\bibinfo
  {journal} {Phys. Rev. Lett.}\ }\textbf {\bibinfo {volume} {121}},\ \bibinfo
  {pages} {251102} (\bibinfo {year} {2018})},\ \Eprint
  {http://arxiv.org/abs/1808.08247} {arXiv:1808.08247 [astro-ph.HE]}
  \BibitemShut {NoStop}%
\bibitem [{\citenamefont {Moore}\ \emph {et~al.}(2019)\citenamefont {Moore},
  \citenamefont {Gerosa},\ and\ \citenamefont {Klein}}]{Moore:2019pke}%
  \BibitemOpen
  \bibfield  {author} {\bibinfo {author} {\bibfnamefont {C.~J.}\ \bibnamefont
  {Moore}}, \bibinfo {author} {\bibfnamefont {D.}~\bibnamefont {Gerosa}}, \
  and\ \bibinfo {author} {\bibfnamefont {A.}~\bibnamefont {Klein}},\ }\href
  {\doibase 10.1093/mnrasl/slz104} {\bibfield  {journal} {\bibinfo  {journal}
  {Mon. Not. Roy. Astron. Soc.}\ }\textbf {\bibinfo {volume} {488}},\ \bibinfo
  {pages} {L94} (\bibinfo {year} {2019})},\ \Eprint
  {http://arxiv.org/abs/1905.11998} {arXiv:1905.11998 [astro-ph.HE]}
  \BibitemShut {NoStop}%
\bibitem [{\citenamefont {Nair}\ \emph {et~al.}(2016)\citenamefont {Nair},
  \citenamefont {Jhingan},\ and\ \citenamefont {Tanaka}}]{Nair:2015bga}%
  \BibitemOpen
  \bibfield  {author} {\bibinfo {author} {\bibfnamefont {R.}~\bibnamefont
  {Nair}}, \bibinfo {author} {\bibfnamefont {S.}~\bibnamefont {Jhingan}}, \
  and\ \bibinfo {author} {\bibfnamefont {T.}~\bibnamefont {Tanaka}},\ }\href
  {\doibase 10.1093/ptep/ptw043} {\bibfield  {journal} {\bibinfo  {journal}
  {PTEP}\ }\textbf {\bibinfo {volume} {2016}},\ \bibinfo {pages} {053E01}
  (\bibinfo {year} {2016})},\ \Eprint {http://arxiv.org/abs/1504.04108}
  {arXiv:1504.04108 [gr-qc]} \BibitemShut {NoStop}%
\bibitem [{\citenamefont {Nair}\ and\ \citenamefont
  {Tanaka}(2018)}]{Nair:2018bxj}%
  \BibitemOpen
  \bibfield  {author} {\bibinfo {author} {\bibfnamefont {R.}~\bibnamefont
  {Nair}}\ and\ \bibinfo {author} {\bibfnamefont {T.}~\bibnamefont {Tanaka}},\
  }\href {\doibase 10.1088/1475-7516/2018/08/033,
  10.1088/1475-7516/2018/11/E01} {\bibfield  {journal} {\bibinfo  {journal}
  {JCAP}\ }\textbf {\bibinfo {volume} {1808}},\ \bibinfo {pages} {033}
  (\bibinfo {year} {2018})},\ \bibinfo {note} {[Erratum:
  JCAP1811,no.11,E01(2018)]},\ \Eprint {http://arxiv.org/abs/1805.08070}
  {arXiv:1805.08070 [gr-qc]} \BibitemShut {NoStop}%
\bibitem [{\citenamefont {Vitale}(2016)}]{Vitale:2016rfr}%
  \BibitemOpen
  \bibfield  {author} {\bibinfo {author} {\bibfnamefont {S.}~\bibnamefont
  {Vitale}},\ }\href {\doibase 10.1103/PhysRevLett.117.051102} {\bibfield
  {journal} {\bibinfo  {journal} {Phys. Rev. Lett.}\ }\textbf {\bibinfo
  {volume} {117}},\ \bibinfo {pages} {051102} (\bibinfo {year} {2016})},\
  \Eprint {http://arxiv.org/abs/1605.01037} {arXiv:1605.01037 [gr-qc]}
  \BibitemShut {NoStop}%
\bibitem [{\citenamefont {Barausse}\ \emph {et~al.}(2016)\citenamefont
  {Barausse}, \citenamefont {Yunes},\ and\ \citenamefont
  {Chamberlain}}]{Barausse:2016eii}%
  \BibitemOpen
  \bibfield  {author} {\bibinfo {author} {\bibfnamefont {E.}~\bibnamefont
  {Barausse}}, \bibinfo {author} {\bibfnamefont {N.}~\bibnamefont {Yunes}}, \
  and\ \bibinfo {author} {\bibfnamefont {K.}~\bibnamefont {Chamberlain}},\
  }\href {\doibase 10.1103/PhysRevLett.116.241104} {\bibfield  {journal}
  {\bibinfo  {journal} {Phys. Rev. Lett.}\ }\textbf {\bibinfo {volume} {116}},\
  \bibinfo {pages} {241104} (\bibinfo {year} {2016})},\ \Eprint
  {http://arxiv.org/abs/1603.04075} {arXiv:1603.04075 [gr-qc]} \BibitemShut
  {NoStop}%
\bibitem [{\citenamefont {Carson}\ and\ \citenamefont
  {Yagi}(2020)}]{Carson_multiBandPRL}%
  \BibitemOpen
  \bibfield  {author} {\bibinfo {author} {\bibfnamefont {Z.}~\bibnamefont
  {Carson}}\ and\ \bibinfo {author} {\bibfnamefont {K.}~\bibnamefont {Yagi}},\
  }\href {\doibase 10.1088/1361-6382/ab5c9a} {\bibfield  {journal} {\bibinfo
  {journal} {Class. Quant. Grav.}\ }\textbf {\bibinfo {volume} {37}},\ \bibinfo
  {pages} {02LT01} (\bibinfo {year} {2020})},\ \Eprint
  {http://arxiv.org/abs/1905.13155} {arXiv:1905.13155 [gr-qc]} \BibitemShut
  {NoStop}%
\bibitem [{\citenamefont {Gnocchi}\ \emph {et~al.}(2019)\citenamefont
  {Gnocchi}, \citenamefont {Maselli}, \citenamefont {Abdelsalhin},
  \citenamefont {Giacobbo},\ and\ \citenamefont {Mapelli}}]{Gnocchi:2019jzp}%
  \BibitemOpen
  \bibfield  {author} {\bibinfo {author} {\bibfnamefont {G.}~\bibnamefont
  {Gnocchi}}, \bibinfo {author} {\bibfnamefont {A.}~\bibnamefont {Maselli}},
  \bibinfo {author} {\bibfnamefont {T.}~\bibnamefont {Abdelsalhin}}, \bibinfo
  {author} {\bibfnamefont {N.}~\bibnamefont {Giacobbo}}, \ and\ \bibinfo
  {author} {\bibfnamefont {M.}~\bibnamefont {Mapelli}},\ }\href@noop {} {\
  (\bibinfo {year} {2019})},\ \Eprint {http://arxiv.org/abs/1905.13460}
  {arXiv:1905.13460 [gr-qc]} \BibitemShut {NoStop}%
\bibitem [{\citenamefont {Yunes}\ and\ \citenamefont
  {Pretorius}(2009{\natexlab{b}})}]{Yunes:2009ke}%
  \BibitemOpen
  \bibfield  {author} {\bibinfo {author} {\bibfnamefont {N.}~\bibnamefont
  {Yunes}}\ and\ \bibinfo {author} {\bibfnamefont {F.}~\bibnamefont
  {Pretorius}},\ }\href {\doibase 10.1103/PhysRevD.80.122003} {\bibfield
  {journal} {\bibinfo  {journal} {Phys. Rev.}\ }\textbf {\bibinfo {volume}
  {D80}},\ \bibinfo {pages} {122003} (\bibinfo {year} {2009}{\natexlab{b}})},\
  \Eprint {http://arxiv.org/abs/0909.3328} {arXiv:0909.3328 [gr-qc]}
  \BibitemShut {NoStop}%
\bibitem [{\citenamefont {Ghosh}\ \emph {et~al.}(2016)\citenamefont {Ghosh}
  \emph {et~al.}}]{Ghosh_IMRcon}%
  \BibitemOpen
  \bibfield  {author} {\bibinfo {author} {\bibfnamefont {A.}~\bibnamefont
  {Ghosh}} \emph {et~al.},\ }\href {\doibase 10.1103/PhysRevD.94.021101}
  {\bibfield  {journal} {\bibinfo  {journal} {Phys. Rev.}\ }\textbf {\bibinfo
  {volume} {D94}},\ \bibinfo {pages} {021101} (\bibinfo {year} {2016})},\
  \Eprint {http://arxiv.org/abs/1602.02453} {arXiv:1602.02453 [gr-qc]}
  \BibitemShut {NoStop}%
\bibitem [{\citenamefont {Ghosh}\ \emph {et~al.}(2018)\citenamefont {Ghosh},
  \citenamefont {Johnson-Mcdaniel}, \citenamefont {Ghosh}, \citenamefont
  {Mishra}, \citenamefont {Ajith}, \citenamefont {Del~Pozzo}, \citenamefont
  {Berry}, \citenamefont {Nielsen},\ and\ \citenamefont
  {London}}]{Ghosh_IMRcon2}%
  \BibitemOpen
  \bibfield  {author} {\bibinfo {author} {\bibfnamefont {A.}~\bibnamefont
  {Ghosh}}, \bibinfo {author} {\bibfnamefont {N.~K.}\ \bibnamefont
  {Johnson-Mcdaniel}}, \bibinfo {author} {\bibfnamefont {A.}~\bibnamefont
  {Ghosh}}, \bibinfo {author} {\bibfnamefont {C.~K.}\ \bibnamefont {Mishra}},
  \bibinfo {author} {\bibfnamefont {P.}~\bibnamefont {Ajith}}, \bibinfo
  {author} {\bibfnamefont {W.}~\bibnamefont {Del~Pozzo}}, \bibinfo {author}
  {\bibfnamefont {C.~P.~L.}\ \bibnamefont {Berry}}, \bibinfo {author}
  {\bibfnamefont {A.~B.}\ \bibnamefont {Nielsen}}, \ and\ \bibinfo {author}
  {\bibfnamefont {L.}~\bibnamefont {London}},\ }\href {\doibase
  10.1088/1361-6382/aa972e} {\bibfield  {journal} {\bibinfo  {journal} {Class.
  Quant. Grav.}\ }\textbf {\bibinfo {volume} {35}},\ \bibinfo {pages} {014002}
  (\bibinfo {year} {2018})},\ \Eprint {http://arxiv.org/abs/1704.06784}
  {arXiv:1704.06784 [gr-qc]} \BibitemShut {NoStop}%
\bibitem [{\citenamefont {Maeda}\ \emph {et~al.}(2009)\citenamefont {Maeda},
  \citenamefont {Ohta},\ and\ \citenamefont {Sasagawa}}]{Maeda:2009uy}%
  \BibitemOpen
  \bibfield  {author} {\bibinfo {author} {\bibfnamefont {K.-i.}\ \bibnamefont
  {Maeda}}, \bibinfo {author} {\bibfnamefont {N.}~\bibnamefont {Ohta}}, \ and\
  \bibinfo {author} {\bibfnamefont {Y.}~\bibnamefont {Sasagawa}},\ }\href
  {\doibase 10.1103/PhysRevD.80.104032} {\bibfield  {journal} {\bibinfo
  {journal} {Phys. Rev.}\ }\textbf {\bibinfo {volume} {D80}},\ \bibinfo {pages}
  {104032} (\bibinfo {year} {2009})},\ \Eprint {http://arxiv.org/abs/0908.4151}
  {arXiv:0908.4151 [hep-th]} \BibitemShut {NoStop}%
\bibitem [{\citenamefont {Yagi}\ \emph
  {et~al.}(2012{\natexlab{a}})\citenamefont {Yagi}, \citenamefont {Stein},
  \citenamefont {Yunes},\ and\ \citenamefont {Tanaka}}]{Yagi_EdGBmap}%
  \BibitemOpen
  \bibfield  {author} {\bibinfo {author} {\bibfnamefont {K.}~\bibnamefont
  {Yagi}}, \bibinfo {author} {\bibfnamefont {L.~C.}\ \bibnamefont {Stein}},
  \bibinfo {author} {\bibfnamefont {N.}~\bibnamefont {Yunes}}, \ and\ \bibinfo
  {author} {\bibfnamefont {T.}~\bibnamefont {Tanaka}},\ }\href {\doibase
  10.1103/PhysRevD.93.029902, 10.1103/PhysRevD.85.064022} {\bibfield  {journal}
  {\bibinfo  {journal} {Phys. Rev.}\ }\textbf {\bibinfo {volume} {D85}},\
  \bibinfo {pages} {064022} (\bibinfo {year} {2012}{\natexlab{a}})},\ \bibinfo
  {note} {[Erratum: Phys. Rev.D93,no.2,029902(2016)]},\ \Eprint
  {http://arxiv.org/abs/1110.5950} {arXiv:1110.5950 [gr-qc]} \BibitemShut
  {NoStop}%
\bibitem [{\citenamefont {Amendola}\ \emph {et~al.}(2007)\citenamefont
  {Amendola}, \citenamefont {Charmousis},\ and\ \citenamefont
  {Davis}}]{Amendola_EdGB}%
  \BibitemOpen
  \bibfield  {author} {\bibinfo {author} {\bibfnamefont {L.}~\bibnamefont
  {Amendola}}, \bibinfo {author} {\bibfnamefont {C.}~\bibnamefont
  {Charmousis}}, \ and\ \bibinfo {author} {\bibfnamefont {S.}~\bibnamefont
  {Davis}},\ }\href {\doibase 10.1088/1475-7516/2007/10/004} {\bibfield
  {journal} {\bibinfo  {journal} {JCAP}\ }\textbf {\bibinfo {volume} {0710}},\
  \bibinfo {pages} {004} (\bibinfo {year} {2007})},\ \Eprint
  {http://arxiv.org/abs/0704.0175} {arXiv:0704.0175 [astro-ph]} \BibitemShut
  {NoStop}%
\bibitem [{\citenamefont {Kanti}\ \emph {et~al.}(1996)\citenamefont {Kanti},
  \citenamefont {Mavromatos}, \citenamefont {Rizos}, \citenamefont {Tamvakis},\
  and\ \citenamefont {Winstanley}}]{Kanti_EdGB}%
  \BibitemOpen
  \bibfield  {author} {\bibinfo {author} {\bibfnamefont {P.}~\bibnamefont
  {Kanti}}, \bibinfo {author} {\bibfnamefont {N.~E.}\ \bibnamefont
  {Mavromatos}}, \bibinfo {author} {\bibfnamefont {J.}~\bibnamefont {Rizos}},
  \bibinfo {author} {\bibfnamefont {K.}~\bibnamefont {Tamvakis}}, \ and\
  \bibinfo {author} {\bibfnamefont {E.}~\bibnamefont {Winstanley}},\ }\href
  {\doibase 10.1103/PhysRevD.54.5049} {\bibfield  {journal} {\bibinfo
  {journal} {Phys. Rev.}\ }\textbf {\bibinfo {volume} {D54}},\ \bibinfo {pages}
  {5049} (\bibinfo {year} {1996})},\ \Eprint
  {http://arxiv.org/abs/hep-th/9511071} {arXiv:hep-th/9511071 [hep-th]}
  \BibitemShut {NoStop}%
\bibitem [{\citenamefont {Pani}\ and\ \citenamefont
  {Cardoso}(2009)}]{Pani_EdGB}%
  \BibitemOpen
  \bibfield  {author} {\bibinfo {author} {\bibfnamefont {P.}~\bibnamefont
  {Pani}}\ and\ \bibinfo {author} {\bibfnamefont {V.}~\bibnamefont {Cardoso}},\
  }\href {\doibase 10.1103/PhysRevD.79.084031} {\bibfield  {journal} {\bibinfo
  {journal} {Phys. Rev.}\ }\textbf {\bibinfo {volume} {D79}},\ \bibinfo {pages}
  {084031} (\bibinfo {year} {2009})},\ \Eprint {http://arxiv.org/abs/0902.1569}
  {arXiv:0902.1569 [gr-qc]} \BibitemShut {NoStop}%
\bibitem [{\citenamefont {Yagi}(2012)}]{Yagi_EdGB}%
  \BibitemOpen
  \bibfield  {author} {\bibinfo {author} {\bibfnamefont {K.}~\bibnamefont
  {Yagi}},\ }\href {\doibase 10.1103/PhysRevD.86.081504} {\bibfield  {journal}
  {\bibinfo  {journal} {Phys. Rev.}\ }\textbf {\bibinfo {volume} {D86}},\
  \bibinfo {pages} {081504} (\bibinfo {year} {2012})},\ \Eprint
  {http://arxiv.org/abs/1204.4524} {arXiv:1204.4524 [gr-qc]} \BibitemShut
  {NoStop}%
\bibitem [{\citenamefont {Nair}\ \emph {et~al.}(2019)\citenamefont {Nair},
  \citenamefont {Perkins}, \citenamefont {Silva},\ and\ \citenamefont
  {Yunes}}]{Nair_dCSMap}%
  \BibitemOpen
  \bibfield  {author} {\bibinfo {author} {\bibfnamefont {R.}~\bibnamefont
  {Nair}}, \bibinfo {author} {\bibfnamefont {S.}~\bibnamefont {Perkins}},
  \bibinfo {author} {\bibfnamefont {H.~O.}\ \bibnamefont {Silva}}, \ and\
  \bibinfo {author} {\bibfnamefont {N.}~\bibnamefont {Yunes}},\ }\href@noop {}
  {\  (\bibinfo {year} {2019})},\ \Eprint {http://arxiv.org/abs/1905.00870}
  {arXiv:1905.00870 [gr-qc]} \BibitemShut {NoStop}%
\bibitem [{\citenamefont {Yamada}\ \emph {et~al.}(2019)\citenamefont {Yamada},
  \citenamefont {Narikawa},\ and\ \citenamefont {Tanaka}}]{Yamada:2019zrb}%
  \BibitemOpen
  \bibfield  {author} {\bibinfo {author} {\bibfnamefont {K.}~\bibnamefont
  {Yamada}}, \bibinfo {author} {\bibfnamefont {T.}~\bibnamefont {Narikawa}}, \
  and\ \bibinfo {author} {\bibfnamefont {T.}~\bibnamefont {Tanaka}},\
  }\href@noop {} {\  (\bibinfo {year} {2019})},\ \Eprint
  {http://arxiv.org/abs/1905.11859} {arXiv:1905.11859 [gr-qc]} \BibitemShut
  {NoStop}%
\bibitem [{\citenamefont {Tahura}\ \emph {et~al.}(2019)\citenamefont {Tahura},
  \citenamefont {Yagi},\ and\ \citenamefont {Carson}}]{Tahura:2019dgr}%
  \BibitemOpen
  \bibfield  {author} {\bibinfo {author} {\bibfnamefont {S.}~\bibnamefont
  {Tahura}}, \bibinfo {author} {\bibfnamefont {K.}~\bibnamefont {Yagi}}, \ and\
  \bibinfo {author} {\bibfnamefont {Z.}~\bibnamefont {Carson}},\ }\href
  {\doibase 10.1103/PhysRevD.100.104001} {\bibfield  {journal} {\bibinfo
  {journal} {Phys. Rev.}\ }\textbf {\bibinfo {volume} {D100}},\ \bibinfo
  {pages} {104001} (\bibinfo {year} {2019})},\ \Eprint
  {http://arxiv.org/abs/1907.10059} {arXiv:1907.10059 [gr-qc]} \BibitemShut
  {NoStop}%
\bibitem [{\citenamefont {Yagi}\ \emph
  {et~al.}(2012{\natexlab{b}})\citenamefont {Yagi}, \citenamefont {Yunes},\
  and\ \citenamefont {Tanaka}}]{Yagi:2012vf}%
  \BibitemOpen
  \bibfield  {author} {\bibinfo {author} {\bibfnamefont {K.}~\bibnamefont
  {Yagi}}, \bibinfo {author} {\bibfnamefont {N.}~\bibnamefont {Yunes}}, \ and\
  \bibinfo {author} {\bibfnamefont {T.}~\bibnamefont {Tanaka}},\ }\href
  {\doibase 10.1103/PhysRevLett.116.169902, 10.1103/PhysRevLett.109.251105}
  {\bibfield  {journal} {\bibinfo  {journal} {Phys. Rev. Lett.}\ }\textbf
  {\bibinfo {volume} {109}},\ \bibinfo {pages} {251105} (\bibinfo {year}
  {2012}{\natexlab{b}})},\ \bibinfo {note} {[Erratum: Phys. Rev.
  Lett.116,no.16,169902(2016)]},\ \Eprint {http://arxiv.org/abs/1208.5102}
  {arXiv:1208.5102 [gr-qc]} \BibitemShut {NoStop}%
\bibitem [{\citenamefont {Ali-Haimoud}\ and\ \citenamefont
  {Chen}(2011)}]{AliHaimoud_dCS}%
  \BibitemOpen
  \bibfield  {author} {\bibinfo {author} {\bibfnamefont {Y.}~\bibnamefont
  {Ali-Haimoud}}\ and\ \bibinfo {author} {\bibfnamefont {Y.}~\bibnamefont
  {Chen}},\ }\href {\doibase 10.1103/PhysRevD.84.124033} {\bibfield  {journal}
  {\bibinfo  {journal} {Phys. Rev.}\ }\textbf {\bibinfo {volume} {D84}},\
  \bibinfo {pages} {124033} (\bibinfo {year} {2011})},\ \Eprint
  {http://arxiv.org/abs/1110.5329} {arXiv:1110.5329 [astro-ph.HE]} \BibitemShut
  {NoStop}%
\bibitem [{\citenamefont {Yagi}\ \emph
  {et~al.}(2012{\natexlab{c}})\citenamefont {Yagi}, \citenamefont {Yunes},\
  and\ \citenamefont {Tanaka}}]{Yagi_dCS}%
  \BibitemOpen
  \bibfield  {author} {\bibinfo {author} {\bibfnamefont {K.}~\bibnamefont
  {Yagi}}, \bibinfo {author} {\bibfnamefont {N.}~\bibnamefont {Yunes}}, \ and\
  \bibinfo {author} {\bibfnamefont {T.}~\bibnamefont {Tanaka}},\ }\href
  {\doibase 10.1103/PhysRevD.89.049902, 10.1103/PhysRevD.86.044037} {\bibfield
  {journal} {\bibinfo  {journal} {Phys. Rev.}\ }\textbf {\bibinfo {volume}
  {D86}},\ \bibinfo {pages} {044037} (\bibinfo {year} {2012}{\natexlab{c}})},\
  \bibinfo {note} {[Erratum: Phys. Rev.D89,049902(2014)]},\ \Eprint
  {http://arxiv.org/abs/1206.6130} {arXiv:1206.6130 [gr-qc]} \BibitemShut
  {NoStop}%
\bibitem [{\citenamefont {Horbatsch}\ and\ \citenamefont
  {Burgess}(2012)}]{Horbatsch_STcosmo}%
  \BibitemOpen
  \bibfield  {author} {\bibinfo {author} {\bibfnamefont {M.~W.}\ \bibnamefont
  {Horbatsch}}\ and\ \bibinfo {author} {\bibfnamefont {C.~P.}\ \bibnamefont
  {Burgess}},\ }\href {\doibase 10.1088/1475-7516/2012/05/010} {\bibfield
  {journal} {\bibinfo  {journal} {JCAP}\ }\textbf {\bibinfo {volume} {1205}},\
  \bibinfo {pages} {010} (\bibinfo {year} {2012})},\ \Eprint
  {http://arxiv.org/abs/1111.4009} {arXiv:1111.4009 [gr-qc]} \BibitemShut
  {NoStop}%
\bibitem [{\citenamefont {Jacobson}(1999)}]{Jacobson_STcosmo}%
  \BibitemOpen
  \bibfield  {author} {\bibinfo {author} {\bibfnamefont {T.}~\bibnamefont
  {Jacobson}},\ }\href {\doibase 10.1103/PhysRevLett.83.2699} {\bibfield
  {journal} {\bibinfo  {journal} {Phys. Rev. Lett.}\ }\textbf {\bibinfo
  {volume} {83}},\ \bibinfo {pages} {2699} (\bibinfo {year} {1999})},\ \Eprint
  {http://arxiv.org/abs/astro-ph/9905303} {arXiv:astro-ph/9905303 [astro-ph]}
  \BibitemShut {NoStop}%
\bibitem [{\citenamefont {Harikumar}\ and\ \citenamefont
  {Rivelles}(2006)}]{Harikumar:2006xf}%
  \BibitemOpen
  \bibfield  {author} {\bibinfo {author} {\bibfnamefont {E.}~\bibnamefont
  {Harikumar}}\ and\ \bibinfo {author} {\bibfnamefont {V.~O.}\ \bibnamefont
  {Rivelles}},\ }\href {\doibase 10.1088/0264-9381/23/24/024} {\bibfield
  {journal} {\bibinfo  {journal} {Class. Quant. Grav.}\ }\textbf {\bibinfo
  {volume} {23}},\ \bibinfo {pages} {7551} (\bibinfo {year} {2006})},\ \Eprint
  {http://arxiv.org/abs/hep-th/0607115} {arXiv:hep-th/0607115 [hep-th]}
  \BibitemShut {NoStop}%
\bibitem [{\citenamefont {Kobakhidze}\ \emph {et~al.}(2016)\citenamefont
  {Kobakhidze}, \citenamefont {Lagger},\ and\ \citenamefont
  {Manning}}]{Kobakhidze:2016cqh}%
  \BibitemOpen
  \bibfield  {author} {\bibinfo {author} {\bibfnamefont {A.}~\bibnamefont
  {Kobakhidze}}, \bibinfo {author} {\bibfnamefont {C.}~\bibnamefont {Lagger}},
  \ and\ \bibinfo {author} {\bibfnamefont {A.}~\bibnamefont {Manning}},\ }\href
  {\doibase 10.1103/PhysRevD.94.064033} {\bibfield  {journal} {\bibinfo
  {journal} {Phys. Rev.}\ }\textbf {\bibinfo {volume} {D94}},\ \bibinfo {pages}
  {064033} (\bibinfo {year} {2016})},\ \Eprint
  {http://arxiv.org/abs/1607.03776} {arXiv:1607.03776 [gr-qc]} \BibitemShut
  {NoStop}%
\bibitem [{\citenamefont {Will}(2014{\natexlab{b}})}]{Will_SEP}%
  \BibitemOpen
  \bibfield  {author} {\bibinfo {author} {\bibfnamefont {C.~M.}\ \bibnamefont
  {Will}},\ }\href {\doibase 10.12942/lrr-2014-4} {\bibfield  {journal}
  {\bibinfo  {journal} {Living Rev. Rel.}\ }\textbf {\bibinfo {volume} {17}},\
  \bibinfo {pages} {4} (\bibinfo {year} {2014}{\natexlab{b}})},\ \Eprint
  {http://arxiv.org/abs/1403.7377} {arXiv:1403.7377 [gr-qc]} \BibitemShut
  {NoStop}%
\bibitem [{\citenamefont {Yunes}\ \emph
  {et~al.}(2010{\natexlab{a}})\citenamefont {Yunes}, \citenamefont
  {Pretorius},\ and\ \citenamefont {Spergel}}]{Yunes_GdotMap}%
  \BibitemOpen
  \bibfield  {author} {\bibinfo {author} {\bibfnamefont {N.}~\bibnamefont
  {Yunes}}, \bibinfo {author} {\bibfnamefont {F.}~\bibnamefont {Pretorius}}, \
  and\ \bibinfo {author} {\bibfnamefont {D.}~\bibnamefont {Spergel}},\ }\href
  {\doibase 10.1103/PhysRevD.81.064018} {\bibfield  {journal} {\bibinfo
  {journal} {Phys. Rev.}\ }\textbf {\bibinfo {volume} {D81}},\ \bibinfo {pages}
  {064018} (\bibinfo {year} {2010}{\natexlab{a}})},\ \Eprint
  {http://arxiv.org/abs/0912.2724} {arXiv:0912.2724 [gr-qc]} \BibitemShut
  {NoStop}%
\bibitem [{\citenamefont {Tahura}\ and\ \citenamefont
  {Yagi}(2018)}]{Tahura_GdotMap}%
  \BibitemOpen
  \bibfield  {author} {\bibinfo {author} {\bibfnamefont {S.}~\bibnamefont
  {Tahura}}\ and\ \bibinfo {author} {\bibfnamefont {K.}~\bibnamefont {Yagi}},\
  }\href {\doibase 10.1103/PhysRevD.98.084042} {\bibfield  {journal} {\bibinfo
  {journal} {Phys. Rev.}\ }\textbf {\bibinfo {volume} {D98}},\ \bibinfo {pages}
  {084042} (\bibinfo {year} {2018})},\ \Eprint
  {http://arxiv.org/abs/1809.00259} {arXiv:1809.00259 [gr-qc]} \BibitemShut
  {NoStop}%
\bibitem [{\citenamefont {Bambi}\ \emph {et~al.}(2005)\citenamefont {Bambi},
  \citenamefont {Giannotti},\ and\ \citenamefont {Villante}}]{Bambi_Gdot}%
  \BibitemOpen
  \bibfield  {author} {\bibinfo {author} {\bibfnamefont {C.}~\bibnamefont
  {Bambi}}, \bibinfo {author} {\bibfnamefont {M.}~\bibnamefont {Giannotti}}, \
  and\ \bibinfo {author} {\bibfnamefont {F.~L.}\ \bibnamefont {Villante}},\
  }\href {\doibase 10.1103/PhysRevD.71.123524} {\bibfield  {journal} {\bibinfo
  {journal} {Phys. Rev.}\ }\textbf {\bibinfo {volume} {D71}},\ \bibinfo {pages}
  {123524} (\bibinfo {year} {2005})},\ \Eprint
  {http://arxiv.org/abs/astro-ph/0503502} {arXiv:astro-ph/0503502 [astro-ph]}
  \BibitemShut {NoStop}%
\bibitem [{\citenamefont {Copi}\ \emph {et~al.}(2004)\citenamefont {Copi},
  \citenamefont {Davis},\ and\ \citenamefont {Krauss}}]{Copi_Gdot}%
  \BibitemOpen
  \bibfield  {author} {\bibinfo {author} {\bibfnamefont {C.~J.}\ \bibnamefont
  {Copi}}, \bibinfo {author} {\bibfnamefont {A.~N.}\ \bibnamefont {Davis}}, \
  and\ \bibinfo {author} {\bibfnamefont {L.~M.}\ \bibnamefont {Krauss}},\
  }\href {\doibase 10.1103/PhysRevLett.92.171301} {\bibfield  {journal}
  {\bibinfo  {journal} {Phys. Rev. Lett.}\ }\textbf {\bibinfo {volume} {92}},\
  \bibinfo {pages} {171301} (\bibinfo {year} {2004})},\ \Eprint
  {http://arxiv.org/abs/astro-ph/0311334} {arXiv:astro-ph/0311334 [astro-ph]}
  \BibitemShut {NoStop}%
\bibitem [{\citenamefont {Manchester}(2015)}]{Manchester_Gdot}%
  \BibitemOpen
  \bibfield  {author} {\bibinfo {author} {\bibfnamefont {R.~N.}\ \bibnamefont
  {Manchester}},\ }\href {\doibase 10.1142/S0218271815300189} {\bibfield
  {journal} {\bibinfo  {journal} {Int. J. Mod. Phys.}\ }\textbf {\bibinfo
  {volume} {D24}},\ \bibinfo {pages} {1530018} (\bibinfo {year} {2015})},\
  \Eprint {http://arxiv.org/abs/1502.05474} {arXiv:1502.05474 [gr-qc]}
  \BibitemShut {NoStop}%
\bibitem [{\citenamefont {S.~Konopliv}\ \emph {et~al.}(2011)\citenamefont
  {S.~Konopliv}, \citenamefont {Asmar}, \citenamefont {M.~Folkner},
  \citenamefont {Karatekin}, \citenamefont {Nunes}, \citenamefont {Smrekar},
  \citenamefont {F.~Yoder},\ and\ \citenamefont {T.~Zuber}}]{Konopliv_Gdot}%
  \BibitemOpen
  \bibfield  {author} {\bibinfo {author} {\bibfnamefont {A.}~\bibnamefont
  {S.~Konopliv}}, \bibinfo {author} {\bibfnamefont {S.}~\bibnamefont {Asmar}},
  \bibinfo {author} {\bibfnamefont {W.}~\bibnamefont {M.~Folkner}}, \bibinfo
  {author} {\bibfnamefont {O.}~\bibnamefont {Karatekin}}, \bibinfo {author}
  {\bibfnamefont {D.}~\bibnamefont {Nunes}}, \bibinfo {author} {\bibfnamefont
  {S.}~\bibnamefont {Smrekar}}, \bibinfo {author} {\bibfnamefont
  {C.}~\bibnamefont {F.~Yoder}}, \ and\ \bibinfo {author} {\bibfnamefont
  {M.}~\bibnamefont {T.~Zuber}},\ }\href {\doibase
  10.1016/j.icarus.2010.10.004} {\bibfield  {journal} {\bibinfo  {journal}
  {Icarus}\ }\textbf {\bibinfo {volume} {211}},\ \bibinfo {pages} {401}
  (\bibinfo {year} {2011})}\BibitemShut {NoStop}%
\bibitem [{\citenamefont {Yagi}\ \emph {et~al.}(2011)\citenamefont {Yagi},
  \citenamefont {Tanahashi},\ and\ \citenamefont {Tanaka}}]{Yagi_EDmap}%
  \BibitemOpen
  \bibfield  {author} {\bibinfo {author} {\bibfnamefont {K.}~\bibnamefont
  {Yagi}}, \bibinfo {author} {\bibfnamefont {N.}~\bibnamefont {Tanahashi}}, \
  and\ \bibinfo {author} {\bibfnamefont {T.}~\bibnamefont {Tanaka}},\ }\href
  {\doibase 10.1103/PhysRevD.83.084036} {\bibfield  {journal} {\bibinfo
  {journal} {Phys. Rev.}\ }\textbf {\bibinfo {volume} {D83}},\ \bibinfo {pages}
  {084036} (\bibinfo {year} {2011})},\ \Eprint {http://arxiv.org/abs/1101.4997}
  {arXiv:1101.4997 [gr-qc]} \BibitemShut {NoStop}%
\bibitem [{\citenamefont {Berti}\ \emph {et~al.}(2018)\citenamefont {Berti},
  \citenamefont {Yagi},\ and\ \citenamefont
  {Yunes}}]{Berti_ModifiedReviewSmall}%
  \BibitemOpen
  \bibfield  {author} {\bibinfo {author} {\bibfnamefont {E.}~\bibnamefont
  {Berti}}, \bibinfo {author} {\bibfnamefont {K.}~\bibnamefont {Yagi}}, \ and\
  \bibinfo {author} {\bibfnamefont {N.}~\bibnamefont {Yunes}},\ }\href
  {\doibase 10.1007/s10714-018-2362-8} {\bibfield  {journal} {\bibinfo
  {journal} {Gen. Rel. Grav.}\ }\textbf {\bibinfo {volume} {50}},\ \bibinfo
  {pages} {46} (\bibinfo {year} {2018})},\ \Eprint
  {http://arxiv.org/abs/1801.03208} {arXiv:1801.03208 [gr-qc]} \BibitemShut
  {NoStop}%
\bibitem [{\citenamefont {Zhang}\ and\ \citenamefont
  {Zhou}(2018)}]{Zhang:2017jze}%
  \BibitemOpen
  \bibfield  {author} {\bibinfo {author} {\bibfnamefont {J.}~\bibnamefont
  {Zhang}}\ and\ \bibinfo {author} {\bibfnamefont {S.-Y.}\ \bibnamefont
  {Zhou}},\ }\href {\doibase 10.1103/PhysRevD.97.081501} {\bibfield  {journal}
  {\bibinfo  {journal} {Phys. Rev.}\ }\textbf {\bibinfo {volume} {D97}},\
  \bibinfo {pages} {081501} (\bibinfo {year} {2018})},\ \Eprint
  {http://arxiv.org/abs/1709.07503} {arXiv:1709.07503 [gr-qc]} \BibitemShut
  {NoStop}%
\bibitem [{\citenamefont {Finn}\ and\ \citenamefont
  {Sutton}(2002)}]{Finn:2001qi}%
  \BibitemOpen
  \bibfield  {author} {\bibinfo {author} {\bibfnamefont {L.~S.}\ \bibnamefont
  {Finn}}\ and\ \bibinfo {author} {\bibfnamefont {P.~J.}\ \bibnamefont
  {Sutton}},\ }\href {\doibase 10.1103/PhysRevD.65.044022} {\bibfield
  {journal} {\bibinfo  {journal} {Phys. Rev.}\ }\textbf {\bibinfo {volume}
  {D65}},\ \bibinfo {pages} {044022} (\bibinfo {year} {2002})},\ \Eprint
  {http://arxiv.org/abs/gr-qc/0109049} {arXiv:gr-qc/0109049 [gr-qc]}
  \BibitemShut {NoStop}%
\bibitem [{\citenamefont {Chung}\ and\ \citenamefont
  {Li}(2018)}]{Chung:2018dxe}%
  \BibitemOpen
  \bibfield  {author} {\bibinfo {author} {\bibfnamefont {K.-W.}\ \bibnamefont
  {Chung}}\ and\ \bibinfo {author} {\bibfnamefont {T.~G.~F.}\ \bibnamefont
  {Li}},\ }\href@noop {} {\  (\bibinfo {year} {2018})},\ \Eprint
  {http://arxiv.org/abs/1808.04050} {arXiv:1808.04050 [gr-qc]} \BibitemShut
  {NoStop}%
\bibitem [{\citenamefont {Miao}\ \emph {et~al.}(2019)\citenamefont {Miao},
  \citenamefont {Shao},\ and\ \citenamefont {Ma}}]{Miao:2019nhf}%
  \BibitemOpen
  \bibfield  {author} {\bibinfo {author} {\bibfnamefont {X.}~\bibnamefont
  {Miao}}, \bibinfo {author} {\bibfnamefont {L.}~\bibnamefont {Shao}}, \ and\
  \bibinfo {author} {\bibfnamefont {B.-Q.}\ \bibnamefont {Ma}},\ }\href@noop {}
  {\  (\bibinfo {year} {2019})},\ \Eprint {http://arxiv.org/abs/1905.12836}
  {arXiv:1905.12836 [astro-ph.CO]} \BibitemShut {NoStop}%
\bibitem [{\citenamefont {Mirshekari}\ \emph {et~al.}(2012)\citenamefont
  {Mirshekari}, \citenamefont {Yunes},\ and\ \citenamefont
  {Will}}]{Mirshekari_MDR}%
  \BibitemOpen
  \bibfield  {author} {\bibinfo {author} {\bibfnamefont {S.}~\bibnamefont
  {Mirshekari}}, \bibinfo {author} {\bibfnamefont {N.}~\bibnamefont {Yunes}}, \
  and\ \bibinfo {author} {\bibfnamefont {C.~M.}\ \bibnamefont {Will}},\ }\href
  {\doibase 10.1103/PhysRevD.85.024041} {\bibfield  {journal} {\bibinfo
  {journal} {Phys. Rev.}\ }\textbf {\bibinfo {volume} {D85}},\ \bibinfo {pages}
  {024041} (\bibinfo {year} {2012})},\ \Eprint {http://arxiv.org/abs/1110.2720}
  {arXiv:1110.2720 [gr-qc]} \BibitemShut {NoStop}%
\bibitem [{\citenamefont {Poisson}\ and\ \citenamefont
  {Will}(1995)}]{Poisson:Fisher}%
  \BibitemOpen
  \bibfield  {author} {\bibinfo {author} {\bibfnamefont {E.}~\bibnamefont
  {Poisson}}\ and\ \bibinfo {author} {\bibfnamefont {C.~M.}\ \bibnamefont
  {Will}},\ }\href {\doibase 10.1103/PhysRevD.52.848} {\bibfield  {journal}
  {\bibinfo  {journal} {Phys. Rev. D}\ }\textbf {\bibinfo {volume} {52}},\
  \bibinfo {pages} {848} (\bibinfo {year} {1995})}\BibitemShut {NoStop}%
\bibitem [{\citenamefont {Berti}\ \emph {et~al.}(2005)\citenamefont {Berti},
  \citenamefont {Buonanno},\ and\ \citenamefont {Will}}]{Berti:Fisher}%
  \BibitemOpen
  \bibfield  {author} {\bibinfo {author} {\bibfnamefont {E.}~\bibnamefont
  {Berti}}, \bibinfo {author} {\bibfnamefont {A.}~\bibnamefont {Buonanno}}, \
  and\ \bibinfo {author} {\bibfnamefont {C.~M.}\ \bibnamefont {Will}},\ }\href
  {\doibase 10.1103/PhysRevD.71.084025} {\bibfield  {journal} {\bibinfo
  {journal} {Phys. Rev.}\ }\textbf {\bibinfo {volume} {D71}},\ \bibinfo {pages}
  {084025} (\bibinfo {year} {2005})},\ \Eprint
  {http://arxiv.org/abs/gr-qc/0411129} {arXiv:gr-qc/0411129 [gr-qc]}
  \BibitemShut {NoStop}%
\bibitem [{\citenamefont {Yagi}\ and\ \citenamefont
  {Tanaka}(2010)}]{Yagi:2009zm}%
  \BibitemOpen
  \bibfield  {author} {\bibinfo {author} {\bibfnamefont {K.}~\bibnamefont
  {Yagi}}\ and\ \bibinfo {author} {\bibfnamefont {T.}~\bibnamefont {Tanaka}},\
  }\href {\doibase 10.1103/PhysRevD.81.109902, 10.1103/PhysRevD.81.064008}
  {\bibfield  {journal} {\bibinfo  {journal} {Phys. Rev.}\ }\textbf {\bibinfo
  {volume} {D81}},\ \bibinfo {pages} {064008} (\bibinfo {year} {2010})},\
  \bibinfo {note} {[Erratum: Phys. Rev.D81,109902(2010)]},\ \Eprint
  {http://arxiv.org/abs/0906.4269} {arXiv:0906.4269 [gr-qc]} \BibitemShut
  {NoStop}%
\bibitem [{\citenamefont {Chamberlain}\ and\ \citenamefont
  {Yunes}(2017)}]{Chamberlain:2017fjl}%
  \BibitemOpen
  \bibfield  {author} {\bibinfo {author} {\bibfnamefont {K.}~\bibnamefont
  {Chamberlain}}\ and\ \bibinfo {author} {\bibfnamefont {N.}~\bibnamefont
  {Yunes}},\ }\href {\doibase 10.1103/PhysRevD.96.084039} {\bibfield  {journal}
  {\bibinfo  {journal} {Phys. Rev.}\ }\textbf {\bibinfo {volume} {D96}},\
  \bibinfo {pages} {084039} (\bibinfo {year} {2017})},\ \Eprint
  {http://arxiv.org/abs/1704.08268} {arXiv:1704.08268 [gr-qc]} \BibitemShut
  {NoStop}%
\bibitem [{\citenamefont {Khan}\ \emph {et~al.}(2016)\citenamefont {Khan},
  \citenamefont {Husa}, \citenamefont {Hannam}, \citenamefont {Ohme},
  \citenamefont {P\"urrer}, \citenamefont {Forteza},\ and\ \citenamefont
  {Boh\'e}}]{PhenomDI}%
  \BibitemOpen
  \bibfield  {author} {\bibinfo {author} {\bibfnamefont {S.}~\bibnamefont
  {Khan}}, \bibinfo {author} {\bibfnamefont {S.}~\bibnamefont {Husa}}, \bibinfo
  {author} {\bibfnamefont {M.}~\bibnamefont {Hannam}}, \bibinfo {author}
  {\bibfnamefont {F.}~\bibnamefont {Ohme}}, \bibinfo {author} {\bibfnamefont
  {M.}~\bibnamefont {P\"urrer}}, \bibinfo {author} {\bibfnamefont {X.~J.}\
  \bibnamefont {Forteza}}, \ and\ \bibinfo {author} {\bibfnamefont
  {A.}~\bibnamefont {Boh\'e}},\ }\href {\doibase 10.1103/PhysRevD.93.044007}
  {\bibfield  {journal} {\bibinfo  {journal} {Phys. Rev. D}\ }\textbf {\bibinfo
  {volume} {93}},\ \bibinfo {pages} {044007} (\bibinfo {year}
  {2016})}\BibitemShut {NoStop}%
\bibitem [{\citenamefont {Husa}\ \emph {et~al.}(2016)\citenamefont {Husa},
  \citenamefont {Khan}, \citenamefont {Hannam}, \citenamefont {P\"urrer},
  \citenamefont {Ohme}, \citenamefont {Forteza},\ and\ \citenamefont
  {Boh\'e}}]{PhenomDII}%
  \BibitemOpen
  \bibfield  {author} {\bibinfo {author} {\bibfnamefont {S.}~\bibnamefont
  {Husa}}, \bibinfo {author} {\bibfnamefont {S.}~\bibnamefont {Khan}}, \bibinfo
  {author} {\bibfnamefont {M.}~\bibnamefont {Hannam}}, \bibinfo {author}
  {\bibfnamefont {M.}~\bibnamefont {P\"urrer}}, \bibinfo {author}
  {\bibfnamefont {F.}~\bibnamefont {Ohme}}, \bibinfo {author} {\bibfnamefont
  {X.~J.}\ \bibnamefont {Forteza}}, \ and\ \bibinfo {author} {\bibfnamefont
  {A.}~\bibnamefont {Boh\'e}},\ }\href {\doibase 10.1103/PhysRevD.93.044006}
  {\bibfield  {journal} {\bibinfo  {journal} {Phys. Rev. D}\ }\textbf {\bibinfo
  {volume} {93}},\ \bibinfo {pages} {044006} (\bibinfo {year}
  {2016})}\BibitemShut {NoStop}%
\bibitem [{\citenamefont {Vallisneri}(2011)}]{Vallisneri:FisherSNR}%
  \BibitemOpen
  \bibfield  {author} {\bibinfo {author} {\bibfnamefont {M.}~\bibnamefont
  {Vallisneri}},\ }\href {\doibase 10.1103/PhysRevLett.107.191104} {\bibfield
  {journal} {\bibinfo  {journal} {Phys. Rev. Lett.}\ }\textbf {\bibinfo
  {volume} {107}},\ \bibinfo {pages} {191104} (\bibinfo {year}
  {2011})}\BibitemShut {NoStop}%
\bibitem [{\citenamefont {Vallisneri}(2008)}]{Vallisneri:FisherSNR2}%
  \BibitemOpen
  \bibfield  {author} {\bibinfo {author} {\bibfnamefont {M.}~\bibnamefont
  {Vallisneri}},\ }\href {\doibase 10.1103/PhysRevD.77.042001} {\bibfield
  {journal} {\bibinfo  {journal} {Phys. Rev. D}\ }\textbf {\bibinfo {volume}
  {77}},\ \bibinfo {pages} {042001} (\bibinfo {year} {2008})}\BibitemShut
  {NoStop}%
\bibitem [{\citenamefont {Cutler}\ and\ \citenamefont
  {Flanagan}(1994)}]{Cutler:Fisher}%
  \BibitemOpen
  \bibfield  {author} {\bibinfo {author} {\bibfnamefont {C.}~\bibnamefont
  {Cutler}}\ and\ \bibinfo {author} {\bibfnamefont {E.~E.}\ \bibnamefont
  {Flanagan}},\ }\href {\doibase 10.1103/PhysRevD.49.2658} {\bibfield
  {journal} {\bibinfo  {journal} {Phys. Rev. D}\ }\textbf {\bibinfo {volume}
  {49}},\ \bibinfo {pages} {2658} (\bibinfo {year} {1994})}\BibitemShut
  {NoStop}%
\bibitem [{aLI()}]{aLIGO}%
  \BibitemOpen
  \href {https://www.advancedligo.mit,.edu/} {\enquote {\bibinfo {title}
  {Advanced {LIGO}},}\ }\bibinfo {howpublished}
  {\url{https://www.advancedligo.mit.edu/}}\BibitemShut {NoStop}%
\bibitem [{\citenamefont {Jani}\ \emph {et~al.}(2019)\citenamefont {Jani},
  \citenamefont {Shoemaker},\ and\ \citenamefont {Cutler}}]{Jani:2019ffg}%
  \BibitemOpen
  \bibfield  {author} {\bibinfo {author} {\bibfnamefont {K.}~\bibnamefont
  {Jani}}, \bibinfo {author} {\bibfnamefont {D.}~\bibnamefont {Shoemaker}}, \
  and\ \bibinfo {author} {\bibfnamefont {C.}~\bibnamefont {Cutler}},\
  }\href@noop {} {\  (\bibinfo {year} {2019})},\ \Eprint
  {http://arxiv.org/abs/1908.04985} {arXiv:1908.04985 [gr-qc]} \BibitemShut
  {NoStop}%
\bibitem [{\citenamefont {Yunes}\ \emph {et~al.}(2016)\citenamefont {Yunes},
  \citenamefont {Yagi},\ and\ \citenamefont
  {Pretorius}}]{Yunes_ModifiedPhysics}%
  \BibitemOpen
  \bibfield  {author} {\bibinfo {author} {\bibfnamefont {N.}~\bibnamefont
  {Yunes}}, \bibinfo {author} {\bibfnamefont {K.}~\bibnamefont {Yagi}}, \ and\
  \bibinfo {author} {\bibfnamefont {F.}~\bibnamefont {Pretorius}},\ }\href
  {\doibase 10.1103/PhysRevD.94.084002} {\bibfield  {journal} {\bibinfo
  {journal} {Phys. Rev. D}\ }\textbf {\bibinfo {volume} {94}},\ \bibinfo
  {pages} {084002} (\bibinfo {year} {2016})}\BibitemShut {NoStop}%
\bibitem [{\citenamefont {Perkins}\ and\ \citenamefont
  {Yunes}(2019)}]{Perkins:2018tir}%
  \BibitemOpen
  \bibfield  {author} {\bibinfo {author} {\bibfnamefont {S.}~\bibnamefont
  {Perkins}}\ and\ \bibinfo {author} {\bibfnamefont {N.}~\bibnamefont
  {Yunes}},\ }\href {\doibase 10.1088/1361-6382/aafce6} {\bibfield  {journal}
  {\bibinfo  {journal} {Class. Quant. Grav.}\ }\textbf {\bibinfo {volume}
  {36}},\ \bibinfo {pages} {055013} (\bibinfo {year} {2019})},\ \Eprint
  {http://arxiv.org/abs/1811.02533} {arXiv:1811.02533 [gr-qc]} \BibitemShut
  {NoStop}%
\bibitem [{\citenamefont {Emparan}\ \emph {et~al.}(2002)\citenamefont
  {Emparan}, \citenamefont {Fabbri},\ and\ \citenamefont
  {Kaloper}}]{Emparan:2002px}%
  \BibitemOpen
  \bibfield  {author} {\bibinfo {author} {\bibfnamefont {R.}~\bibnamefont
  {Emparan}}, \bibinfo {author} {\bibfnamefont {A.}~\bibnamefont {Fabbri}}, \
  and\ \bibinfo {author} {\bibfnamefont {N.}~\bibnamefont {Kaloper}},\ }\href
  {\doibase 10.1088/1126-6708/2002/08/043} {\bibfield  {journal} {\bibinfo
  {journal} {JHEP}\ }\textbf {\bibinfo {volume} {08}},\ \bibinfo {pages} {043}
  (\bibinfo {year} {2002})},\ \Eprint {http://arxiv.org/abs/hep-th/0206155}
  {arXiv:hep-th/0206155 [hep-th]} \BibitemShut {NoStop}%
\bibitem [{\citenamefont {Tanaka}(2003)}]{Tanaka:2002rb}%
  \BibitemOpen
  \bibfield  {author} {\bibinfo {author} {\bibfnamefont {T.}~\bibnamefont
  {Tanaka}},\ }\bibfield  {booktitle} {\emph {\bibinfo {booktitle} {{Brane
  world: New perspective in cosmology. Proceedings, 2nd Workshop on
  relativistic and cosmological aspects of the brane world, Kyoto, Japan,
  January 15-18, 2002}}},\ }\href {\doibase 10.1143/PTPS.148.307} {\bibfield
  {journal} {\bibinfo  {journal} {Prog. Theor. Phys. Suppl.}\ }\textbf
  {\bibinfo {volume} {148}},\ \bibinfo {pages} {307} (\bibinfo {year}
  {2003})},\ \Eprint {http://arxiv.org/abs/gr-qc/0203082} {arXiv:gr-qc/0203082
  [gr-qc]} \BibitemShut {NoStop}%
\bibitem [{\citenamefont {Babichev}\ \emph {et~al.}(2013)\citenamefont
  {Babichev}, \citenamefont {Dokuchaev},\ and\ \citenamefont
  {Eroshenko}}]{Babichev:2014lda}%
  \BibitemOpen
  \bibfield  {author} {\bibinfo {author} {\bibfnamefont {E.~O.}\ \bibnamefont
  {Babichev}}, \bibinfo {author} {\bibfnamefont {V.~I.}\ \bibnamefont
  {Dokuchaev}}, \ and\ \bibinfo {author} {\bibfnamefont {Y.~N.}\ \bibnamefont
  {Eroshenko}},\ }\href {\doibase 10.3367/UFNr.0183.201312a.1257,
  10.3367/UFNe.0183.201312a.1257} {\bibfield  {journal} {\bibinfo  {journal}
  {Phys. Usp.}\ }\textbf {\bibinfo {volume} {56}},\ \bibinfo {pages} {1155}
  (\bibinfo {year} {2013})},\ \bibinfo {note} {[Usp. Fiz.
  Nauk189,no.12,1257(2013)]},\ \Eprint {http://arxiv.org/abs/1406.0841}
  {arXiv:1406.0841 [gr-qc]} \BibitemShut {NoStop}%
\bibitem [{\citenamefont {Babichev}\ \emph {et~al.}(2005)\citenamefont
  {Babichev}, \citenamefont {Dokuchaev},\ and\ \citenamefont
  {Eroshenko}}]{Babichev:2005py}%
  \BibitemOpen
  \bibfield  {author} {\bibinfo {author} {\bibfnamefont {E.}~\bibnamefont
  {Babichev}}, \bibinfo {author} {\bibfnamefont {V.}~\bibnamefont {Dokuchaev}},
  \ and\ \bibinfo {author} {\bibfnamefont {Y.}~\bibnamefont {Eroshenko}},\
  }\href {\doibase 10.1134/1.1901765} {\bibfield  {journal} {\bibinfo
  {journal} {J. Exp. Theor. Phys.}\ }\textbf {\bibinfo {volume} {100}},\
  \bibinfo {pages} {528} (\bibinfo {year} {2005})},\ \bibinfo {note} {[Zh.
  Eksp. Teor. Fiz.127,597(2005)]},\ \Eprint
  {http://arxiv.org/abs/astro-ph/0505618} {arXiv:astro-ph/0505618 [astro-ph]}
  \BibitemShut {NoStop}%
\bibitem [{\citenamefont {Babichev}\ \emph {et~al.}(2004)\citenamefont
  {Babichev}, \citenamefont {Dokuchaev},\ and\ \citenamefont
  {Eroshenko}}]{Babichev:2004yx}%
  \BibitemOpen
  \bibfield  {author} {\bibinfo {author} {\bibfnamefont {E.}~\bibnamefont
  {Babichev}}, \bibinfo {author} {\bibfnamefont {V.}~\bibnamefont {Dokuchaev}},
  \ and\ \bibinfo {author} {\bibfnamefont {{\relax Yu}.}~\bibnamefont
  {Eroshenko}},\ }\href {\doibase 10.1103/PhysRevLett.93.021102} {\bibfield
  {journal} {\bibinfo  {journal} {Phys. Rev. Lett.}\ }\textbf {\bibinfo
  {volume} {93}},\ \bibinfo {pages} {021102} (\bibinfo {year} {2004})},\
  \Eprint {http://arxiv.org/abs/gr-qc/0402089} {arXiv:gr-qc/0402089 [gr-qc]}
  \BibitemShut {NoStop}%
\bibitem [{\citenamefont {Caputo}\ \emph {et~al.}(2020)\citenamefont {Caputo},
  \citenamefont {Sberna}, \citenamefont {Toubiana}, \citenamefont {Babak},
  \citenamefont {Barausse}, \citenamefont {Marsat},\ and\ \citenamefont
  {Pani}}]{Caputo:2020irr}%
  \BibitemOpen
  \bibfield  {author} {\bibinfo {author} {\bibfnamefont {A.}~\bibnamefont
  {Caputo}}, \bibinfo {author} {\bibfnamefont {L.}~\bibnamefont {Sberna}},
  \bibinfo {author} {\bibfnamefont {A.}~\bibnamefont {Toubiana}}, \bibinfo
  {author} {\bibfnamefont {S.}~\bibnamefont {Babak}}, \bibinfo {author}
  {\bibfnamefont {E.}~\bibnamefont {Barausse}}, \bibinfo {author}
  {\bibfnamefont {S.}~\bibnamefont {Marsat}}, \ and\ \bibinfo {author}
  {\bibfnamefont {P.}~\bibnamefont {Pani}},\ }\href@noop {} {\  (\bibinfo
  {year} {2020})},\ \Eprint {http://arxiv.org/abs/2001.03620} {arXiv:2001.03620
  [astro-ph.HE]} \BibitemShut {NoStop}%
\bibitem [{\citenamefont {Will}(1998)}]{Will_mg}%
  \BibitemOpen
  \bibfield  {author} {\bibinfo {author} {\bibfnamefont {C.~M.}\ \bibnamefont
  {Will}},\ }\href {\doibase 10.1103/PhysRevD.57.2061} {\bibfield  {journal}
  {\bibinfo  {journal} {Phys. Rev.}\ }\textbf {\bibinfo {volume} {D57}},\
  \bibinfo {pages} {2061} (\bibinfo {year} {1998})},\ \Eprint
  {http://arxiv.org/abs/gr-qc/9709011} {arXiv:gr-qc/9709011 [gr-qc]}
  \BibitemShut {NoStop}%
\bibitem [{\citenamefont {Berti}\ \emph {et~al.}(2011)\citenamefont {Berti},
  \citenamefont {Gair},\ and\ \citenamefont {Sesana}}]{Berti:2011jz}%
  \BibitemOpen
  \bibfield  {author} {\bibinfo {author} {\bibfnamefont {E.}~\bibnamefont
  {Berti}}, \bibinfo {author} {\bibfnamefont {J.}~\bibnamefont {Gair}}, \ and\
  \bibinfo {author} {\bibfnamefont {A.}~\bibnamefont {Sesana}},\ }\href
  {\doibase 10.1103/PhysRevD.84.101501} {\bibfield  {journal} {\bibinfo
  {journal} {Phys. Rev.}\ }\textbf {\bibinfo {volume} {D84}},\ \bibinfo {pages}
  {101501} (\bibinfo {year} {2011})},\ \Eprint {http://arxiv.org/abs/1107.3528}
  {arXiv:1107.3528 [gr-qc]} \BibitemShut {NoStop}%
\bibitem [{\citenamefont {Cornish}\ \emph {et~al.}(2011)\citenamefont
  {Cornish}, \citenamefont {Sampson}, \citenamefont {Yunes},\ and\
  \citenamefont {Pretorius}}]{Cornish:2011ys}%
  \BibitemOpen
  \bibfield  {author} {\bibinfo {author} {\bibfnamefont {N.}~\bibnamefont
  {Cornish}}, \bibinfo {author} {\bibfnamefont {L.}~\bibnamefont {Sampson}},
  \bibinfo {author} {\bibfnamefont {N.}~\bibnamefont {Yunes}}, \ and\ \bibinfo
  {author} {\bibfnamefont {F.}~\bibnamefont {Pretorius}},\ }\href {\doibase
  10.1103/PhysRevD.84.062003} {\bibfield  {journal} {\bibinfo  {journal} {Phys.
  Rev.}\ }\textbf {\bibinfo {volume} {D84}},\ \bibinfo {pages} {062003}
  (\bibinfo {year} {2011})},\ \Eprint {http://arxiv.org/abs/1105.2088}
  {arXiv:1105.2088 [gr-qc]} \BibitemShut {NoStop}%
\bibitem [{\citenamefont {Keppel}\ and\ \citenamefont
  {Ajith}(2010)}]{Keppel:2010qu}%
  \BibitemOpen
  \bibfield  {author} {\bibinfo {author} {\bibfnamefont {D.}~\bibnamefont
  {Keppel}}\ and\ \bibinfo {author} {\bibfnamefont {P.}~\bibnamefont {Ajith}},\
  }\href@noop {} {\  (\bibinfo {year} {2010})},\ \Eprint
  {http://arxiv.org/abs/1004.0284} {arXiv:1004.0284 [gr-qc]} \BibitemShut
  {NoStop}%
\bibitem [{\citenamefont {Will}(2018)}]{Will:2018gku}%
  \BibitemOpen
  \bibfield  {author} {\bibinfo {author} {\bibfnamefont {C.~M.}\ \bibnamefont
  {Will}},\ }\href {\doibase 10.1088/1361-6382/aad13c} {\bibfield  {journal}
  {\bibinfo  {journal} {Class. Quant. Grav.}\ }\textbf {\bibinfo {volume}
  {35}},\ \bibinfo {pages} {17LT01} (\bibinfo {year} {2018})},\ \Eprint
  {http://arxiv.org/abs/1805.10523} {arXiv:1805.10523 [gr-qc]} \BibitemShut
  {NoStop}%
\bibitem [{\citenamefont {Chandrasekhar}\ and\ \citenamefont
  {Detweiler}(1975)}]{Chandrasekhar_QNM}%
  \BibitemOpen
  \bibfield  {author} {\bibinfo {author} {\bibfnamefont {S.}~\bibnamefont
  {Chandrasekhar}}\ and\ \bibinfo {author} {\bibfnamefont {S.}~\bibnamefont
  {Detweiler}},\ }\href {http://www.jstor.org/stable/78902} {\bibfield
  {journal} {\bibinfo  {journal} {Proceedings of the Royal Society of London.
  Series A, Mathematical and Physical Sciences}\ }\textbf {\bibinfo {volume}
  {344}},\ \bibinfo {pages} {441} (\bibinfo {year} {1975})}\BibitemShut
  {NoStop}%
\bibitem [{\citenamefont {Vishveshwara}(1970)}]{Vishveshwara_QNM}%
  \BibitemOpen
  \bibfield  {author} {\bibinfo {author} {\bibfnamefont {C.~V.}\ \bibnamefont
  {Vishveshwara}},\ }\href {\doibase 10.1103/PhysRevD.1.2870} {\bibfield
  {journal} {\bibinfo  {journal} {Phys. Rev. D}\ }\textbf {\bibinfo {volume}
  {1}},\ \bibinfo {pages} {2870} (\bibinfo {year} {1970})}\BibitemShut
  {NoStop}%
\bibitem [{\citenamefont {Ghosh}\ \emph {et~al.}(2017)\citenamefont {Ghosh},
  \citenamefont {Johnson-McDaniel}, \citenamefont {Ghosh}, \citenamefont
  {Mishra}, \citenamefont {Ajith}, \citenamefont {Pozzo}, \citenamefont
  {Berry}, \citenamefont {Nielsen},\ and\ \citenamefont {London}}]{Ghosh_2017}%
  \BibitemOpen
  \bibfield  {author} {\bibinfo {author} {\bibfnamefont {A.}~\bibnamefont
  {Ghosh}}, \bibinfo {author} {\bibfnamefont {N.~K.}\ \bibnamefont
  {Johnson-McDaniel}}, \bibinfo {author} {\bibfnamefont {A.}~\bibnamefont
  {Ghosh}}, \bibinfo {author} {\bibfnamefont {C.~K.}\ \bibnamefont {Mishra}},
  \bibinfo {author} {\bibfnamefont {P.}~\bibnamefont {Ajith}}, \bibinfo
  {author} {\bibfnamefont {W.~D.}\ \bibnamefont {Pozzo}}, \bibinfo {author}
  {\bibfnamefont {C.~P.~L.}\ \bibnamefont {Berry}}, \bibinfo {author}
  {\bibfnamefont {A.~B.}\ \bibnamefont {Nielsen}}, \ and\ \bibinfo {author}
  {\bibfnamefont {L.}~\bibnamefont {London}},\ }\href {\doibase
  10.1088/1361-6382/aa972e} {\bibfield  {journal} {\bibinfo  {journal}
  {Classical and Quantum Gravity}\ }\textbf {\bibinfo {volume} {35}},\ \bibinfo
  {pages} {014002} (\bibinfo {year} {2017})}\BibitemShut {NoStop}%
\bibitem [{\citenamefont {Hughes}\ and\ \citenamefont
  {Menou}(2005)}]{Hughes:2004vw}%
  \BibitemOpen
  \bibfield  {author} {\bibinfo {author} {\bibfnamefont {S.~A.}\ \bibnamefont
  {Hughes}}\ and\ \bibinfo {author} {\bibfnamefont {K.}~\bibnamefont {Menou}},\
  }\href {\doibase 10.1086/428826} {\bibfield  {journal} {\bibinfo  {journal}
  {Astrophys. J.}\ }\textbf {\bibinfo {volume} {623}},\ \bibinfo {pages} {689}
  (\bibinfo {year} {2005})},\ \Eprint {http://arxiv.org/abs/astro-ph/0410148}
  {arXiv:astro-ph/0410148 [astro-ph]} \BibitemShut {NoStop}%
\bibitem [{\citenamefont {Alexander}\ \emph {et~al.}(2008)\citenamefont
  {Alexander}, \citenamefont {Finn},\ and\ \citenamefont
  {Yunes}}]{Alexander:2007kv}%
  \BibitemOpen
  \bibfield  {author} {\bibinfo {author} {\bibfnamefont {S.}~\bibnamefont
  {Alexander}}, \bibinfo {author} {\bibfnamefont {L.~S.}\ \bibnamefont {Finn}},
  \ and\ \bibinfo {author} {\bibfnamefont {N.}~\bibnamefont {Yunes}},\ }\href
  {\doibase 10.1103/PhysRevD.78.066005} {\bibfield  {journal} {\bibinfo
  {journal} {Phys. Rev.}\ }\textbf {\bibinfo {volume} {D78}},\ \bibinfo {pages}
  {066005} (\bibinfo {year} {2008})},\ \Eprint {http://arxiv.org/abs/0712.2542}
  {arXiv:0712.2542 [gr-qc]} \BibitemShut {NoStop}%
\bibitem [{\citenamefont {Yunes}\ and\ \citenamefont
  {Finn}(2009)}]{Yunes:2008bu}%
  \BibitemOpen
  \bibfield  {author} {\bibinfo {author} {\bibfnamefont {N.}~\bibnamefont
  {Yunes}}\ and\ \bibinfo {author} {\bibfnamefont {L.~S.}\ \bibnamefont
  {Finn}},\ }\href {\doibase 10.1088/1742-6596/154/1/012041} {\bibfield
  {journal} {\bibinfo  {journal} {J. Phys. Conf. Ser.}\ }\textbf {\bibinfo
  {volume} {154}},\ \bibinfo {pages} {012041} (\bibinfo {year} {2009})},\
  \Eprint {http://arxiv.org/abs/0811.0181} {arXiv:0811.0181 [gr-qc]}
  \BibitemShut {NoStop}%
\bibitem [{\citenamefont {Yunes}\ \emph
  {et~al.}(2010{\natexlab{b}})\citenamefont {Yunes}, \citenamefont
  {O'Shaughnessy}, \citenamefont {Owen},\ and\ \citenamefont
  {Alexander}}]{Yunes:2010yf}%
  \BibitemOpen
  \bibfield  {author} {\bibinfo {author} {\bibfnamefont {N.}~\bibnamefont
  {Yunes}}, \bibinfo {author} {\bibfnamefont {R.}~\bibnamefont
  {O'Shaughnessy}}, \bibinfo {author} {\bibfnamefont {B.~J.}\ \bibnamefont
  {Owen}}, \ and\ \bibinfo {author} {\bibfnamefont {S.}~\bibnamefont
  {Alexander}},\ }\href {\doibase 10.1103/PhysRevD.82.064017} {\bibfield
  {journal} {\bibinfo  {journal} {Phys. Rev.}\ }\textbf {\bibinfo {volume}
  {D82}},\ \bibinfo {pages} {064017} (\bibinfo {year} {2010}{\natexlab{b}})},\
  \Eprint {http://arxiv.org/abs/1005.3310} {arXiv:1005.3310 [gr-qc]}
  \BibitemShut {NoStop}%
\bibitem [{\citenamefont {Yagi}\ and\ \citenamefont
  {Yang}(2018)}]{Yagi:2017zhb}%
  \BibitemOpen
  \bibfield  {author} {\bibinfo {author} {\bibfnamefont {K.}~\bibnamefont
  {Yagi}}\ and\ \bibinfo {author} {\bibfnamefont {H.}~\bibnamefont {Yang}},\
  }\href {\doibase 10.1103/PhysRevD.97.104018} {\bibfield  {journal} {\bibinfo
  {journal} {Phys. Rev.}\ }\textbf {\bibinfo {volume} {D97}},\ \bibinfo {pages}
  {104018} (\bibinfo {year} {2018})},\ \Eprint
  {http://arxiv.org/abs/1712.00682} {arXiv:1712.00682 [gr-qc]} \BibitemShut
  {NoStop}%
\bibitem [{\citenamefont {Berti}\ \emph {et~al.}(2015)\citenamefont {Berti}
  \emph {et~al.}}]{Berti_ModifiedReviewLarge}%
  \BibitemOpen
  \bibfield  {author} {\bibinfo {author} {\bibfnamefont {E.}~\bibnamefont
  {Berti}} \emph {et~al.},\ }\href {\doibase 10.1088/0264-9381/32/24/243001}
  {\bibfield  {journal} {\bibinfo  {journal} {Class. Quant. Grav.}\ }\textbf
  {\bibinfo {volume} {32}},\ \bibinfo {pages} {243001} (\bibinfo {year}
  {2015})},\ \Eprint {http://arxiv.org/abs/1501.07274} {arXiv:1501.07274
  [gr-qc]} \BibitemShut {NoStop}%
\bibitem [{\citenamefont {Prabhu}\ and\ \citenamefont
  {Stein}(2018)}]{Prabhu:2018aun}%
  \BibitemOpen
  \bibfield  {author} {\bibinfo {author} {\bibfnamefont {K.}~\bibnamefont
  {Prabhu}}\ and\ \bibinfo {author} {\bibfnamefont {L.~C.}\ \bibnamefont
  {Stein}},\ }\href {\doibase 10.1103/PhysRevD.98.021503} {\bibfield  {journal}
  {\bibinfo  {journal} {Phys. Rev.}\ }\textbf {\bibinfo {volume} {D98}},\
  \bibinfo {pages} {021503} (\bibinfo {year} {2018})},\ \Eprint
  {http://arxiv.org/abs/1805.02668} {arXiv:1805.02668 [gr-qc]} \BibitemShut
  {NoStop}%
\bibitem [{\citenamefont {Berti}\ \emph {et~al.}(2013)\citenamefont {Berti},
  \citenamefont {Cardoso}, \citenamefont {Gualtieri}, \citenamefont
  {Horbatsch},\ and\ \citenamefont {Sperhake}}]{Berti_STcosmo}%
  \BibitemOpen
  \bibfield  {author} {\bibinfo {author} {\bibfnamefont {E.}~\bibnamefont
  {Berti}}, \bibinfo {author} {\bibfnamefont {V.}~\bibnamefont {Cardoso}},
  \bibinfo {author} {\bibfnamefont {L.}~\bibnamefont {Gualtieri}}, \bibinfo
  {author} {\bibfnamefont {M.}~\bibnamefont {Horbatsch}}, \ and\ \bibinfo
  {author} {\bibfnamefont {U.}~\bibnamefont {Sperhake}},\ }\href {\doibase
  10.1103/PhysRevD.87.124020} {\bibfield  {journal} {\bibinfo  {journal} {Phys.
  Rev.}\ }\textbf {\bibinfo {volume} {D87}},\ \bibinfo {pages} {124020}
  (\bibinfo {year} {2013})},\ \Eprint {http://arxiv.org/abs/1304.2836}
  {arXiv:1304.2836 [gr-qc]} \BibitemShut {NoStop}%
\bibitem [{\citenamefont {Snyder}(1947)}]{Snyder:QuantizedST}%
  \BibitemOpen
  \bibfield  {author} {\bibinfo {author} {\bibfnamefont {H.~S.}\ \bibnamefont
  {Snyder}},\ }\href {\doibase 10.1103/PhysRev.71.38} {\bibfield  {journal}
  {\bibinfo  {journal} {Phys. Rev.}\ }\textbf {\bibinfo {volume} {71}},\
  \bibinfo {pages} {38} (\bibinfo {year} {1947})}\BibitemShut {NoStop}%
\bibitem [{\citenamefont {Wolf}\ and\ \citenamefont
  {Lagos}(2019)}]{Wolf:2019hun}%
  \BibitemOpen
  \bibfield  {author} {\bibinfo {author} {\bibfnamefont {W.~J.}\ \bibnamefont
  {Wolf}}\ and\ \bibinfo {author} {\bibfnamefont {M.}~\bibnamefont {Lagos}},\
  }\href@noop {} {\  (\bibinfo {year} {2019})},\ \Eprint
  {http://arxiv.org/abs/1910.10580} {arXiv:1910.10580 [gr-qc]} \BibitemShut
  {NoStop}%
\bibitem [{\citenamefont {Arkani-Hamed}\ \emph {et~al.}(1998)\citenamefont
  {Arkani-Hamed}, \citenamefont {Dimopoulos},\ and\ \citenamefont
  {Dvali}}]{ArkaniHamed:1998rs}%
  \BibitemOpen
  \bibfield  {author} {\bibinfo {author} {\bibfnamefont {N.}~\bibnamefont
  {Arkani-Hamed}}, \bibinfo {author} {\bibfnamefont {S.}~\bibnamefont
  {Dimopoulos}}, \ and\ \bibinfo {author} {\bibfnamefont {G.~R.}\ \bibnamefont
  {Dvali}},\ }\href {\doibase 10.1016/S0370-2693(98)00466-3} {\bibfield
  {journal} {\bibinfo  {journal} {Phys. Lett.}\ }\textbf {\bibinfo {volume}
  {B429}},\ \bibinfo {pages} {263} (\bibinfo {year} {1998})},\ \Eprint
  {http://arxiv.org/abs/hep-ph/9803315} {arXiv:hep-ph/9803315 [hep-ph]}
  \BibitemShut {NoStop}%
\bibitem [{\citenamefont {Arkani-Hamed}\ \emph {et~al.}(1999)\citenamefont
  {Arkani-Hamed}, \citenamefont {Dimopoulos},\ and\ \citenamefont
  {Dvali}}]{ArkaniHamed:1998nn}%
  \BibitemOpen
  \bibfield  {author} {\bibinfo {author} {\bibfnamefont {N.}~\bibnamefont
  {Arkani-Hamed}}, \bibinfo {author} {\bibfnamefont {S.}~\bibnamefont
  {Dimopoulos}}, \ and\ \bibinfo {author} {\bibfnamefont {G.~R.}\ \bibnamefont
  {Dvali}},\ }\href {\doibase 10.1103/PhysRevD.59.086004} {\bibfield  {journal}
  {\bibinfo  {journal} {Phys. Rev.}\ }\textbf {\bibinfo {volume} {D59}},\
  \bibinfo {pages} {086004} (\bibinfo {year} {1999})},\ \Eprint
  {http://arxiv.org/abs/hep-ph/9807344} {arXiv:hep-ph/9807344 [hep-ph]}
  \BibitemShut {NoStop}%
\bibitem [{\citenamefont {Randall}\ and\ \citenamefont
  {Sundrum}(1999{\natexlab{a}})}]{Randall:1999ee}%
  \BibitemOpen
  \bibfield  {author} {\bibinfo {author} {\bibfnamefont {L.}~\bibnamefont
  {Randall}}\ and\ \bibinfo {author} {\bibfnamefont {R.}~\bibnamefont
  {Sundrum}},\ }\href {\doibase 10.1103/PhysRevLett.83.3370} {\bibfield
  {journal} {\bibinfo  {journal} {Phys. Rev. Lett.}\ }\textbf {\bibinfo
  {volume} {83}},\ \bibinfo {pages} {3370} (\bibinfo {year}
  {1999}{\natexlab{a}})},\ \Eprint {http://arxiv.org/abs/hep-ph/9905221}
  {arXiv:hep-ph/9905221 [hep-ph]} \BibitemShut {NoStop}%
\bibitem [{\citenamefont {Randall}\ and\ \citenamefont
  {Sundrum}(1999{\natexlab{b}})}]{Randall:Braneworld}%
  \BibitemOpen
  \bibfield  {author} {\bibinfo {author} {\bibfnamefont {L.}~\bibnamefont
  {Randall}}\ and\ \bibinfo {author} {\bibfnamefont {R.}~\bibnamefont
  {Sundrum}},\ }\href {\doibase 10.1103/PhysRevLett.83.4690} {\bibfield
  {journal} {\bibinfo  {journal} {Phys. Rev. Lett.}\ }\textbf {\bibinfo
  {volume} {83}},\ \bibinfo {pages} {4690} (\bibinfo {year}
  {1999}{\natexlab{b}})},\ \Eprint {http://arxiv.org/abs/hep-th/9906064}
  {arXiv:hep-th/9906064 [hep-th]} \BibitemShut {NoStop}%
\bibitem [{\citenamefont {Figueras}\ and\ \citenamefont
  {Wiseman}(2011)}]{Figueras:2011gd}%
  \BibitemOpen
  \bibfield  {author} {\bibinfo {author} {\bibfnamefont {P.}~\bibnamefont
  {Figueras}}\ and\ \bibinfo {author} {\bibfnamefont {T.}~\bibnamefont
  {Wiseman}},\ }\href {\doibase 10.1103/PhysRevLett.107.081101} {\bibfield
  {journal} {\bibinfo  {journal} {Phys. Rev. Lett.}\ }\textbf {\bibinfo
  {volume} {107}},\ \bibinfo {pages} {081101} (\bibinfo {year} {2011})},\
  \Eprint {http://arxiv.org/abs/1105.2558} {arXiv:1105.2558 [hep-th]}
  \BibitemShut {NoStop}%
\bibitem [{\citenamefont {Abdolrahimi}\ \emph {et~al.}(2013)\citenamefont
  {Abdolrahimi}, \citenamefont {Cattoen}, \citenamefont {Page},\ and\
  \citenamefont {Yaghoobpour-Tari}}]{Abdolrahimi:2012qi}%
  \BibitemOpen
  \bibfield  {author} {\bibinfo {author} {\bibfnamefont {S.}~\bibnamefont
  {Abdolrahimi}}, \bibinfo {author} {\bibfnamefont {C.}~\bibnamefont
  {Cattoen}}, \bibinfo {author} {\bibfnamefont {D.~N.}\ \bibnamefont {Page}}, \
  and\ \bibinfo {author} {\bibfnamefont {S.}~\bibnamefont {Yaghoobpour-Tari}},\
  }\href {\doibase 10.1016/j.physletb.2013.02.034} {\bibfield  {journal}
  {\bibinfo  {journal} {Phys. Lett.}\ }\textbf {\bibinfo {volume} {B720}},\
  \bibinfo {pages} {405} (\bibinfo {year} {2013})},\ \Eprint
  {http://arxiv.org/abs/1206.0708} {arXiv:1206.0708 [hep-th]} \BibitemShut
  {NoStop}%
\bibitem [{\citenamefont {Adelberger}\ \emph {et~al.}(2007)\citenamefont
  {Adelberger}, \citenamefont {Heckel}, \citenamefont {Hoedl}, \citenamefont
  {Hoyle}, \citenamefont {Kapner},\ and\ \citenamefont
  {Upadhye}}]{Adelberger:2006dh}%
  \BibitemOpen
  \bibfield  {author} {\bibinfo {author} {\bibfnamefont {E.~G.}\ \bibnamefont
  {Adelberger}}, \bibinfo {author} {\bibfnamefont {B.~R.}\ \bibnamefont
  {Heckel}}, \bibinfo {author} {\bibfnamefont {S.~A.}\ \bibnamefont {Hoedl}},
  \bibinfo {author} {\bibfnamefont {C.~D.}\ \bibnamefont {Hoyle}}, \bibinfo
  {author} {\bibfnamefont {D.~J.}\ \bibnamefont {Kapner}}, \ and\ \bibinfo
  {author} {\bibfnamefont {A.}~\bibnamefont {Upadhye}},\ }\href {\doibase
  10.1103/PhysRevLett.98.131104} {\bibfield  {journal} {\bibinfo  {journal}
  {Phys. Rev. Lett.}\ }\textbf {\bibinfo {volume} {98}},\ \bibinfo {pages}
  {131104} (\bibinfo {year} {2007})},\ \Eprint
  {http://arxiv.org/abs/hep-ph/0611223} {arXiv:hep-ph/0611223 [hep-ph]}
  \BibitemShut {NoStop}%
\bibitem [{\citenamefont {Johannsen}\ \emph {et~al.}(2009)\citenamefont
  {Johannsen}, \citenamefont {Psaltis},\ and\ \citenamefont
  {McClintock}}]{Johannsen_ED}%
  \BibitemOpen
  \bibfield  {author} {\bibinfo {author} {\bibfnamefont {T.}~\bibnamefont
  {Johannsen}}, \bibinfo {author} {\bibfnamefont {D.}~\bibnamefont {Psaltis}},
  \ and\ \bibinfo {author} {\bibfnamefont {J.~E.}\ \bibnamefont {McClintock}},\
  }\href {\doibase 10.1088/0004-637X/691/2/997} {\bibfield  {journal} {\bibinfo
   {journal} {Astrophys. J.}\ }\textbf {\bibinfo {volume} {691}},\ \bibinfo
  {pages} {997} (\bibinfo {year} {2009})},\ \Eprint
  {http://arxiv.org/abs/0803.1835} {arXiv:0803.1835 [astro-ph]} \BibitemShut
  {NoStop}%
\bibitem [{\citenamefont {Johannsen}(2009)}]{Johannsen_ED2}%
  \BibitemOpen
  \bibfield  {author} {\bibinfo {author} {\bibfnamefont {T.}~\bibnamefont
  {Johannsen}},\ }\href {\doibase 10.1051/0004-6361/200912803} {\bibfield
  {journal} {\bibinfo  {journal} {Astron. Astrophys.}\ }\textbf {\bibinfo
  {volume} {507}},\ \bibinfo {pages} {617} (\bibinfo {year} {2009})},\ \Eprint
  {http://arxiv.org/abs/0812.0809} {arXiv:0812.0809 [astro-ph]} \BibitemShut
  {NoStop}%
\bibitem [{\citenamefont {Psaltis}(2007)}]{Psaltis_ED}%
  \BibitemOpen
  \bibfield  {author} {\bibinfo {author} {\bibfnamefont {D.}~\bibnamefont
  {Psaltis}},\ }\href {\doibase 10.1103/PhysRevLett.98.181101} {\bibfield
  {journal} {\bibinfo  {journal} {Phys. Rev. Lett.}\ }\textbf {\bibinfo
  {volume} {98}},\ \bibinfo {pages} {181101} (\bibinfo {year} {2007})},\
  \Eprint {http://arxiv.org/abs/astro-ph/0612611} {arXiv:astro-ph/0612611
  [astro-ph]} \BibitemShut {NoStop}%
\bibitem [{\citenamefont {Gnedin}\ \emph {et~al.}(2009)\citenamefont {Gnedin},
  \citenamefont {Maccarone}, \citenamefont {Psaltis},\ and\ \citenamefont
  {Zepf}}]{Gnedin_ED}%
  \BibitemOpen
  \bibfield  {author} {\bibinfo {author} {\bibfnamefont {O.~Y.}\ \bibnamefont
  {Gnedin}}, \bibinfo {author} {\bibfnamefont {T.~J.}\ \bibnamefont
  {Maccarone}}, \bibinfo {author} {\bibfnamefont {D.}~\bibnamefont {Psaltis}},
  \ and\ \bibinfo {author} {\bibfnamefont {S.~E.}\ \bibnamefont {Zepf}},\
  }\href {\doibase 10.1088/0004-637X/705/2/L168} {\bibfield  {journal}
  {\bibinfo  {journal} {Astrophys. J.}\ }\textbf {\bibinfo {volume} {705}},\
  \bibinfo {pages} {L168} (\bibinfo {year} {2009})},\ \Eprint
  {http://arxiv.org/abs/0906.5351} {arXiv:0906.5351 [astro-ph.CO]} \BibitemShut
  {NoStop}%
\bibitem [{\citenamefont {Emparan}\ \emph {et~al.}(2003)\citenamefont
  {Emparan}, \citenamefont {Garcia-Bellido},\ and\ \citenamefont
  {Kaloper}}]{Emparan_Mdot}%
  \BibitemOpen
  \bibfield  {author} {\bibinfo {author} {\bibfnamefont {R.}~\bibnamefont
  {Emparan}}, \bibinfo {author} {\bibfnamefont {J.}~\bibnamefont
  {Garcia-Bellido}}, \ and\ \bibinfo {author} {\bibfnamefont {N.}~\bibnamefont
  {Kaloper}},\ }\href {\doibase 10.1088/1126-6708/2003/01/079} {\bibfield
  {journal} {\bibinfo  {journal} {JHEP}\ }\textbf {\bibinfo {volume} {01}},\
  \bibinfo {pages} {079} (\bibinfo {year} {2003})},\ \Eprint
  {http://arxiv.org/abs/hep-th/0212132} {arXiv:hep-th/0212132 [hep-th]}
  \BibitemShut {NoStop}%
\bibitem [{\citenamefont {de~Rham}\ \emph {et~al.}(2017)\citenamefont
  {de~Rham}, \citenamefont {Deskins}, \citenamefont {Tolley},\ and\
  \citenamefont {Zhou}}]{deRham:2016nuf}%
  \BibitemOpen
  \bibfield  {author} {\bibinfo {author} {\bibfnamefont {C.}~\bibnamefont
  {de~Rham}}, \bibinfo {author} {\bibfnamefont {J.~T.}\ \bibnamefont
  {Deskins}}, \bibinfo {author} {\bibfnamefont {A.~J.}\ \bibnamefont {Tolley}},
  \ and\ \bibinfo {author} {\bibfnamefont {S.-Y.}\ \bibnamefont {Zhou}},\
  }\href {\doibase 10.1103/RevModPhys.89.025004} {\bibfield  {journal}
  {\bibinfo  {journal} {Rev. Mod. Phys.}\ }\textbf {\bibinfo {volume} {89}},\
  \bibinfo {pages} {025004} (\bibinfo {year} {2017})},\ \Eprint
  {http://arxiv.org/abs/1606.08462} {arXiv:1606.08462 [astro-ph.CO]}
  \BibitemShut {NoStop}%
\bibitem [{\citenamefont {Aghanim}\ \emph {et~al.}(2018)\citenamefont {Aghanim}
  \emph {et~al.}}]{Aghanim:2018eyx}%
  \BibitemOpen
  \bibfield  {author} {\bibinfo {author} {\bibfnamefont {N.}~\bibnamefont
  {Aghanim}} \emph {et~al.} (\bibinfo {collaboration} {Planck}),\ }\href@noop
  {} {\  (\bibinfo {year} {2018})},\ \Eprint {http://arxiv.org/abs/1807.06209}
  {arXiv:1807.06209 [astro-ph.CO]} \BibitemShut {NoStop}%
\bibitem [{\citenamefont {de~Rham}(2014)}]{deRham_mg}%
  \BibitemOpen
  \bibfield  {author} {\bibinfo {author} {\bibfnamefont {C.}~\bibnamefont
  {de~Rham}},\ }\href {\doibase 10.12942/lrr-2014-7} {\bibfield  {journal}
  {\bibinfo  {journal} {Living Rev. Rel.}\ }\textbf {\bibinfo {volume} {17}},\
  \bibinfo {pages} {7} (\bibinfo {year} {2014})},\ \Eprint
  {http://arxiv.org/abs/1401.4173} {arXiv:1401.4173 [hep-th]} \BibitemShut
  {NoStop}%
\bibitem [{\citenamefont {Hinterbichler}(2012)}]{Hinterbichler_mg}%
  \BibitemOpen
  \bibfield  {author} {\bibinfo {author} {\bibfnamefont {K.}~\bibnamefont
  {Hinterbichler}},\ }\href {\doibase 10.1103/RevModPhys.84.671} {\bibfield
  {journal} {\bibinfo  {journal} {Rev. Mod. Phys.}\ }\textbf {\bibinfo {volume}
  {84}},\ \bibinfo {pages} {671} (\bibinfo {year} {2012})},\ \Eprint
  {http://arxiv.org/abs/1105.3735} {arXiv:1105.3735 [hep-th]} \BibitemShut
  {NoStop}%
\bibitem [{\citenamefont {Rubakov}\ and\ \citenamefont
  {Tinyakov}(2008)}]{Rubakov_mg}%
  \BibitemOpen
  \bibfield  {author} {\bibinfo {author} {\bibfnamefont {V.~A.}\ \bibnamefont
  {Rubakov}}\ and\ \bibinfo {author} {\bibfnamefont {P.~G.}\ \bibnamefont
  {Tinyakov}},\ }\href {\doibase 10.1070/PU2008v051n08ABEH006600} {\bibfield
  {journal} {\bibinfo  {journal} {Phys. Usp.}\ }\textbf {\bibinfo {volume}
  {51}},\ \bibinfo {pages} {759} (\bibinfo {year} {2008})},\ \Eprint
  {http://arxiv.org/abs/0802.4379} {arXiv:0802.4379 [hep-th]} \BibitemShut
  {NoStop}%
\bibitem [{\citenamefont {Desai}(2018)}]{Desai:2017dwg}%
  \BibitemOpen
  \bibfield  {author} {\bibinfo {author} {\bibfnamefont {S.}~\bibnamefont
  {Desai}},\ }\href {\doibase 10.1016/j.physletb.2018.01.052} {\bibfield
  {journal} {\bibinfo  {journal} {Phys. Lett.}\ }\textbf {\bibinfo {volume}
  {B778}},\ \bibinfo {pages} {325} (\bibinfo {year} {2018})},\ \Eprint
  {http://arxiv.org/abs/1708.06502} {arXiv:1708.06502 [astro-ph.CO]}
  \BibitemShut {NoStop}%
\bibitem [{\citenamefont {Gupta}\ and\ \citenamefont
  {Desai}(2018)}]{Gupta:2018hgm}%
  \BibitemOpen
  \bibfield  {author} {\bibinfo {author} {\bibfnamefont {S.}~\bibnamefont
  {Gupta}}\ and\ \bibinfo {author} {\bibfnamefont {S.}~\bibnamefont {Desai}},\
  }\href {\doibase 10.1016/j.aop.2018.09.017} {\bibfield  {journal} {\bibinfo
  {journal} {Annals Phys.}\ }\textbf {\bibinfo {volume} {399}},\ \bibinfo
  {pages} {85} (\bibinfo {year} {2018})},\ \Eprint
  {http://arxiv.org/abs/1810.00198} {arXiv:1810.00198 [astro-ph.CO]}
  \BibitemShut {NoStop}%
\bibitem [{\citenamefont {Magueijo}\ and\ \citenamefont
  {Smolin}(2002)}]{Magueijo_dsr}%
  \BibitemOpen
  \bibfield  {author} {\bibinfo {author} {\bibfnamefont {J.}~\bibnamefont
  {Magueijo}}\ and\ \bibinfo {author} {\bibfnamefont {L.}~\bibnamefont
  {Smolin}},\ }\href {\doibase 10.1103/PhysRevLett.88.190403} {\bibfield
  {journal} {\bibinfo  {journal} {Phys. Rev. Lett.}\ }\textbf {\bibinfo
  {volume} {88}},\ \bibinfo {pages} {190403} (\bibinfo {year} {2002})},\
  \Eprint {http://arxiv.org/abs/hep-th/0112090} {arXiv:hep-th/0112090 [hep-th]}
  \BibitemShut {NoStop}%
\bibitem [{\citenamefont {Amelino-Camelia}(2001)}]{AmelinoCamelia_dsr}%
  \BibitemOpen
  \bibfield  {author} {\bibinfo {author} {\bibfnamefont {G.}~\bibnamefont
  {Amelino-Camelia}},\ }\href {\doibase 10.1016/S0370-2693(01)00506-8}
  {\bibfield  {journal} {\bibinfo  {journal} {Phys. Lett.}\ }\textbf {\bibinfo
  {volume} {B510}},\ \bibinfo {pages} {255} (\bibinfo {year} {2001})},\ \Eprint
  {http://arxiv.org/abs/hep-th/0012238} {arXiv:hep-th/0012238 [hep-th]}
  \BibitemShut {NoStop}%
\bibitem [{\citenamefont {Amelino-Camelia}(2002)}]{AmelinoCamelia_dsr2}%
  \BibitemOpen
  \bibfield  {author} {\bibinfo {author} {\bibfnamefont {G.}~\bibnamefont
  {Amelino-Camelia}},\ }\href {\doibase 10.1038/418034a} {\bibfield  {journal}
  {\bibinfo  {journal} {Nature}\ }\textbf {\bibinfo {volume} {418}},\ \bibinfo
  {pages} {34} (\bibinfo {year} {2002})},\ \Eprint
  {http://arxiv.org/abs/gr-qc/0207049} {arXiv:gr-qc/0207049 [gr-qc]}
  \BibitemShut {NoStop}%
\bibitem [{\citenamefont {Amelino-Camelia}(2010)}]{AmelinoCamelia_dsr3}%
  \BibitemOpen
  \bibfield  {author} {\bibinfo {author} {\bibfnamefont {G.}~\bibnamefont
  {Amelino-Camelia}},\ }\href {\doibase 10.3390/sym2010230,
  10.1142/9789814287333_0006} {\bibfield  {journal} {\bibinfo  {journal}
  {Symmetry}\ }\textbf {\bibinfo {volume} {2}},\ \bibinfo {pages} {230}
  (\bibinfo {year} {2010})},\ \Eprint {http://arxiv.org/abs/1003.3942}
  {arXiv:1003.3942 [gr-qc]} \BibitemShut {NoStop}%
\bibitem [{\citenamefont {Sefiedgar}\ \emph {et~al.}(2011)\citenamefont
  {Sefiedgar}, \citenamefont {Nozari},\ and\ \citenamefont
  {Sepangi}}]{Sefiedgar_edt}%
  \BibitemOpen
  \bibfield  {author} {\bibinfo {author} {\bibfnamefont {A.~S.}\ \bibnamefont
  {Sefiedgar}}, \bibinfo {author} {\bibfnamefont {K.}~\bibnamefont {Nozari}}, \
  and\ \bibinfo {author} {\bibfnamefont {H.~R.}\ \bibnamefont {Sepangi}},\
  }\href {\doibase 10.1016/j.physletb.2010.11.067} {\bibfield  {journal}
  {\bibinfo  {journal} {Phys. Lett.}\ }\textbf {\bibinfo {volume} {B696}},\
  \bibinfo {pages} {119} (\bibinfo {year} {2011})},\ \Eprint
  {http://arxiv.org/abs/1012.1406} {arXiv:1012.1406 [gr-qc]} \BibitemShut
  {NoStop}%
\bibitem [{\citenamefont {Vacaru}(2012)}]{Vacaru_horava}%
  \BibitemOpen
  \bibfield  {author} {\bibinfo {author} {\bibfnamefont {S.~I.}\ \bibnamefont
  {Vacaru}},\ }\href {\doibase 10.1007/s10714-011-1324-1} {\bibfield  {journal}
  {\bibinfo  {journal} {Gen. Rel. Grav.}\ }\textbf {\bibinfo {volume} {44}},\
  \bibinfo {pages} {1015} (\bibinfo {year} {2012})},\ \Eprint
  {http://arxiv.org/abs/1010.5457} {arXiv:1010.5457 [math-ph]} \BibitemShut
  {NoStop}%
\bibitem [{\citenamefont {Blas}\ and\ \citenamefont
  {Sanctuary}(2011)}]{blas_horava}%
  \BibitemOpen
  \bibfield  {author} {\bibinfo {author} {\bibfnamefont {D.}~\bibnamefont
  {Blas}}\ and\ \bibinfo {author} {\bibfnamefont {H.}~\bibnamefont
  {Sanctuary}},\ }\href {\doibase 10.1103/PhysRevD.84.064004} {\bibfield
  {journal} {\bibinfo  {journal} {Phys. Rev.}\ }\textbf {\bibinfo {volume}
  {D84}},\ \bibinfo {pages} {064004} (\bibinfo {year} {2011})},\ \Eprint
  {http://arxiv.org/abs/1105.5149} {arXiv:1105.5149 [gr-qc]} \BibitemShut
  {NoStop}%
\bibitem [{\citenamefont {Horava}(2009{\natexlab{a}})}]{Horava}%
  \BibitemOpen
  \bibfield  {author} {\bibinfo {author} {\bibfnamefont {P.}~\bibnamefont
  {Horava}},\ }\href {\doibase 10.1088/1126-6708/2009/03/020} {\bibfield
  {journal} {\bibinfo  {journal} {JHEP}\ }\textbf {\bibinfo {volume} {03}},\
  \bibinfo {pages} {020} (\bibinfo {year} {2009}{\natexlab{a}})},\ \Eprint
  {http://arxiv.org/abs/0812.4287} {arXiv:0812.4287 [hep-th]} \BibitemShut
  {NoStop}%
\bibitem [{\citenamefont {Horava}(2009{\natexlab{b}})}]{Horava_2}%
  \BibitemOpen
  \bibfield  {author} {\bibinfo {author} {\bibfnamefont {P.}~\bibnamefont
  {Horava}},\ }\href {\doibase 10.1103/PhysRevD.79.084008} {\bibfield
  {journal} {\bibinfo  {journal} {Phys. Rev.}\ }\textbf {\bibinfo {volume}
  {D79}},\ \bibinfo {pages} {084008} (\bibinfo {year} {2009}{\natexlab{b}})},\
  \Eprint {http://arxiv.org/abs/0901.3775} {arXiv:0901.3775 [hep-th]}
  \BibitemShut {NoStop}%
\bibitem [{\citenamefont {Calcagni}(2010)}]{Calcagni_mf}%
  \BibitemOpen
  \bibfield  {author} {\bibinfo {author} {\bibfnamefont {G.}~\bibnamefont
  {Calcagni}},\ }\href {\doibase 10.1103/PhysRevLett.104.251301} {\bibfield
  {journal} {\bibinfo  {journal} {Phys. Rev. Lett.}\ }\textbf {\bibinfo
  {volume} {104}},\ \bibinfo {pages} {251301} (\bibinfo {year} {2010})},\
  \Eprint {http://arxiv.org/abs/0912.3142} {arXiv:0912.3142 [hep-th]}
  \BibitemShut {NoStop}%
\bibitem [{\citenamefont {Calcagni}(2012{\natexlab{a}})}]{Calcagni_mf2}%
  \BibitemOpen
  \bibfield  {author} {\bibinfo {author} {\bibfnamefont {G.}~\bibnamefont
  {Calcagni}},\ }\href {\doibase 10.4310/ATMP.2012.v16.n2.a5} {\bibfield
  {journal} {\bibinfo  {journal} {Adv. Theor. Math. Phys.}\ }\textbf {\bibinfo
  {volume} {16}},\ \bibinfo {pages} {549} (\bibinfo {year}
  {2012}{\natexlab{a}})},\ \Eprint {http://arxiv.org/abs/1106.5787}
  {arXiv:1106.5787 [hep-th]} \BibitemShut {NoStop}%
\bibitem [{\citenamefont {Calcagni}(2012{\natexlab{b}})}]{Calcagni_mf3}%
  \BibitemOpen
  \bibfield  {author} {\bibinfo {author} {\bibfnamefont {G.}~\bibnamefont
  {Calcagni}},\ }\href {\doibase 10.1007/JHEP01(2012)065} {\bibfield  {journal}
  {\bibinfo  {journal} {JHEP}\ }\textbf {\bibinfo {volume} {01}},\ \bibinfo
  {pages} {065} (\bibinfo {year} {2012}{\natexlab{b}})},\ \Eprint
  {http://arxiv.org/abs/1107.5041} {arXiv:1107.5041 [hep-th]} \BibitemShut
  {NoStop}%
\bibitem [{\citenamefont {Calcagni}(2017)}]{Calcagni_mf4}%
  \BibitemOpen
  \bibfield  {author} {\bibinfo {author} {\bibfnamefont {G.}~\bibnamefont
  {Calcagni}},\ }\href {\doibase 10.1140/epjc/s10052-017-4841-6} {\bibfield
  {journal} {\bibinfo  {journal} {Eur. Phys. J.}\ }\textbf {\bibinfo {volume}
  {C77}},\ \bibinfo {pages} {291} (\bibinfo {year} {2017})},\ \Eprint
  {http://arxiv.org/abs/1603.03046} {arXiv:1603.03046 [gr-qc]} \BibitemShut
  {NoStop}%
\end{thebibliography}%
\end{document}